\documentclass[a4paper,fleqn,usenatbib]{mnras}
\usepackage{txfonts} 
\usepackage[T1]{fontenc}
\usepackage{ae,aecompl}

\usepackage{graphicx}	

\usepackage{amsmath}	
\usepackage{amssymb}	
\usepackage{natbib}
\usepackage{subfig}
\usepackage{balance}
\usepackage{tablefootnote}
\usepackage{caption}
\usepackage{hyperref}
\usepackage{booktabs}
\usepackage{multirow}

\title[New Brown Dwarfs from SDSS-III MARVELS]{Exploring the Brown Dwarf Desert: New Substellar Companions from the SDSS-III MARVELS Survey}

\author[N. Grieves et al.]{Nolan Grieves,$^{1}$
\thanks{E-mail: ngrieves@ufl.edu}
Jian Ge$^{1}$,
Neil Thomas$^{1}$,
Bo Ma$^{1}$,
Sirinrat Sithajan$^{1}$,
Luan Ghezzi$^{2}$,
\newauthor
Ben Kimock$^{1}$,
Kevin Willis$^{1}$,
Nathan De Lee$^{3,4}$,
Brian Lee$^{1}$,
Scott W. Fleming$^{5,6}$,
\newauthor
Eric Agol$^{7}$,
Nicholas Troup$^{8}$,
Martin Paegert$^{4}$,
Donald P. Schneider$^{9,10}$,
\newauthor
Keivan Stassun$^{4}$,
Frank Varosi$^{1}$,
Bo Zhao$^{1}$,
Jian Liu$^{1}$,
Rui Li$^{1}$,
\newauthor
Gustavo F. Porto de Mello$^{11,12}$,
Dmitry Bizyaev$^{13,14}$,
Kaike Pan$^{13}$,
\newauthor
Let\'{\i}cia Dutra-Ferreira$^{11,12}$,
Diego Lorenzo-Oliveira$^{11,12}$,
Bas\'{\i}lio X. Santiago$^{11,15}$,
\newauthor
Luiz N. da Costa$^{2,11}$,
Marcio A. G. Maia$^{2,11}$,
Ricardo L. C. Ogando$^{2,11}$,
\newauthor
and E. F. del Peloso$^{11}$ \\
$^1$Department of Astronomy, University of Florida, Gainesville, FL 32611, USA \\
$^2$Observat\'orio Nacional, Rua General Jos\'e Cristino 77, S\~ao Crist\'ov\~ao, Rio de Janeiro, RJ 20921-400, Brazil \\
$^3$Department of Physics, Geology, and Engineering Tech, Northern Kentucky University, Highland Heights, KY 41099, USA \\
$^4$Department of Physics and Astronomy, Vanderbilt University, Nashville, TN 37235, USA \\
$^5$Space Telescope Science Institute, 3700 San Marin Dr, Baltimore, MD 21218, USA \\
$^6$CSRA Inc., 3700 San Martin Dr, Baltimore, MD 21218, USA \\
$^7$Department of Astronomy, University of Washington, Seattle, WA 98195, USA \\
$^8$Department of Astronomy, University of Virginia, Charlottesville, VA 22904-4325, USA \\
$^9$Department of Astronomy and Astrophysics, The Pennsylvania State University, University Park, PA 16802, USA \\
$^{10}$Center for Exoplanets and Habitable Worlds, The Pennsylvania State University, University Park, PA 16802, USA \\
$^{11}$Laborat\'orio Interinstitucional de e-Astronomia - LIneA, Rua Gereral Jos\'e Cristino 77, S\~ao Crist\'ov\~ao, Rio de Janeiro, RJ 20921-400, Brazil \\
$^{12}$Observat\'orio do Valongo, Universidade Federal do Rio de Janeiro,  Ladeira do Pedro Ant\^onio 43, Rio de Janeiro, RJ 20080-090, Brazil \\
$^{13}$Apache Point Observatory and New Mexico State University, P.O. Box 59, Sunspot, NM, 88349-0059, USA \\
$^{14}$Sternberg Astronomical Institute, Moscow State University, Moscow, 119992, Russia \\
$^{15}$Instituto de F\'{\i}sica, UFRGS, Caixa Postal 15051, Porto Alegre, RS 91501-970, Brazil}
\date{Accepted 2017 February 6. Received 2017 January 24; in original form 2016 September 6}

\pubyear{2017}

\begin{document}
\label{firstpage}
\pagerange{\pageref{firstpage}--\pageref{lastpage}}
\maketitle

\begin{abstract}
Planet searches using the radial velocity technique show a paucity of companions to solar-type stars within $\sim$5 AU in the mass range of $\sim$10 - 80 M$_{\text{Jup}}$. This deficit, known as the brown dwarf desert, currently has no conclusive explanation. New substellar companions in this region help asses the reality of the desert and provide insight to the formation and evolution of these objects. Here we present 10 new brown dwarf and two low-mass stellar companion candidates around solar-type stars from the Multi-object APO Radial-Velocity Exoplanet Large-Area Survey (MARVELS) of the Sloan Digital Sky Survey III (SDSS-III). These companions were selected from processed MARVELS data using the latest University of Florida Two Dimensional (UF2D) pipeline, which shows significant improvement and reduction of systematic errors over previous pipelines. The 10 brown dwarf companions range in mass from $\sim$13 to 76 M$_{\text{Jup}}$ and have orbital radii of less than 1 AU. The two stellar companions have minimum masses of $\sim$98 and 100 M$_{\text{Jup}}$. The host stars of the MARVELS brown dwarf sample have a mean metallicity of [Fe/H] = 0.03 $\pm$ 0.08 dex. Given our stellar sample we estimate the brown dwarf occurrence rate around solar-type stars with periods less than $\sim$300 days to be $\sim$0.56\%. 
\end{abstract}



\section{Introduction}

Brown dwarfs (BDs) are substellar objects ranging in mass between low-mass main-sequence stars and giant planets. They are generally defined as having masses above $\sim$13 M$_{\text{Jup}}$, the minimum mass required for an object to ignite deuterium burning in its core, and below $\sim$80 M$_{\text{Jup}}$, or 0.075 M$_\odot$, above which a dense object becomes sufficiently massive to fuse atomic hydrogen into helium in its core \citep{Boss2003}. The existence of BDs was first suggested by \citet{Kumar1962}, on the basis of theoretical calculations, and the realization that as stars of sufficiently low mass condense out of the interstellar medium, they will become dense enough to be supported by electron degeneracy pressure before they become hot enough to fuse hydrogen in their core. However, there was not a confirmed observation of a BD until 1995 \citep{Rebolo1995,Nakajima1995}. Technological advancements continue to produce new BD discoveries with increasing detail. These new data give a statistical sense of these objects and place observational constraints on their theoretical properties and proposed formation mechanisms.

One of the most important observational constraints on BD formation is known as the ``brown dwarf desert'', or the lack of companions within the mass range of  $\sim$10-80 M$_{\text{Jup}}$ with orbital radii less than $\sim$5 AU around solar-type stars \citep{MarcyButler2000}. \citet{GretherLineweaver2006} investigated the significance of the brown dwarf desert by analyzing companions with orbital periods less than 5 years around Sun-like stars. \citet{GretherLineweaver2006} found that $\sim$16\% of Sun-like stars have close (P $<$ 5 yr) companions more massive than Jupiter, but $<$1\% are BDs. The minimum number of companions per unit interval in log mass, or the driest part of the desert, is at M $\sim$ 31 M$_{\text{Jup}}$ \citep{GretherLineweaver2006}. \citet{MaGe2014} created a catalogue of all known brown dwarf companions around solar-type stars, which totaled 62 BDs. Through statistical analysis \citet{MaGe2014} identified a depletion of brown dwarfs with masses between 35 and 55 M$_{\text{Jup}}$ and periods $<$ 100 days, representing the `driest' part in the brown dwarf desert. 

\citet{Troup2016} consider stellar and substellar companion candidates in the Sloan Digital Sky Survey III (SDSS-III; \citet{Eisenstein2011}) Apache Point Observatory Galactic Evolution Experiment (APOGEE-1; \citet{Majewski2015}), which observed 146,000 stars with typical radial velocity (RV) precision of $\sim$100-200 m s$^{-1}$. Of these, 14,840 had at least eight visits; this subset had a median baseline of 384 days, median of 13 visits, and a median Signal-to-Noise ratio (S/N) of 12.2 per visit. With a final count of 382 candidate companions (including 112 BD companion candidates, 71 orbiting evolved stars), \citet{Troup2016} find evidence for the brown dwarf desert, but only out to $\sim$0.5-1 AU, with no indication of a desert beyond. However, of the 382 stars in their sample, only 36 are solar-type (main-sequence with 5000K $<$ T$_{\text{eff}}$ $<$ 6000K) stars. \citet{Troup2016} argue that this result agrees with the work of \citet{Guillot2014} who find that the the BD desert may not be as ``dry" as initially thought when considering stars more massive than the Sun, as well as evidence that the BD desert is a special case for solar-mass stars \citep{DucheneKraus2013}. \citet{DucheneKraus2013} suggest tidal interaction with host stars may shape the brown dwarf desert. While assessing the characteristic timescale for orbital decay, \citet{DamianiDiaz2016} find that F-type stars may host massive companions for a significantly longer time than G-type stars for orbital periods less than five days; however, they conclude that brown dwarf occurrence should largely be unaffected by tidal decay for orbital periods greater than ten days, regardless of host mass.

Debate continues on valid formation mechanisms of BDs and whether or not different formation mechanisms dominate in different mass, period, and eccentricity regimes. Three main mechanisms are advocated in the field: (i) the ejection of protostellar embryos from their natal cloud cores \citep{Boss2001,Reipurth2001}, (ii) the fragmentation of protostellar discs \citep{StamatellosWhitworth2009,Stamatellos2011}, or (iii) collapsing molecular cloud cores \citep{HennebelleChabrier2013,Hopkins2012}, i.e., the same process as Sun-like stars. 

Through their statistical study \citet{MaGe2014} reported that BD companions with minimum masses below 42.5 M$_{\text{Jup}}$ have an eccentricity distribution consistent with that of massive planets, while those with masses above 42.5 M$_{\text{Jup}}$ have an eccentricity distribution consistent with that of binaries. \citet{MaGe2014} also find that host stars of BD companions are not metal rich and have a significantly different metallicity distribution compared to host stars of giant planets, suggesting a formation mechanism different from the core-accretion \citep{IdaLin2004} scenario. \citet{MaGe2014} conclude that BD companions with minimum masses below 42.5 M$_{\text{Jup}}$ likely form in a protoplanetary disc through the disc gravitational instability scenario, and their eccentricity is excited through scattering, while BD companions with minimum masses above 42.5 M$_{\text{Jup}}$ likely form in the same way as stars, through molecular cloud fragmentation. \citet{Troup2016} find 14 candidate BD companions around metal-poor stars ([Fe/H] $<$ -0.5), also challenging the core-accretion formation model for BDs, which they find unsurprising given that \citet{Carney2003} observed that lower metallicity populations generally have a higher number of binaries. 

Here we present new BD candidate companions using data from the Sloan Digital Sky Survey III (SDSS-III) Multi-object APO Radial Velocity Exoplanet Large-area Survey (MARVELS; \citet{Ge2008}). MARVELS observed $\sim$3300 stars with $\geq$10 RV measurements. Unlike APOGEE, MARVELS primarily targeted solar-like dwarfs, and thus, is complementary to the work done by APOGEE, which primarily targets giant stars. The latest University of Florida Two Dimensional (UF2D) pipeline has RV precisions of $\sim$60-100 m s$^{-1}$ \citep{Thomas2015UFphd}. With this new pipeline we find 12 new low-mass companion candidates around main-sequence stars, 10 of which are in the BD mass regime. 

In $\S$ \ref{sec:marvdat} we discuss the detailed processing of the MARVELS data required to obtain RV measurements, including the history of its data pipelines. In $\S$ \ref{sec:vet} we describe our companion vetting procedure designed to minimize false positive selection. In $\S$ \ref{sec:host} we describe our procedure for obtaining host star parameters. In $\S$ \ref{sec:res} we present our results and conclusions including our brown dwarf companion parameters.

\section{MARVELS DATA}\label{sec:marvdat}

\begin{figure*}
	\centering
	 \subfloat[HIP 45859]{
	 \includegraphics[width=0.66\columnwidth]{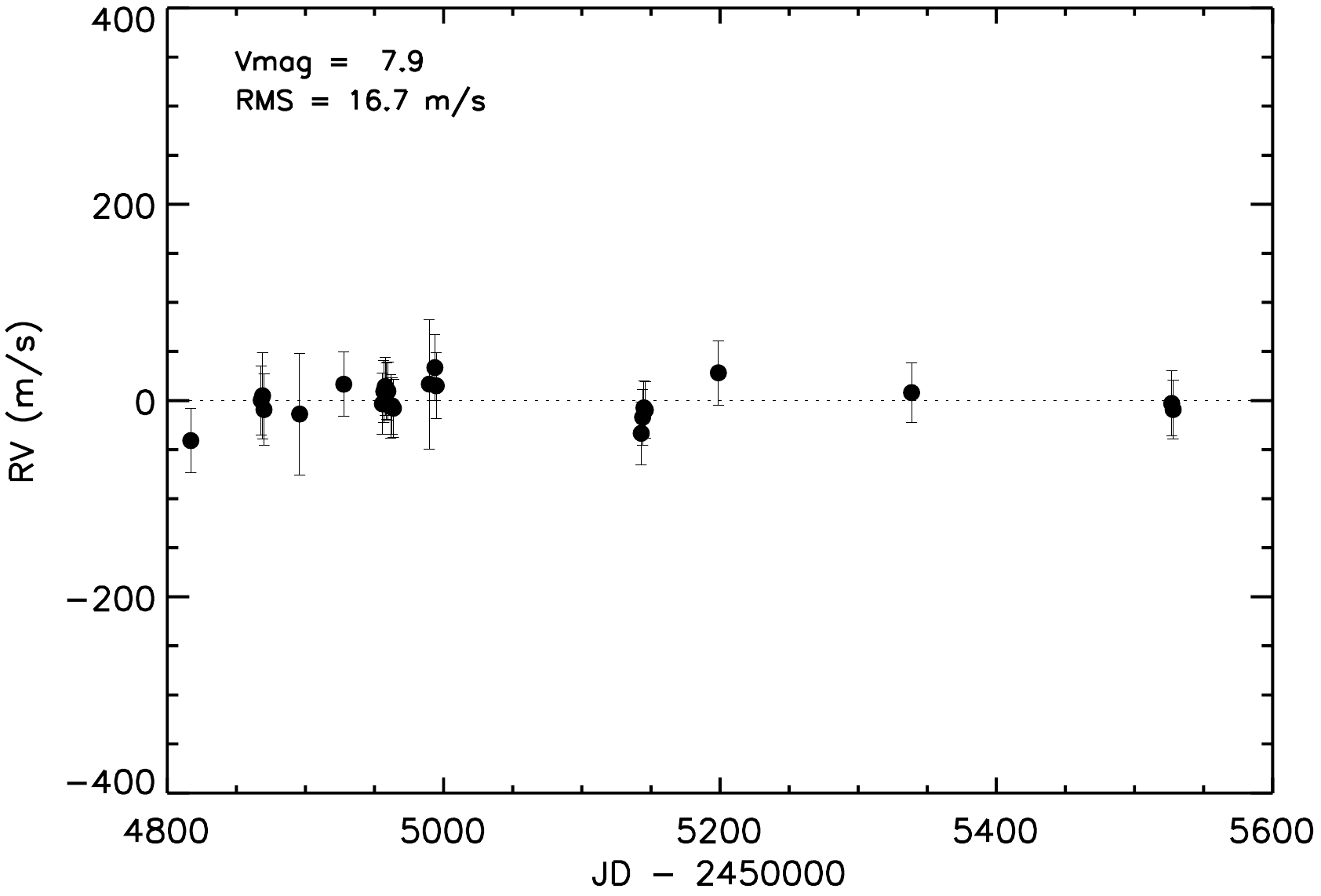}
	 }
	\hfill
        	\centering
   	 \subfloat[TYC 3970-01165-1]{
	 \includegraphics[width=0.66\columnwidth]{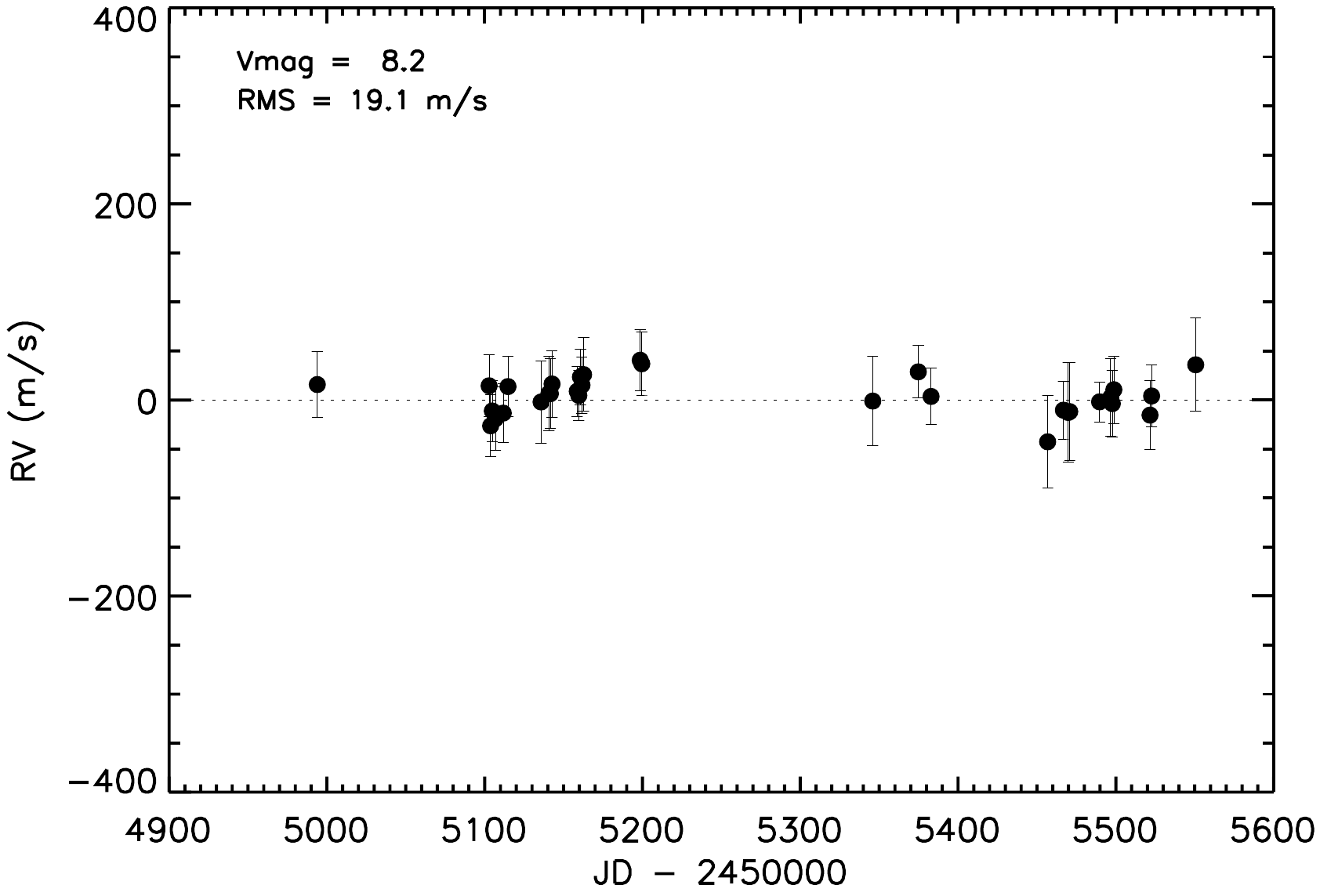}
	 }
	\hfill
	\centering
	 \subfloat[HIP 91739]{
	 \includegraphics[width=0.66\columnwidth]{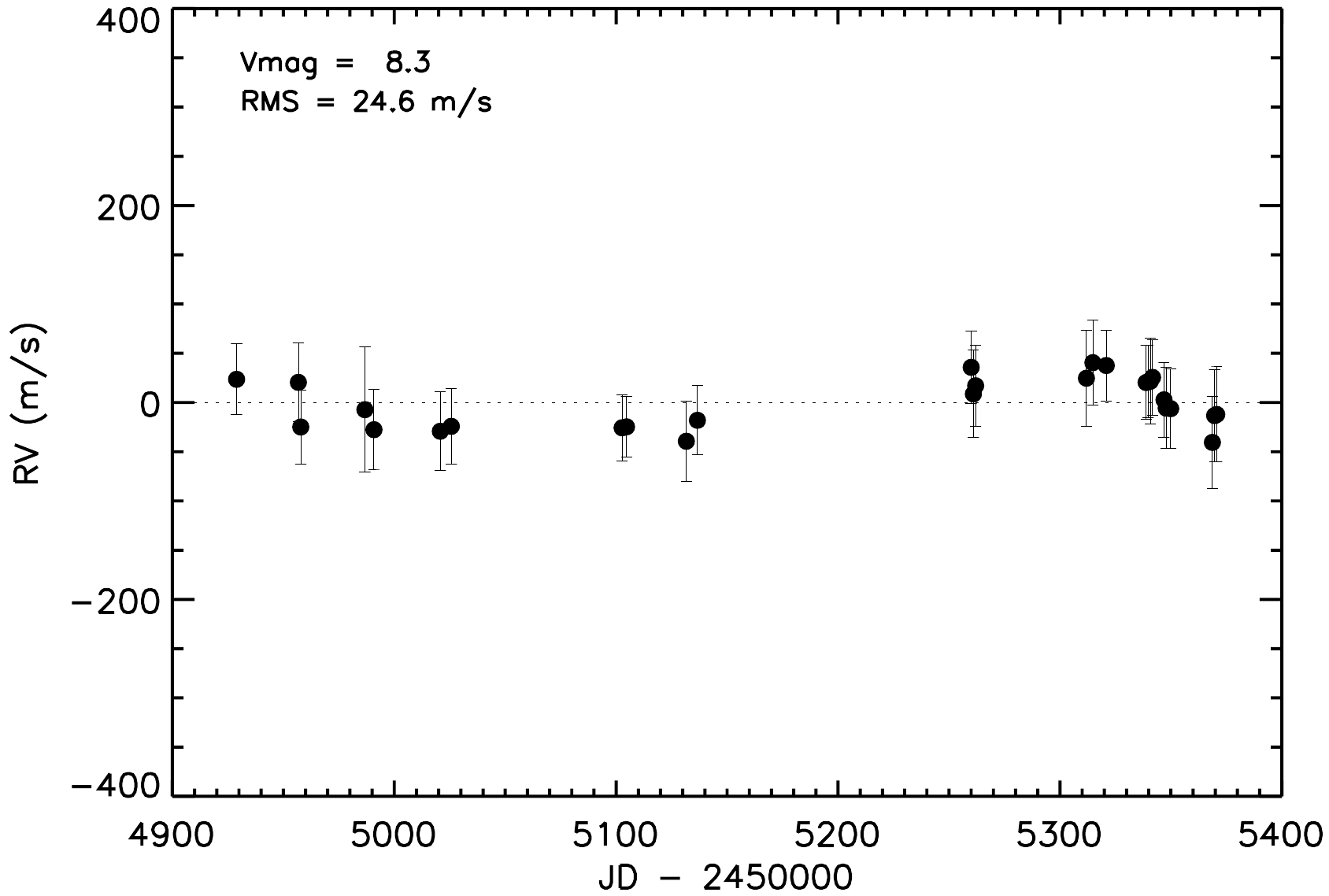}
	 }
	\hfill
        	\centering
   	 \subfloat[HIP 86151]{
	 \includegraphics[width=0.66\columnwidth]{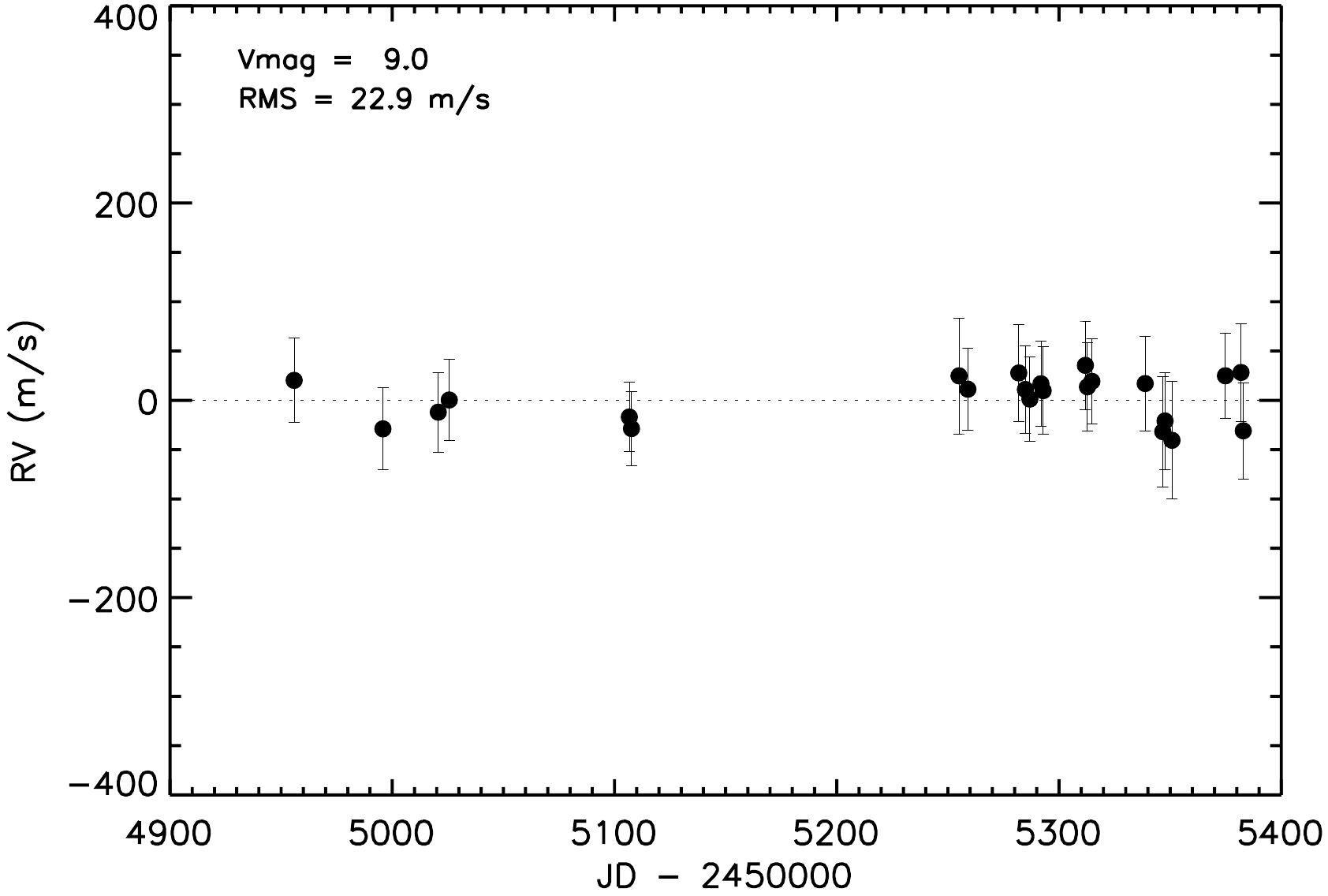}
	 }	 
	\hfill
	 \centering
	  \subfloat[TYC 3134-00564-1]{
	 \includegraphics[width=0.66\columnwidth]{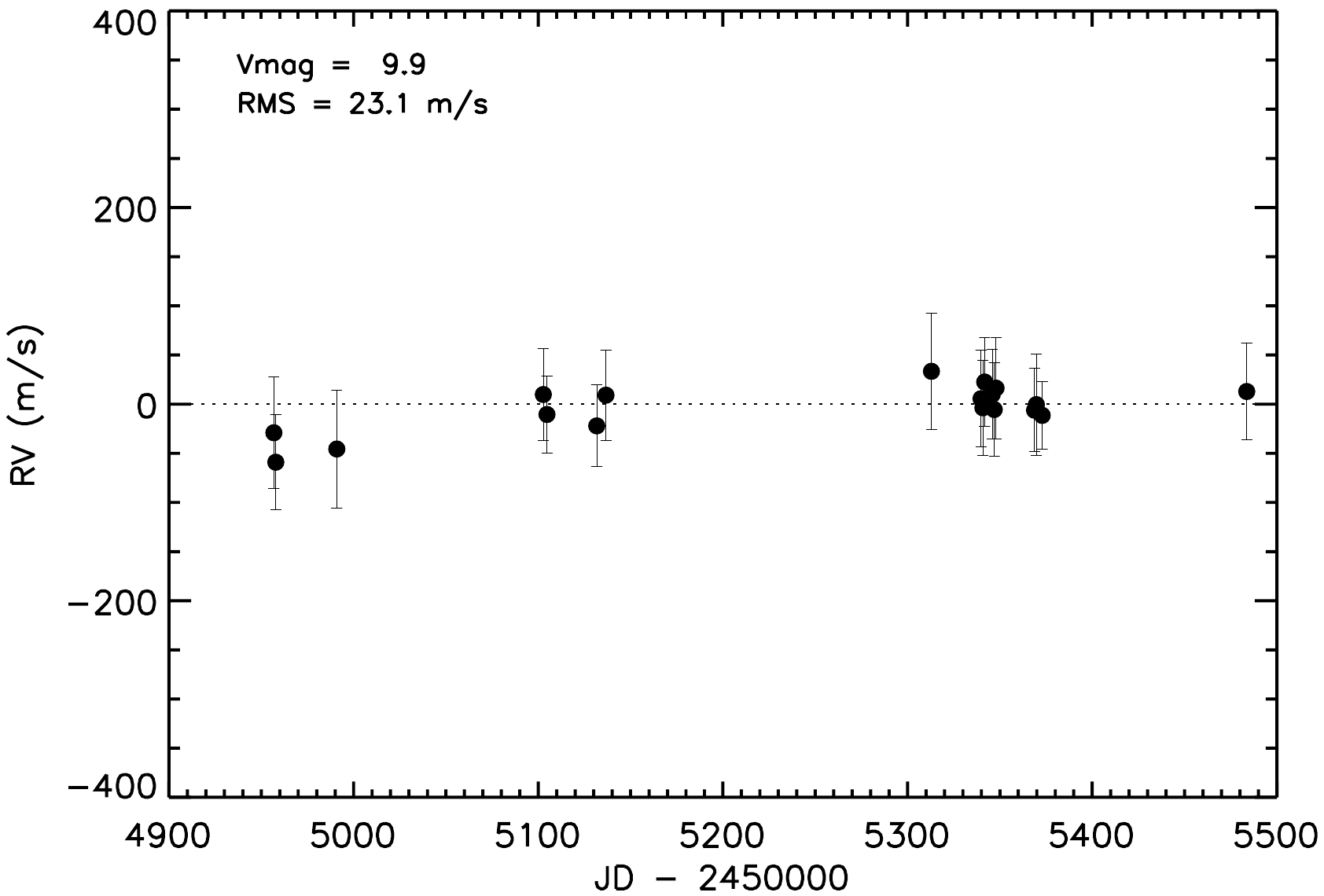}
	 }	 
	 \hfill
	 \centering
	  \subfloat[TYC 3010-01139-1]{
	 \includegraphics[width=0.66\columnwidth]{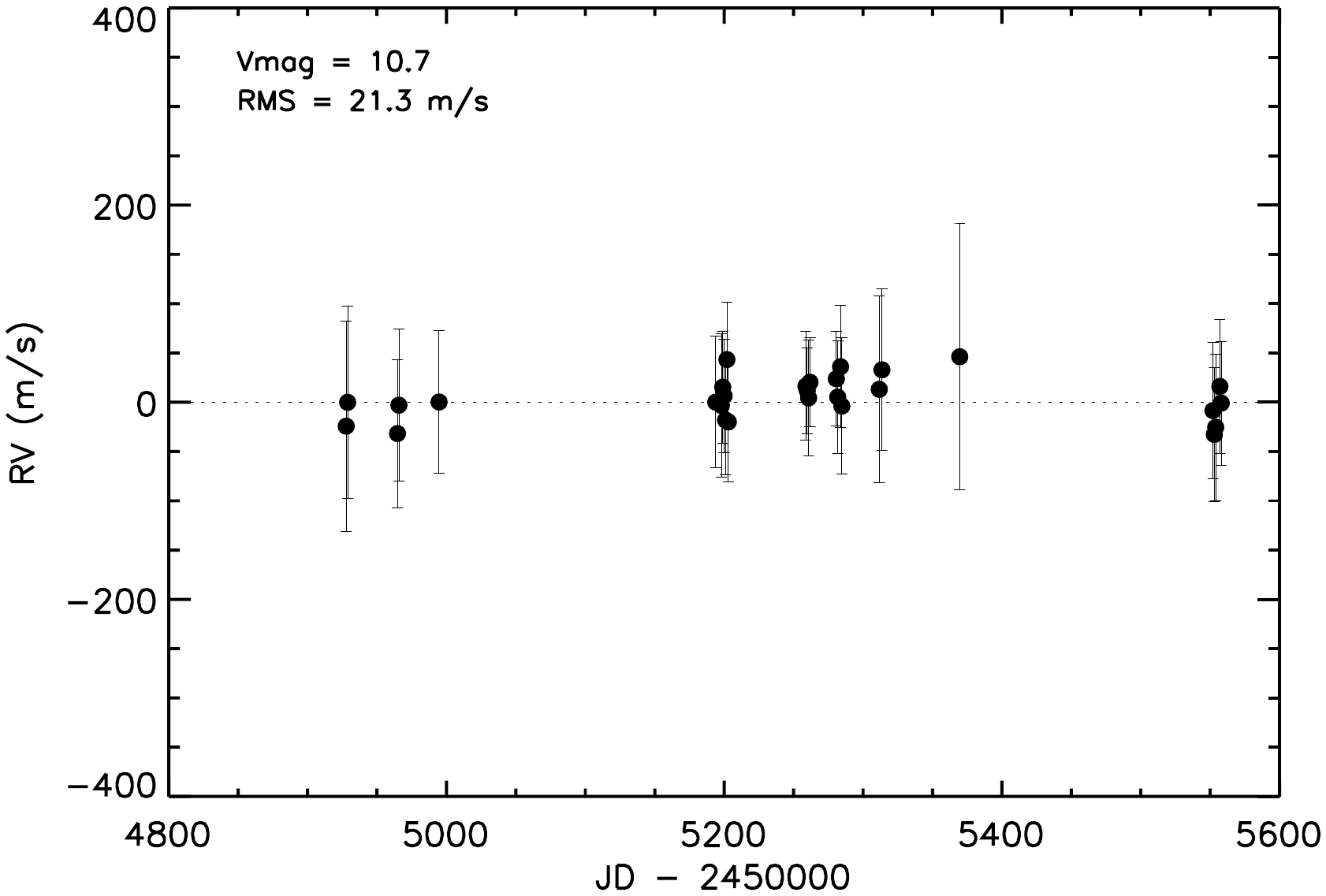}
	 }	 
	 \hfill
	 \centering
	  \subfloat[TYC 3542-01419-1]{
	 \includegraphics[width=0.66\columnwidth]{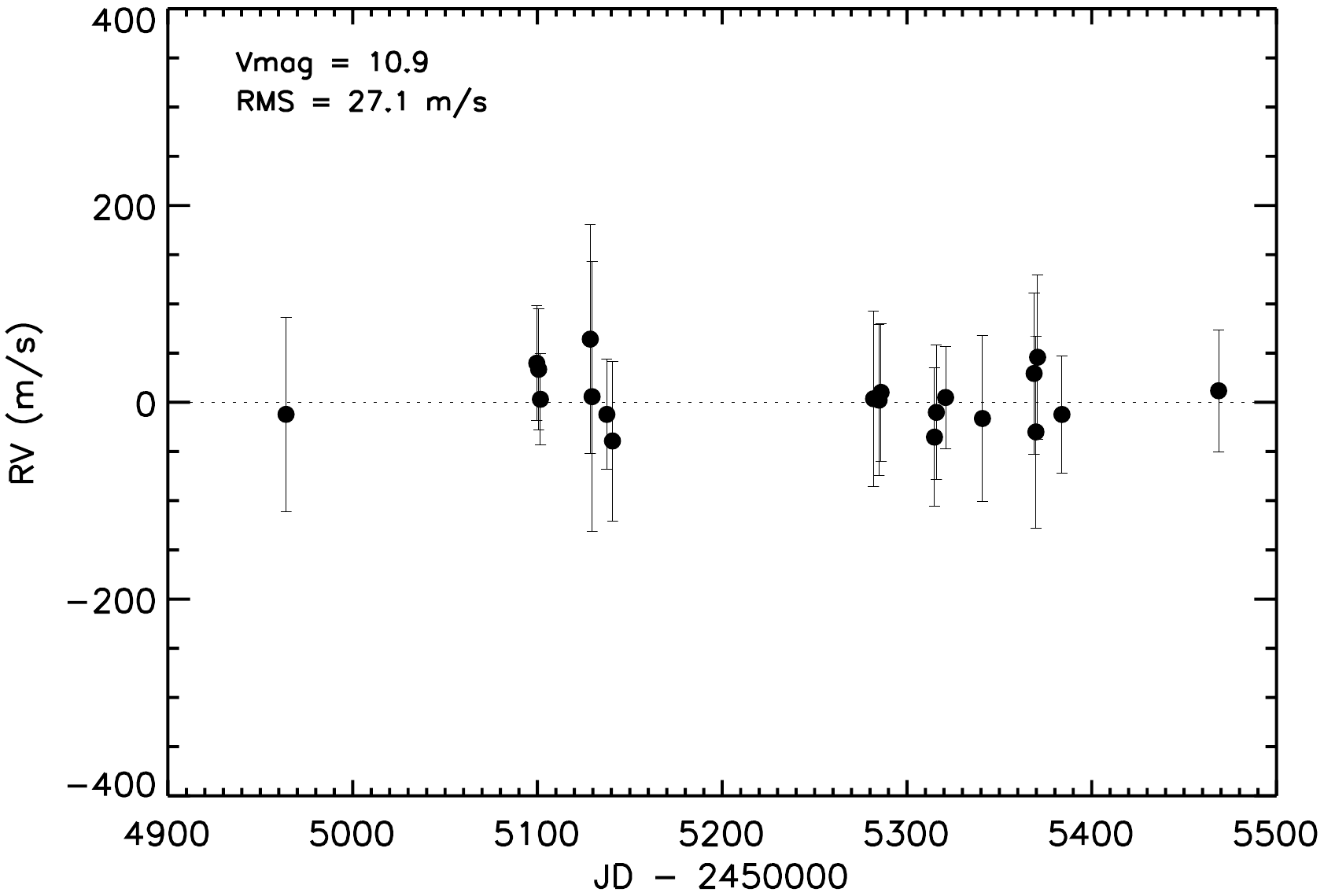}
	 }	 
	\hfill
	 \centering
	 \subfloat[TYC 2950-01319-1]{
	 \includegraphics[width=0.66\columnwidth]{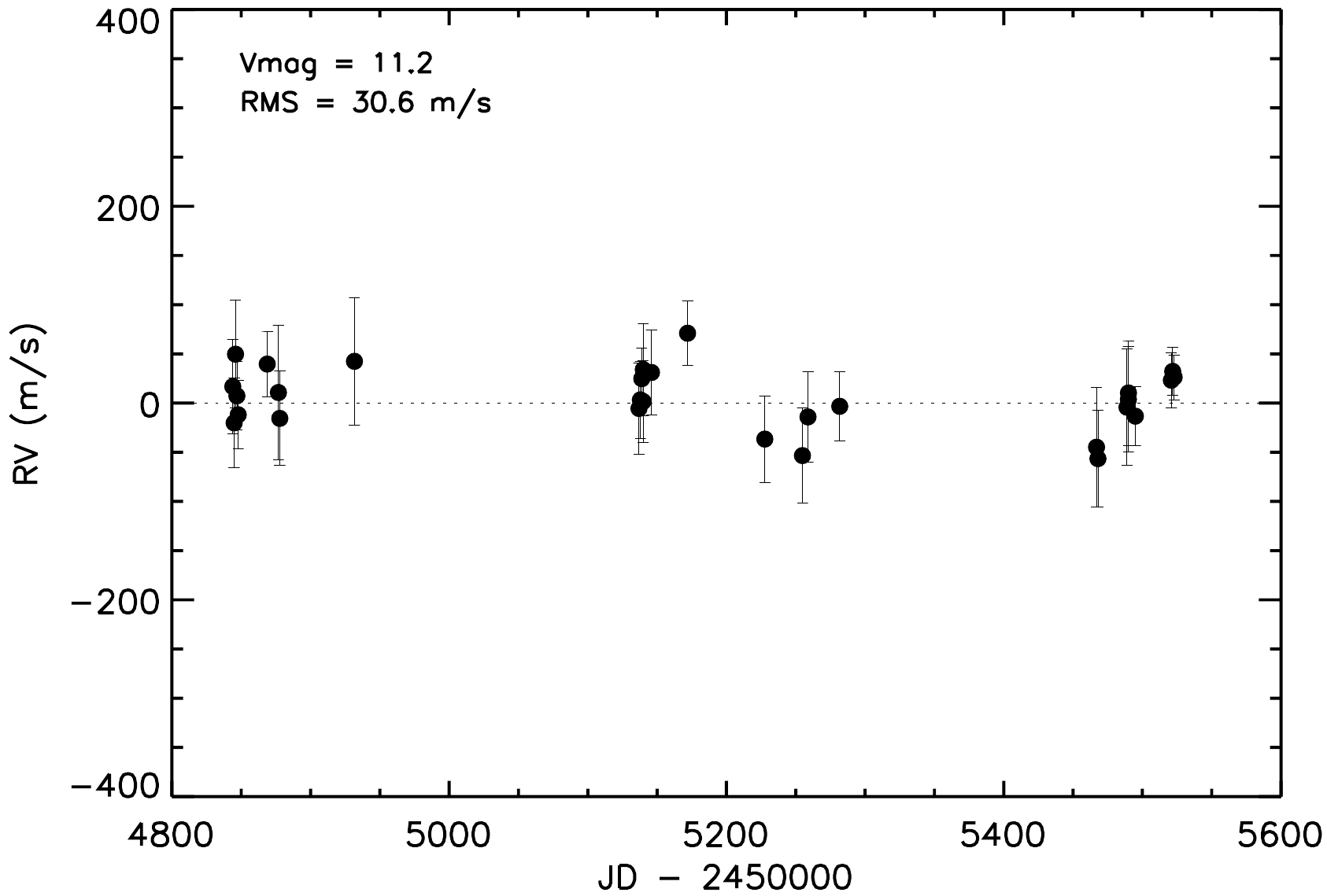}
	 }	 
	\hfill
	 \centering
	 \subfloat[TYC 2013-00057-1]{
	 \includegraphics[width=0.66\columnwidth]{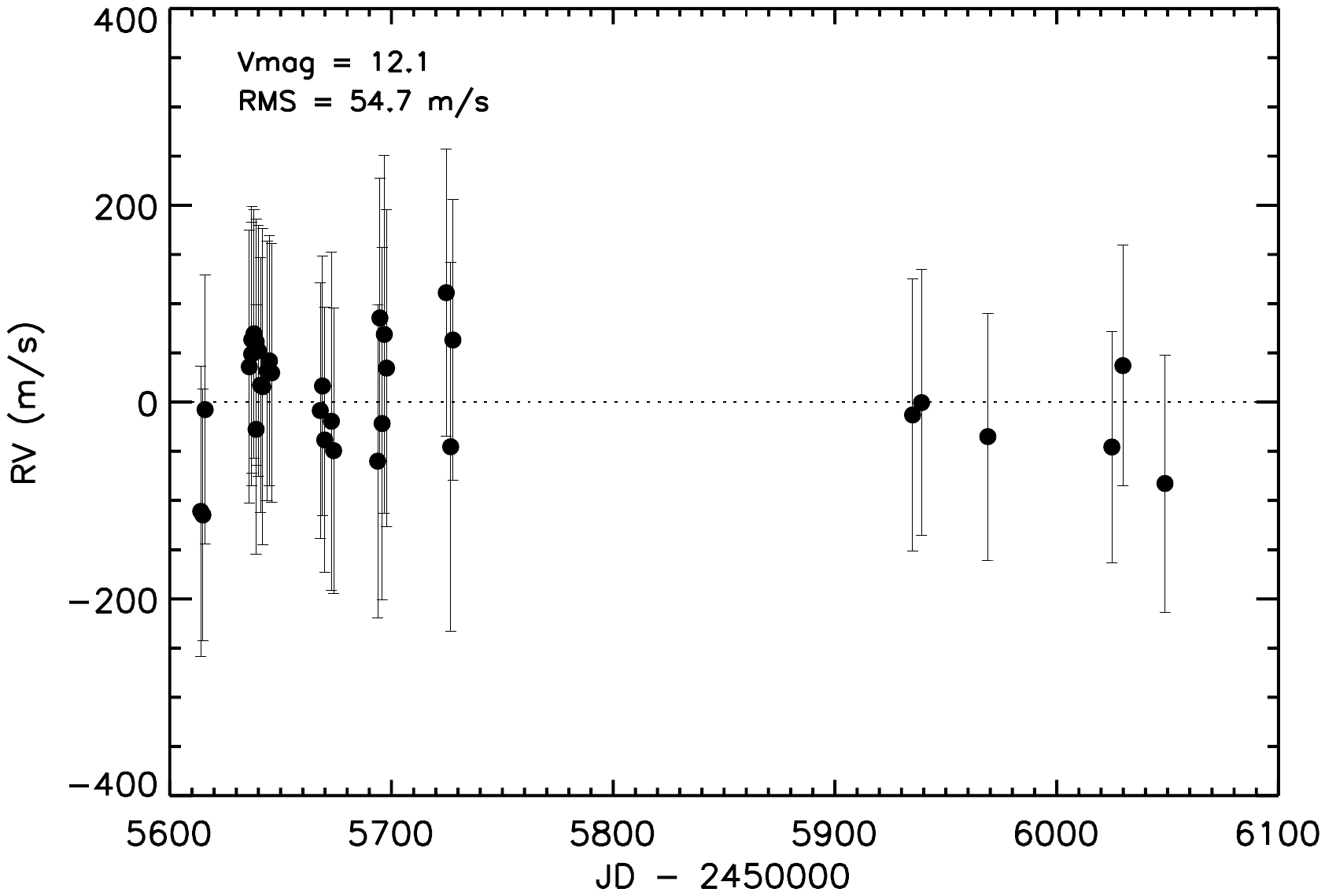}
	 }	 
    	\caption{RV plots showing stable stars of various magnitudes with data from the new UF2D pipeline. Guide Star Catolog 2.3 \citep[GSC 2.3;][]{Lasker2008} V magnitudes are shown in each RV plot for the corresponding star. Each plot displays RV RMS, exhibiting typical RV precision of the new UF2D pipeline for some of the brightest (V $\sim$ 8) and dimmest (V $\sim$ 12) stars in the MARVELS survey. The horizontal dashed line shows 0 m s$^{-1}$ RV as a reference.}
	\label{fig:rms}
\end{figure*}

MARVELS was in operation from 2008 - 2012 as a part of the 6-year SDSS-III program \citep{Eisenstein2011}, a continuation of the Sloan Digital Sky Survey \citep[SDSS;][]{York2000}. The goal of MARVELS is to detect a large sample of short-to-intermediate period giant planets obtained from a large, well-characterized, and homogeneous survey of stars with known properties. Radial velocity (RV) observations were taken using the SDSS 2.5 m telescope at Apache Point Observatory \citep{Gunn2006}. MARVELS measures RVs using a fiber-fed dispersed fixed delay interferometer (DFDI) with a moderate dispersion spectrograph. Specifically, this set-up combines a Michelson interferometer and a medium resolution (R $\sim$ 11,000) spectrograph, which overlays interferometer fringes on a long-slit stellar spectrum \citep{Ge2009,Eisenstein2011}. Unlike echelle spectrographs, which require many orders and a large detector area to achieve high RV precision, this unique setup allows simultaneous observation of 60 stars with moderate to high RV precision.

\citet{vanEyken2010} presents a complete overview of the principles and physics behind the DFDI method. The general creation process of the MARVELS DFDI spectra is explained by \citet{Ge2002}. Starlight from the telescope or light from a calibration source, a Tungsten lamp shining through a temperature-stabilized Iodine gas cell (TIO), is fed into an optical fiber. The incoming beam is split in two by a Michelson-type interferometer. A parallel glass plate creates a fixed optical delay between the two plates of the interferometer. A virtual white fringe pattern is created and re-imaged onto the entrance slit of the spectrometer, and is then dispersed by the medium-resolution grating \citep{Ge2002}. The dispersed fringes are recorded on a detector. The resulting fringes overlaid on the 2D spectra contain high spatial frequency Doppler information allowing for precise RV measurements. \citet{Ge2006} displayed the ability of the DFDI method with the detection of a giant planet of minimum mass $\sim$0.488 M$_{\text{Jup}}$ (HD 102195b) using the precursor to the MARVELS instrument, the Exoplanet Tracker.

Early analysis of MARVELS data showed good instrumental performance over timescales of less than a month. By observing a RV stable star or a known planet hosting star, \citet{Ge2009} found the root mean square (RMS) errors vary between only 1-2 times the photon noise for RV measurements within one month. However, when analyzing reference stars over multiple month periods, \citet{Ge2009} found RMS errors as large as a factor of 5 times higher than the photon limiting error. They acknowledged the need for further consideration of long-term effects such as distortion, illumination, and large barycentric velocity. During MARVELS data pipeline development, \citet{Lee2011} found systematic RV errors in the form of month-to-month offsets at the level of tens to hundreds of m s$^{-1}$. \citet{Lee2011} inferred that the systematic errors may be due to imperfections in the detailed preprocessing of the images. 

\citet{Alam2015} provide an overview of the history of the MARVELS instrument and the associated data reduction. The original pipeline results are presented in the SDSS Data Release 11 (DR11) as the ``cross correlation function" (CCF) and DFDI reductions, which revealed the large instrumental errors and the requirement of a new analysis method. This discovery prompted the creation of a new refurbished pipeline designed to understand and correct the systematic errors \citep{Thomas2016}. This refurbished pipeline is presented in the SDSS Data Release 12 (DR12) as the ``University of Florida One Dimensional" (UF1D) reductions. \citet{Alam2015} summarize the differences between the CCF+DFDI and UF1D reductions including changes in the calibration of the wavelength solution on the detector, identification and extraction of each spectrum, and the measurement of the slant of the interferometric comb and of the resulting interference pattern of the absorption-line features. The DR12 UF1D reductions do not show great improvements over the DR11 CCF+DFDI reductions; however, the UF1D reductions resulted in a better understanding of the systematic errors of the instrument and thus allowed more accurate error estimations for individual targets.

\citet{Thomas2016} improved upon the DR12 UF1D results by adopting a `Lucky RV' method, where statistically preferred regions of the spectra  are used in the determination of overall RV and poorer regions are discarded once performance deteriorates for a given target (each star goes through $\sim$1,000 iterations to randomly select a specific combination and number of regions that create the best RV performance). Using this new method combined with two new filters, which result in an overall rejection of $\sim$20\% of all observations, \citet{Thomas2016} was able to greatly reduce large long-term RV errors that plagued previous pipelines. The improved UF1D pipeline with Lucky RV and filters results in a RV precision of $\sim$66-114 m s$^{-1}$ (photon limits $\sim$27-90 m s$^{-1}$) for 8-12 magnitude stable stars, respectively. The UF1D pipeline reductions identified that optical fiber illumination changes, likely isolated to fiber couplings in the gang connector, caused variations between the stellar and calibration fiber illuminations over the lifetime of the survey, causing the largest measurement error in the MARVELS RVs. However, the final UF1D pipeline still does not completely remove the systematic offsets from all targets, with an overall influence from offset errors having a median of 64.3 m s$^{-1}$; the Lucky RV method also has the disadvantage of excluding $\sim$20\% of observations \citet{Thomas2016}.

\begin{figure}
\center
\includegraphics[width=\columnwidth]{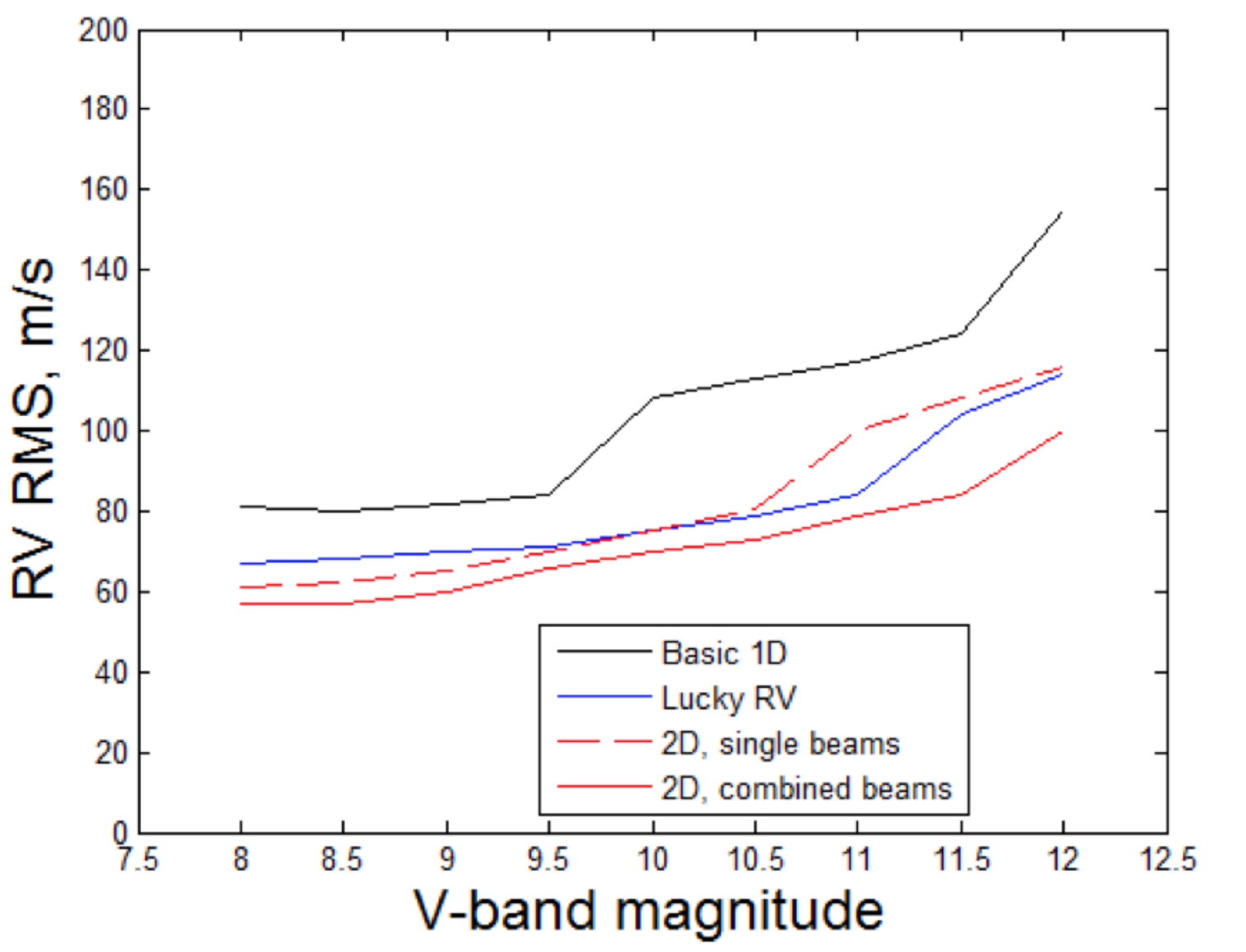}
\caption{MARVELS pipeline performances for RV stable stars across the entire survey. The Lucky RV pipeline improves upon the basic UF1D pipeline by choosing only specific regions in each spectra that produce the best RV performance and excluding $\sim$20\% of observations altogether. The UF2D single beam pipeline performance rivals the Lucky RV pipeline performance, but has the advantage of including all portions of each spectra, increasing signal-to-noise. While previous pipelines were too unstable to combine adjacent beams of the same star due to systematic errors, the UF2D pipeline has the capability to combine beams resulting in the overall best MARVELS pipeline performance.}
\label{fig:pipelinerv}
\end{figure}

After the refinement of the UF1D pipeline, a new ``University of Florida Two Dimensional" (UF2D) pipeline was developed in order to use the interferometric power of MARVELS and take full advantage of the instrument's capabilities. Rather than relying on the underlying physics of interferometry, as previous 2D DFDI pipelines generally did \citep{Ge2002,vanEyken2010,Wang2011}, the UF2D pipeline uses a simpler and more visually understandable method, which is outlined by \citet{Thomas2015UFphd} and \citet{Wang2012j}. Two new innovations, specifically 2D modeling of the underlying fringe density distortion and the correction of temporal distortions, have led to current RV performance of $\sim$60-96 m s$^{-1}$ for 8-12 magnitude RV stable stars, respectively. Figure \ref{fig:rms} shows UF2D RV results for stable stars of various magnitudes. However, the main improvement of the UF2D pipeline is the lack of large errors found over months and years that have previously plagued the survey. Most notably, combining the results from adjacent beams of the same star improves RV accuracy, which was not the case with 1D data due to systematic errors dominating both beams. Figure \ref{fig:pipelinerv} displays the overall RV performance for each MARVELS pipeline. We use these new UF2D RV results for our current search of substellar companions.

\section{Candidate Vetting Procedure}\label{sec:vet}

The MARVELS survey has already identified two low-mass stellar companions to solar-type stars (MARVELS-2b, \citet{Fleming2012}; MARVELS-3b, \citet{Wisniewski2012}), as well as three BD companions (MARVELS-4b, \citet{Ma2013}; MARVELS-5b, \citet{Jiang2013}; MARVELS-6b, \citet{DeLee2013}). \citet{Ma2016} announced the discovery of a system with both a giant planet (MARVELS-7b) and BD (MARVELS-7c) orbiting the primary star of a close binary system. The MARVELS Pilot Project also showed success in this domain with the detection of a BD \citep{Fleming2010} and two new eclipsing binaries \citep{Fleming2011}. However, the MARVELS survey has had difficulties in this area as well. \citet{Lee2011} announced the discovery of a $\sim$28.0 M$_{\text{Jup}}$ companion (MARVELS-1b) to TYC 1240-00945-1, but this was later determined to be a false positive in the form of a face-on double-lined binary star \citep{Wright2013}. \citet{Mack2013} warns of a similar situation where the MARVELS target TYC 3010-01494-1, a highly eccentric spectroscopic binary, initially appeared as a single star with a substellar companion with minimum mass of $\sim$50 M$_{\text{Jup}}$ despite an extensive amount of observation and analysis. 

MARVELS observed $\sim$5,500 stars during its 4-year survey, and produced 3,257 stars with 10 or more valid RV observations in our latest UF2D pipeline. Previous results show that finding substellar companions in this dataset is susceptible to error. Examining each star can also be extremely time consuming and will likely lead to biases due to expectations of the person examining the individual target. We created an automatic method to search for substellar companions that would minimize bias and time spent on each target. We focus on finding companions around solar-type stars (dwarfs/subgiants) and apply several selection criteria to avoid false positives.

\subsection{Defining a Stellar Sample}

We determine a giant or dwarf/subgiant designation for each star by a $J$-band reduced proper motion (RPM$_{J}$) cut, detailed in \citet{Paegert2015}. Here we use the same method as \citet{Paegert2015} and compute RPM$_{J}$ as follows:
\begin{equation}
\mu = \sqrt{(cos\ d *  \mu_{r})^{2} + \mu_{d}^{2}}
\end{equation}
\begin{equation}
\text{RPM}_{J} = J + 5\ log(\mu),
\end{equation}
where $J$ is the star's 2MASS Survey \citep{Skrutskie2006} $J$-band magnitude and $\mu_{r}$, $\mu_{d}$, and $d$ are Guide Star Catolog 2.3 \citep[GSC 2.3;][]{Lasker2008} proper motions in right ascension and declination (in arc seconds per year) and declination, respectively. An empirical RPM$_{J}$ cut described in \citet{Collier2007} is applied:
\begin{equation}
\begin{split}
y = -58 + 313.42(J - H) - 583.6(J - H)^{2} \\
+ 473.18(J - H^{3} - 141.25(J - H)^{4},
\end{split}
\end{equation} 
where $H$ is the star's 2MASS Survey $H$-band magnitude.

\begin{figure}
\center
\includegraphics[width=\columnwidth]{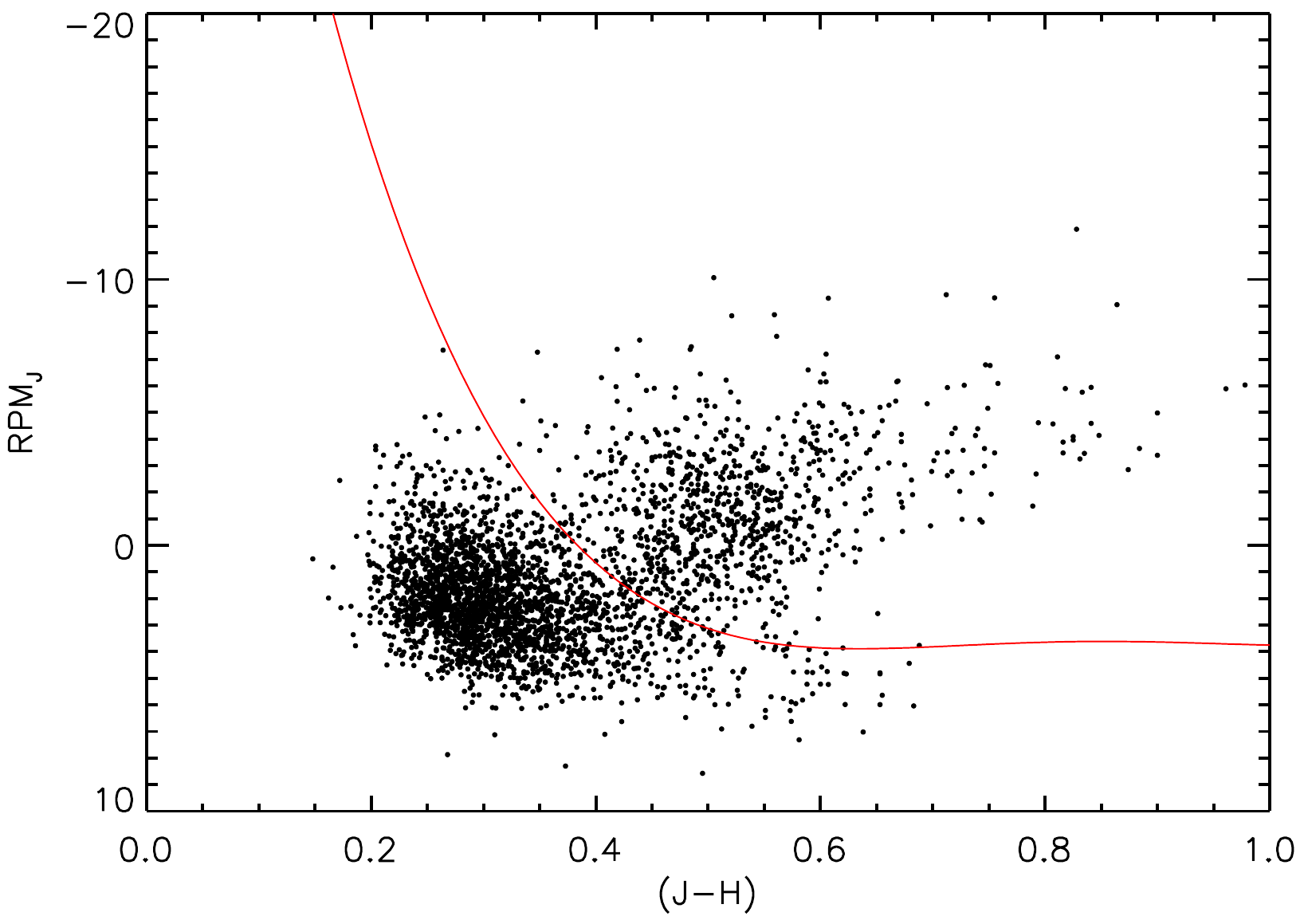}
\caption{$J$-band reduced proper motion (RPM$_{J}$) vs J-H magnitude for the 3,257 stars in the MARVELS survey that have $\ge$10 valid RV observations in the UF2D pipeline. The red line shows the RPM$_{J}$ cut to determine giant or dwarf/subgiant designation. Stars above the red line are designated as RPM$_{J}$ giants and those below as dwarf/subgiants. In this sample we classify 2,378 stars as dwarf/subgiants and 879 as giants.}
\label{fig:rpmcut}
\end{figure}

Stars with RPM$_{J}$ $\leq$ $y$ are regarded as RPM$_{J}$-dwarfs and stars with RPM$_{J}$ $>$ $y$ as RPM$_{J}$-giants. With MARVELS data this method has a giant contamination rate of $\sim$4\% and subgiants are mixed in with the ``dwarf" sample at a level of 20\%-40\% \citep{Paegert2015}. Figure \ref{fig:rpmcut} shows the RPM$_{J}$ cut on our sample of 3,257 stars with $\ge$10 RV observations, resulting in 2,378 stars classified as dwarf/subgiant. Our final stellar sample consists of 2,340 stars that have $\ge$10 RV observations, are RPM$_{J}$ dwarf/subgiants, and have a time baseline longer than 300 days. These 2,340 stars have a mean baseline of $\sim$582 days and a median of 26 RV observations. 

\subsection{Initial Companion Selection} \label{sec:initcomsec}
We first analyze our sample of 2,340 stars by determining if they are RV variable using the criteria defined in \citet{Troup2016}:

\begin{equation}
\Sigma_{RV} = \text{stddev}\Bigg(\frac{\textbf{v} - \widetilde{\text{v}}}{\sigma_{\textbf{v}}}\Bigg) \ge 2.5,
\end{equation}
where $\textbf{v}$ and $\sigma_{\textbf{v}}$ are the RV measurements and their uncertainties, and $\widetilde{\text{v}}$ is the median RV measurement for the star. 653 stars in our sample are RV variable with these criteria. We find a False Alarm Probability (FAP) of $\sim$0.2\% for these 653 stars having real RV variation. We find a FAP by simulating RVs for each star based on a Gaussian distribution with a sigma equal to the mean RV error for that star. We calculate $\Sigma_{RV}$ for each star and find the number of simulated stars with $\Sigma_{RV}$ $\ge$ 2.5. We run this simulation 100 times and find a mean of 1.3 stars with $\Sigma_{RV}$ $\ge$ 2.5, corresponding to a FAP of (1.3/653)$\sim$0.2\%. The MARVELS UF2D data has $\Sigma_{RV}$ values that peak at $\sim$0.6, which is lower than the expected value of 1 given correct error bars, suggesting a rather conservative estimate of our errors. The UF2D RV error designation for this study is similar to the UF1D pipeline error designation, detailed in \citep{Thomas2016}. The RV errors are based on photon noise, general pipeline errors, and illumination profile (or \textit{Y}-profile, see \citet{Thomas2016}) variation errors. We find that these errors are somewhat conservative; however, using more conservative errors ensures that only stars with strong RV signals will be considered as new candidates for this study.

We search for periodic signals in these 653 RV variable stars using a Lomb-Scargle (LS) periodogram \citep{Lomb1976,Scargle1982}, which is well suited to find periodic signals in unevenly spaced RV data. For each RV variable star we find the period with the max LS periodogram power between one and 750 days ($\sim$max temporal baseline of MARVELS data), which we use as an initial guess for our Keplerian fitting. We also fit a basic linear trend to each star's RV data, subtract this trend, and find the max LS periodogram power for this data. For Keplerian fitting we use the RV orbit fitting software, \texttt{RVLIN}, an IDL\footnote{IDL is a commercial programming language and environment by Harris Geospatial Solutions: \url{http://www.harrisgeospatial.com/IDL}} package developed by \citet{WrightHoward2012}. This software uses a Levenberg-Marquardt algorithm to fit RV data and find the best-fitting orbital parameters. Each RV variable star has eight initial fits. We input four different eccentricities (0.05, 0.3, 0.6, and 0.9) for both of the max LS periodogram periods found with and without a linear trend removed. We include a linear component in the Keplerian fit for the max LS periodogram periods found after removing a linear trend in the RV. 

All fits are run through numerous selection criteria designed to eliminate poor fits to the data. We use several criteria from \citet{Troup2016}, who find 382 stellar and substellar companions in the APOGEE data. Some of these criteria use $U_{N}$, the uniformity index \citep{MadoreFreedman2005}, which is used to quantify the phase coverage of the data; as well as $V_{N}$, the ``velocity" uniformity index \citep{Troup2016}, which quantifies the velocity coverage and prevents a selection program from settling on a very eccentric orbit that is unsupported by the data. $U_{N}$ is defined as:
\begin{equation}
U_{N} = \frac{N}{N-1} \bigg[1 - \sum_{i=1}^{N} (\phi_{i+1} - \phi_{i})^{2} \bigg],
\end{equation}
where the values $\phi_{i}$ are the sorted phases associated with the corresponding Modified Julian Date (MJD) of the measurement \textit{i}, and $\phi_{N+1}$ = $\phi_{1}$ + 1. $U_{N}$ is normalized such that $0 \le U_{N} \le 1$, where $U_{N}$ = 1 indicates a curve evenly sampled in phase space. $V_{N}$ is defined as:
\begin{equation}
V_{N} = \frac{N}{N-1} \bigg[1 - \sum_{i=1}^{N} (\nu_{i+1} - \nu_{i})^{2} \bigg],
\end{equation}
where $\nu_{i}$ = (v$_{\text{i}}$ - v$_{\text{min}}$)/(v$_{\text{max}}$ - v$_{\text{min}}$). The values of v$_{\text{i}}$ are the radial velocity measurements, sorted by their value. If a fit has a linear trend applied, then that trend is subtracted from the v$_{\text{i}}$ values. For the fits with no linear trend applied, we subtract the average of the raw velocities: v$_{\text{i}}$ = v$_{\text{raw,i}}$ - $\bar{\text{v}}_{\text{raw}}$. We also make use of the modified $\chi^{2}$ statistic, which is defined by \citet{Troup2016}:
\begin{equation}
\chi^{2}_{\text{mod}} = \frac{\chi^{2}_{\text{red}}}{\sqrt{U_{N}V_{N}}},
\end{equation}
where $\chi^{2}_{\text{red}}$ is the traditional reduced $\chi^{2}$ goodness-of-fit statistic.

A fit must meet these 11 criteria to be considered valid:

\renewcommand{\labelenumi}{(\roman{enumi})}
\renewcommand{\labelitemi}{\textendash}
\begin{enumerate}
\item Periods cannot be within 5\% of 3, 2, 1, 1/2, or 1/3 \\
\hspace{\labelwidth}\phantom{\texttt{type}--} days (to avoid one-day aliased systems) \\ \label{cri1} 
\item Eccentricities cannot be extremely high ($e > 0.934$) \\ \label{cri2} 
\item Fits must have good phase and velocity coverage: \\
\hspace{\labelwidth}\phantom{\texttt{type}---} ($U_{N}V_{N} \ge  0.5 $) \\ \label{cri3} 
\item $\frac{K}{\widetilde{\sigma}_{\text{v}}} \ge 3 + 3(1 - V_{N})e$ \\ \label{cri4} 
\item $\frac{K}{|\Delta \widetilde{\text{v}}|} \ge 3 + 3(1 - V_{N})e$ \\ \label{cri5} 
\item $\chi^{2}_{\text{mod}} \le \frac{K / \widetilde{\sigma}_{\text{v}}}{3 + 3(1 - V_{N})e}$ \\ \label{cri6} 
\item $\chi^{2}_{\text{mod}} \le \frac{K / |\Delta \widetilde{\text{v}}|}{3 + 3(1 - V_{N})e}$ \\ \label{cri7} 
\item $\frac{K}{\mbox{stddev}(|\textbf{v} - \widetilde{\text{v}}|)} < 4$ \\ \label{cri8}
\item If $e$ $>$ 0.4 then there must be an RV point in the \\
\hspace{\labelwidth}\phantom{\texttt{type}---} phase space ranges  $\phi$ $<$ 0.1 and $\phi$ $>$ 0.9. \\ \label{cri9} 
\item $\Delta T \ge 2P$ \\ \label{cri10} 
\item $\frac{K}{\text{max}(\textbf{v}) - \text{min}(\textbf{v})} \ge 1/3$ \\ \label{cri11} 
\end{enumerate}
where $\widetilde{\text{v}}$ is the median RV measurement for the star, $\widetilde{\sigma}_{\text{v}}$ is the median RV uncertainty, $K$ is the RV semi-amplitude of the best-fit model, $|\Delta \widetilde{\text{v}}|$ is the median absolute residuals of the best-fit model, $\textbf{v}$ are the RV measurements, $\Delta T$ is the temporal baseline of the star, P the period of the fit, $U_{N}$ is the uniformity index, $V_{N}$ is the velocity uniformity index, and $\chi^{2}_{\text{mod}}$ is the modified $\chi^{2}$ statistic.

Criteria \ref{cri1} - \ref{cri7} are detailed in \citet{Troup2016}. Criteria \ref{cri8} - \ref{cri11} are additional criteria to those in \citet{Troup2016}, which further eliminate poor fits to the RV data. Criteria \ref{cri8} and \ref{cri9} are designed to eliminate highly eccentric fits with large RV semi-amplitudes that are clearly not supported by the data. We apply a strict period coverage with criterion \ref{cri10}, which ensures that the RV data covers at least two full phases of the fitted period. We eliminate unlikely fits with linear trends much larger than the semi-amplitude using criterion \ref{cri11}.

Of the initial 653 RV variable MARVELS stars, 84 have at least one Keplerian fit that meets all 11 selection criteria. If a star has multiple Keplerian fits that match all 11 criteria, we use the fit with the lowest $\chi^{2}_{\text{mod}}$ and no linear trend component as our fit for the remainder of the vetting procedure. We use a Keplerian fit with a linear trend component if only fits with a linear trend passed all the criteria. Assuming a 1 M$_{\odot}$ host star, 27 of these companion candidates have M sin$\textit{i}$ values within 10 M$_{\text{Jup}}$ - 110 M$_{\text{Jup}}$. We use these 27 candidates as our final BD candidates from this initial selection process. In $\S$ \ref{sec:cridep} we determine rough dependencies on our results for each of the 11 criteria. Given the low BD occurrence rates of previous studies  of $\lesssim$1\% \citep{Vogt2002,Patel2007,Wittenmyer2009,Sahlmann2011,Santerne2016}, we would expect to find $\lesssim$ 23 BDs in our stellar sample. Occurrence rate expectations combined with prior studies, which exhibit the likelihood of false positives in the MARVELS data, deems further analysis necessary to validate these candidates.

\subsection{Periodogram False Alarm Probability}

A Lomb-Scargle (LS) periodogram is well suited to find periodic signals in unevenly spaced RV data, but periodogram peaks do not necessarily signify real substellar companions. Although the periodogram contains information on how well an oscillating signal at any grid frequency fits the observed data, it also includes signs of other undesirable effects including aliases, parasite frequencies, and spectral leakage \citep{Suveges2012}. The false alarm probability (FAP) evaluates the statistical significance of a periodogram peak by testing the zero hypothesis of the observed time series (pure white noise) against a periodic deterministic signal.

\begin{figure}
      	\centering
	\includegraphics[width=0.49\linewidth]{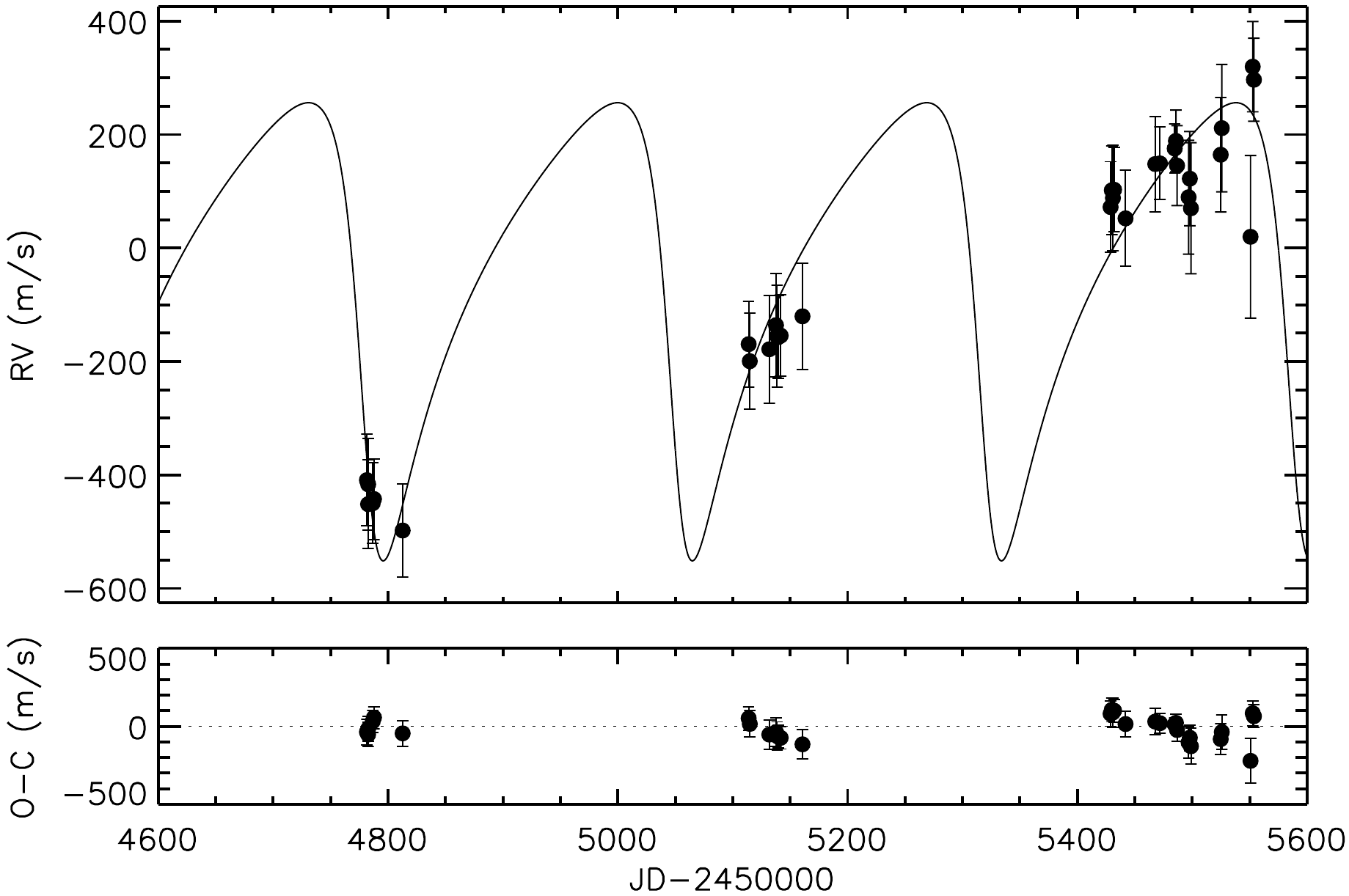}
	\hfill
	\centering
	\includegraphics[width=0.45\linewidth]{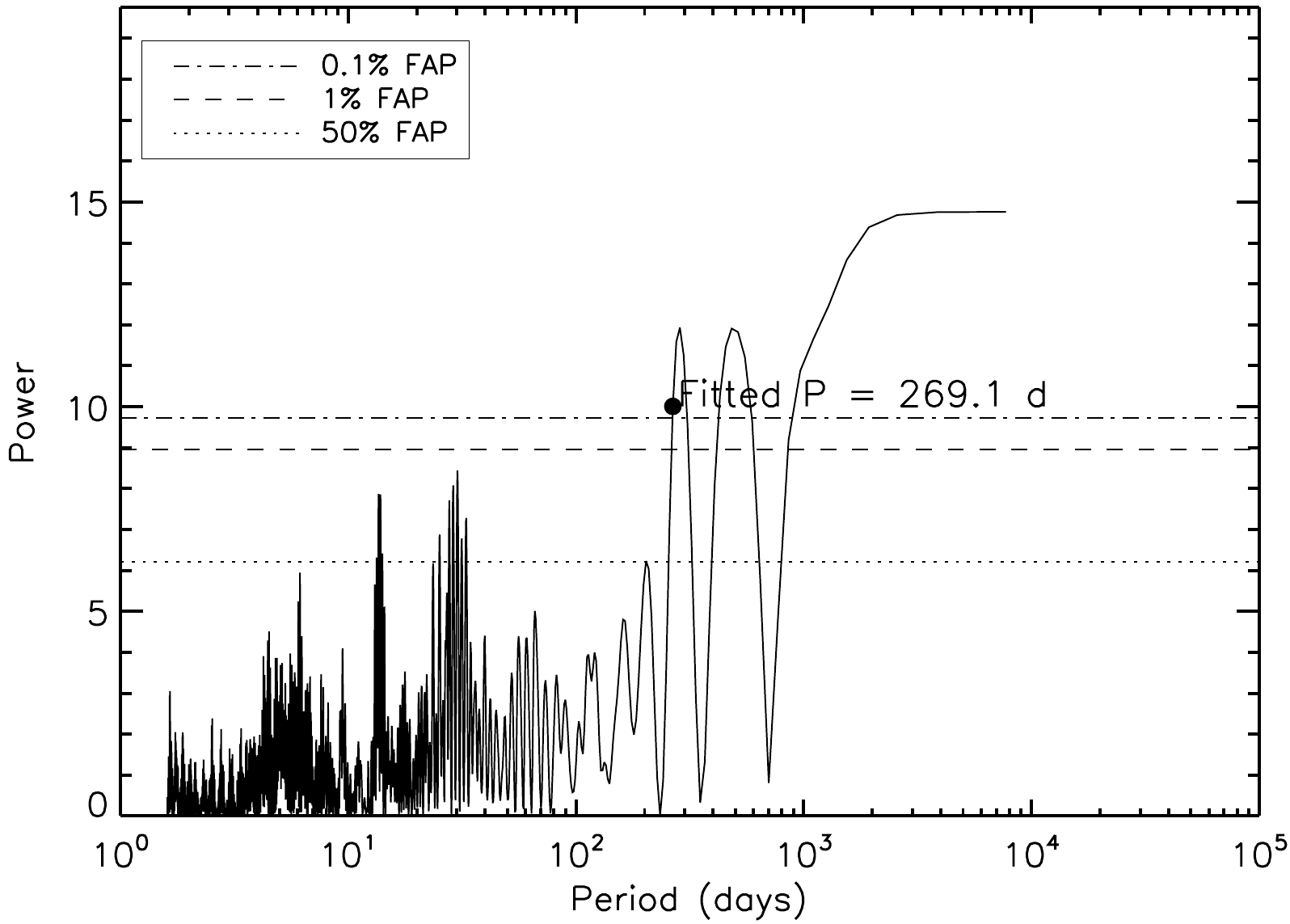}
\caption{\textbf{TYC 1717-02014-1}: When analyzing both the RV measuerments and LS periodogram of this candidate we find it likely that this candidate may be a long period stellar companion rather than the low-mass (M sin$\textit{i}$ $\sim$11.4 M$_{\text{Jup}}$) object our initial fitting finds. We remove this candidate from our list of potential BDs.}
\label{fig:tyc1717}
\end{figure}

We assess the FAP for a given orbit fit by using a randomized bootstrapping method to compare how the likelihood of the observed RVs compare to a random set with the same sampling \citep{Suveges2012}. We shuffle all of the RV results among the existing observation times and then record the peak power in the LS periodogram from this scrambled data set. This is repeated for 10,000 iterations. If more than 0.1\% of the peak powers in the randomized data exceed the power of the model period fit (0.1\% FAP), then we consider the candidate to be a likely false positive. However, if a companion's model fit has an eccentricity $>$ 0.5, the model period's peak must only be within 90\% of the max peak of the bootstrapped data. If a candidate's fit has a linear trend, we first remove the linear trend from the RV before computing the LS periodogram values. Figures \ref{fig:marv-8} -  \ref{fig:marv-21} show periodograms for each of the final BD candidates. This test eliminates four of the potential BD candidates, leaving a total of 23 BD candidates. A less strenuous 1\% FAP requirement still excludes these four candidates. 

\begin{figure}
      	\centering
         \subfloat[TYC 3545-00813-1]{	
	\includegraphics[width=0.49\linewidth]{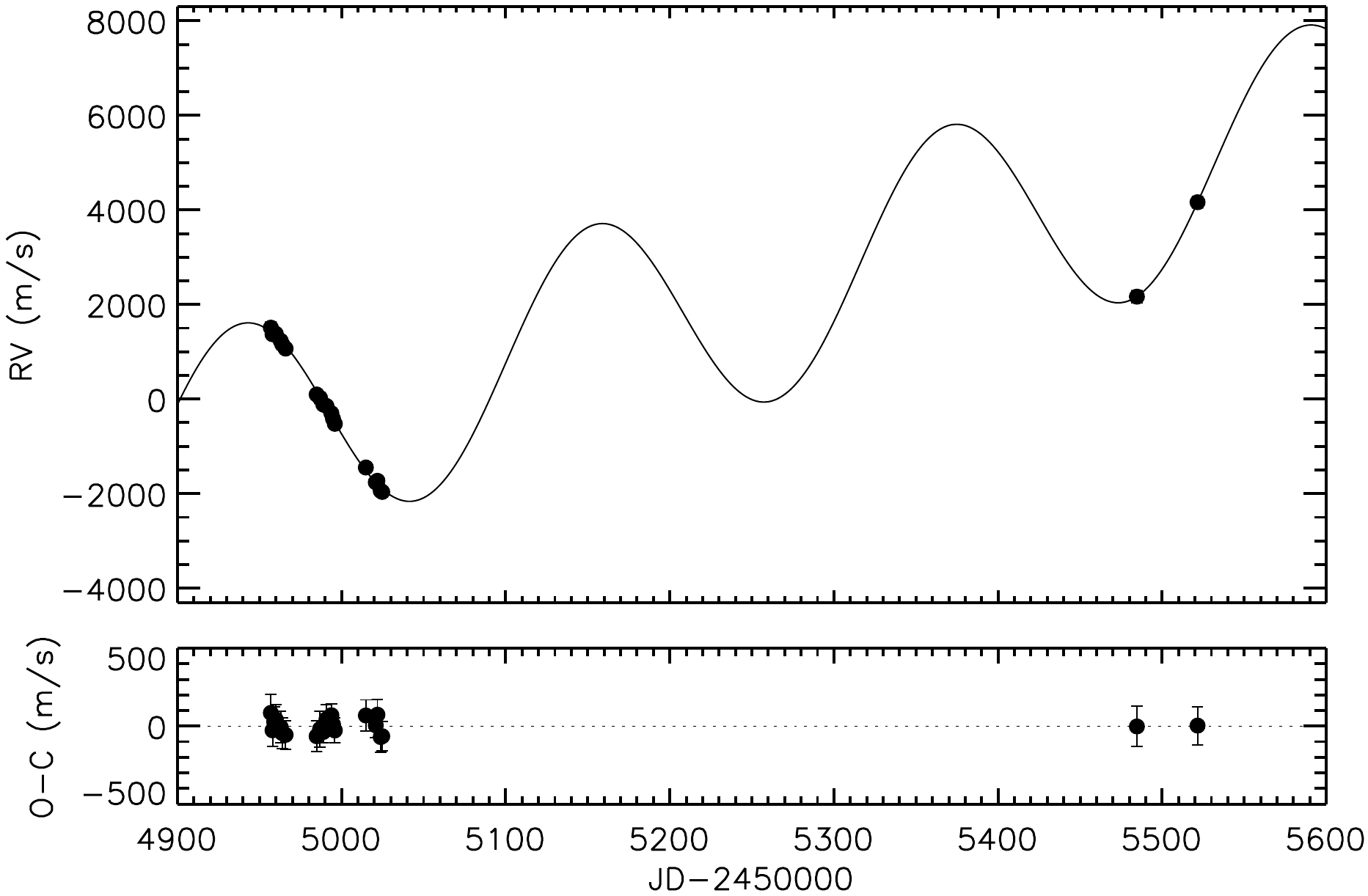}
	\hfill
	\centering
	\includegraphics[width=0.45\linewidth]{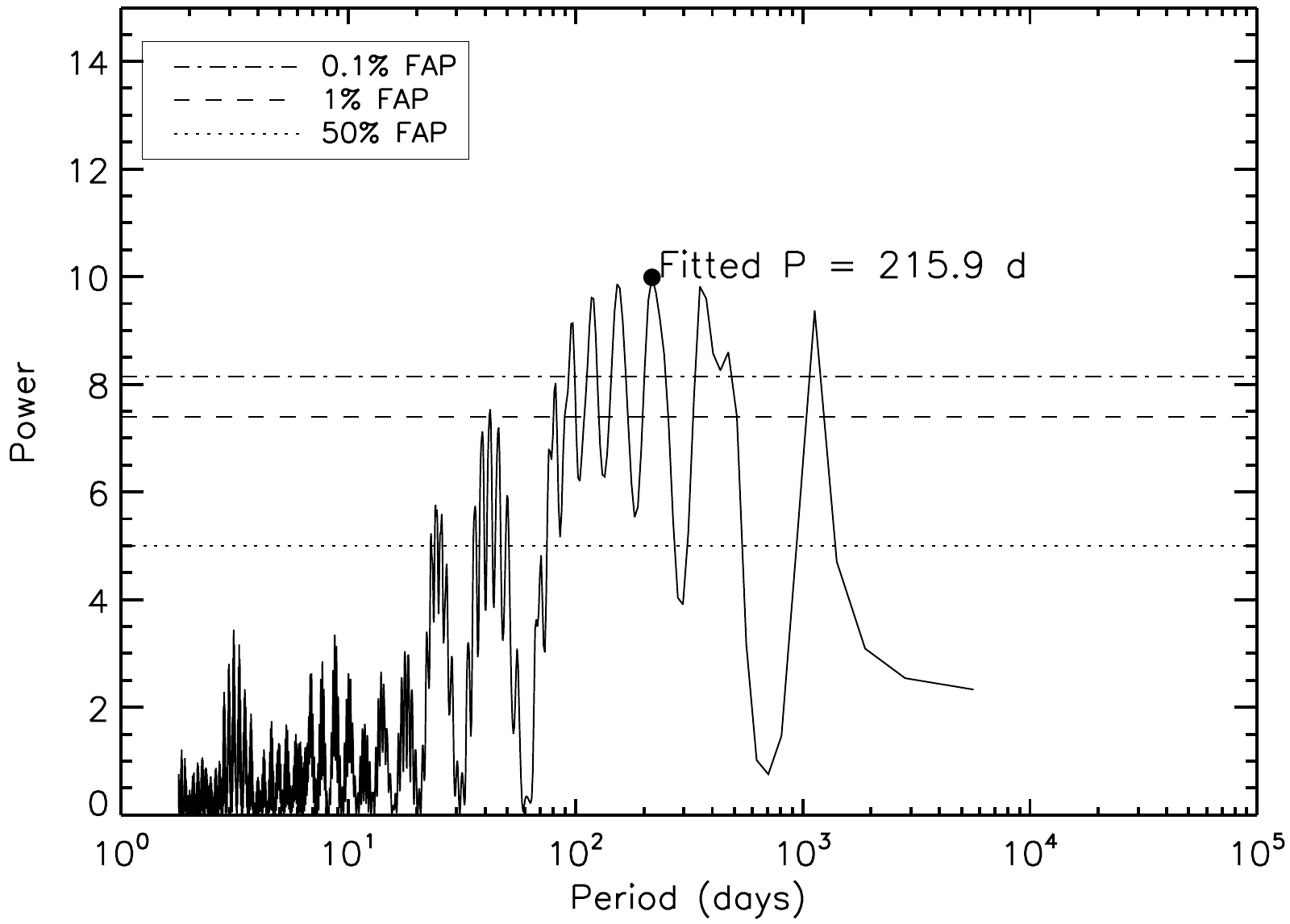}
	}
	\hfill
      	\centering
	         \subfloat[TYC 3549-00411-1]{	
	\includegraphics[width=0.49\linewidth]{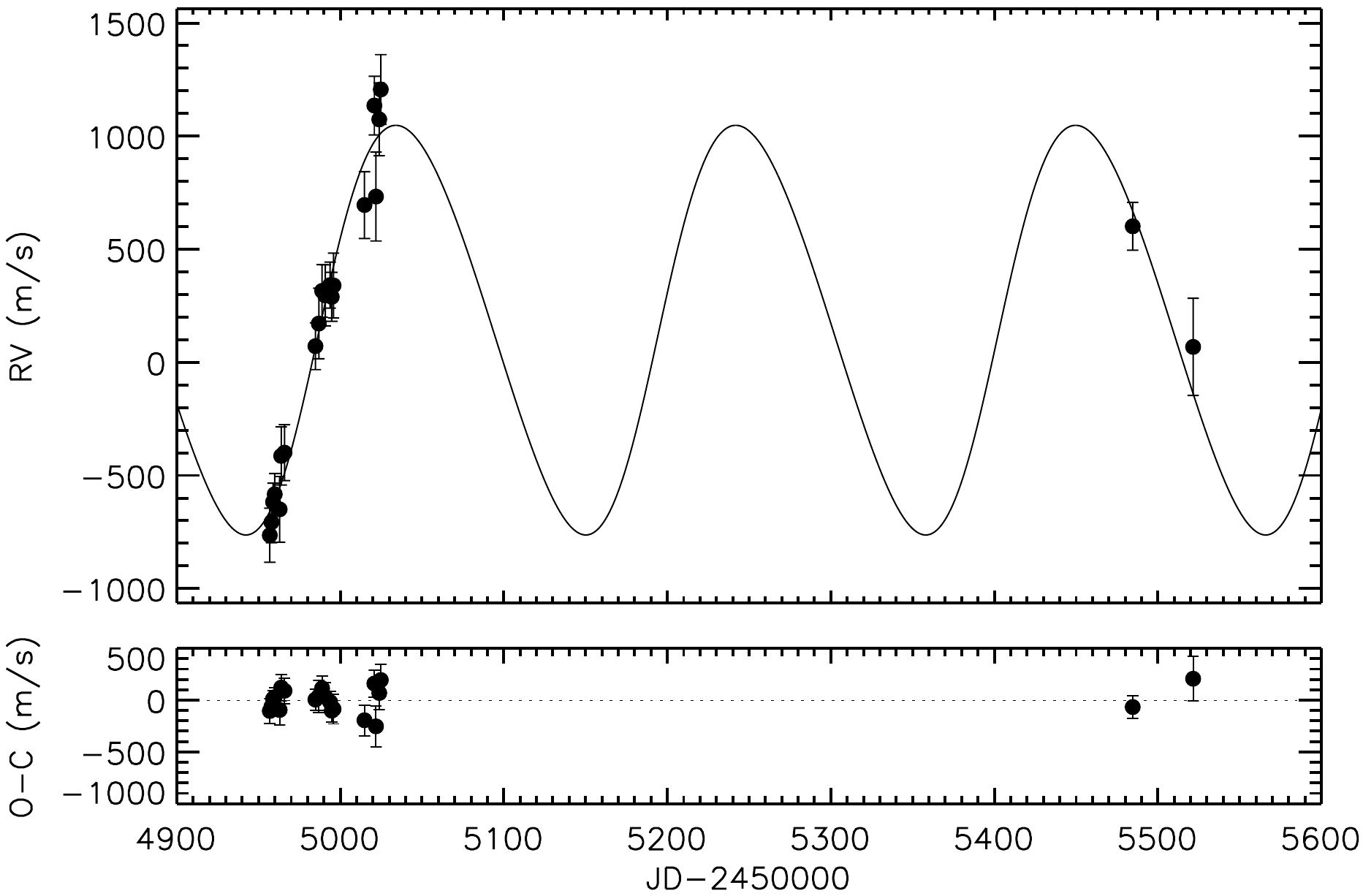}
	\hfill
	\centering
	\includegraphics[width=0.45\linewidth]{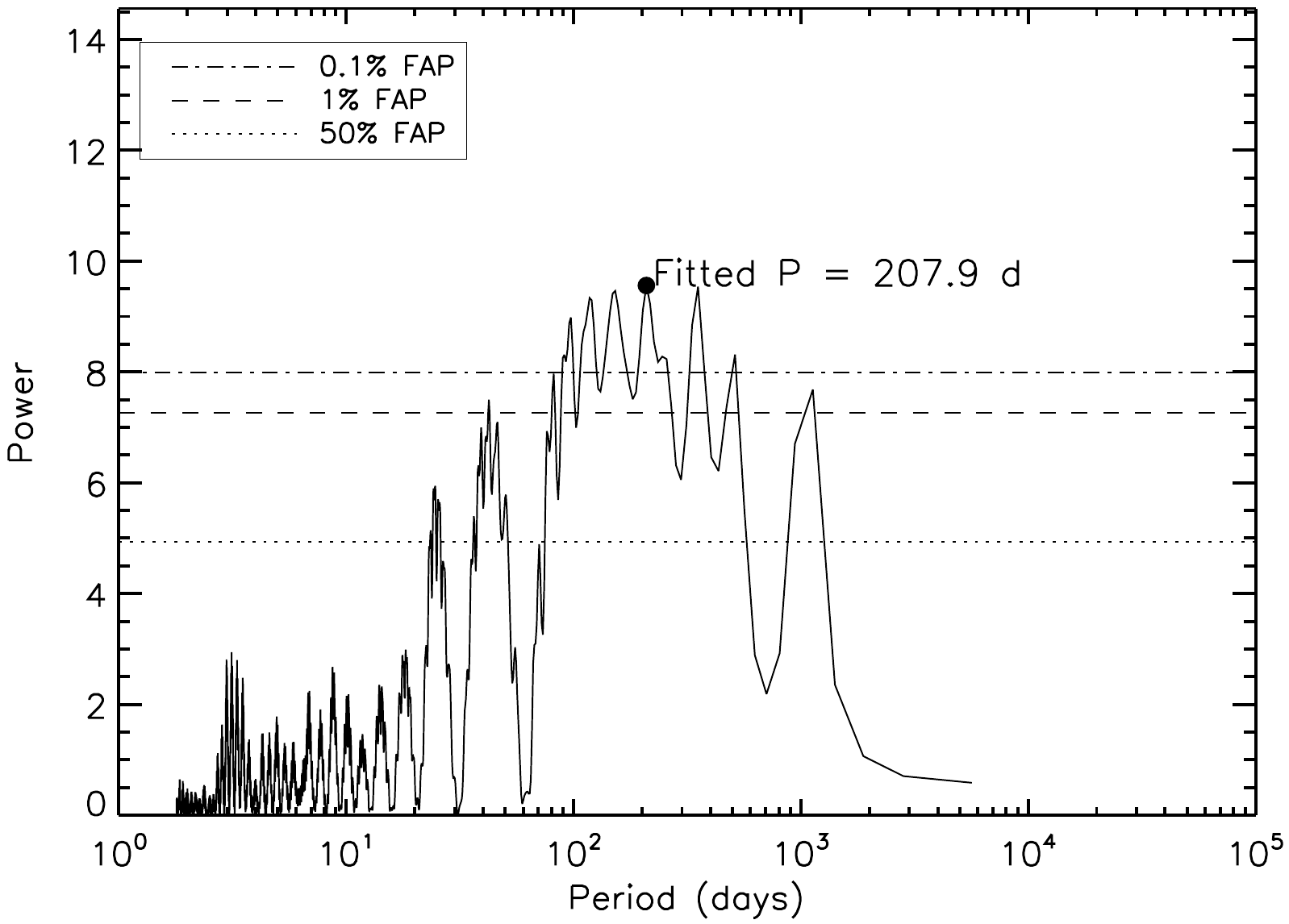}
	}
	\hfill
      	\centering
	         \subfloat[TYC 3545-02661-1]{	
	\includegraphics[width=0.49\linewidth]{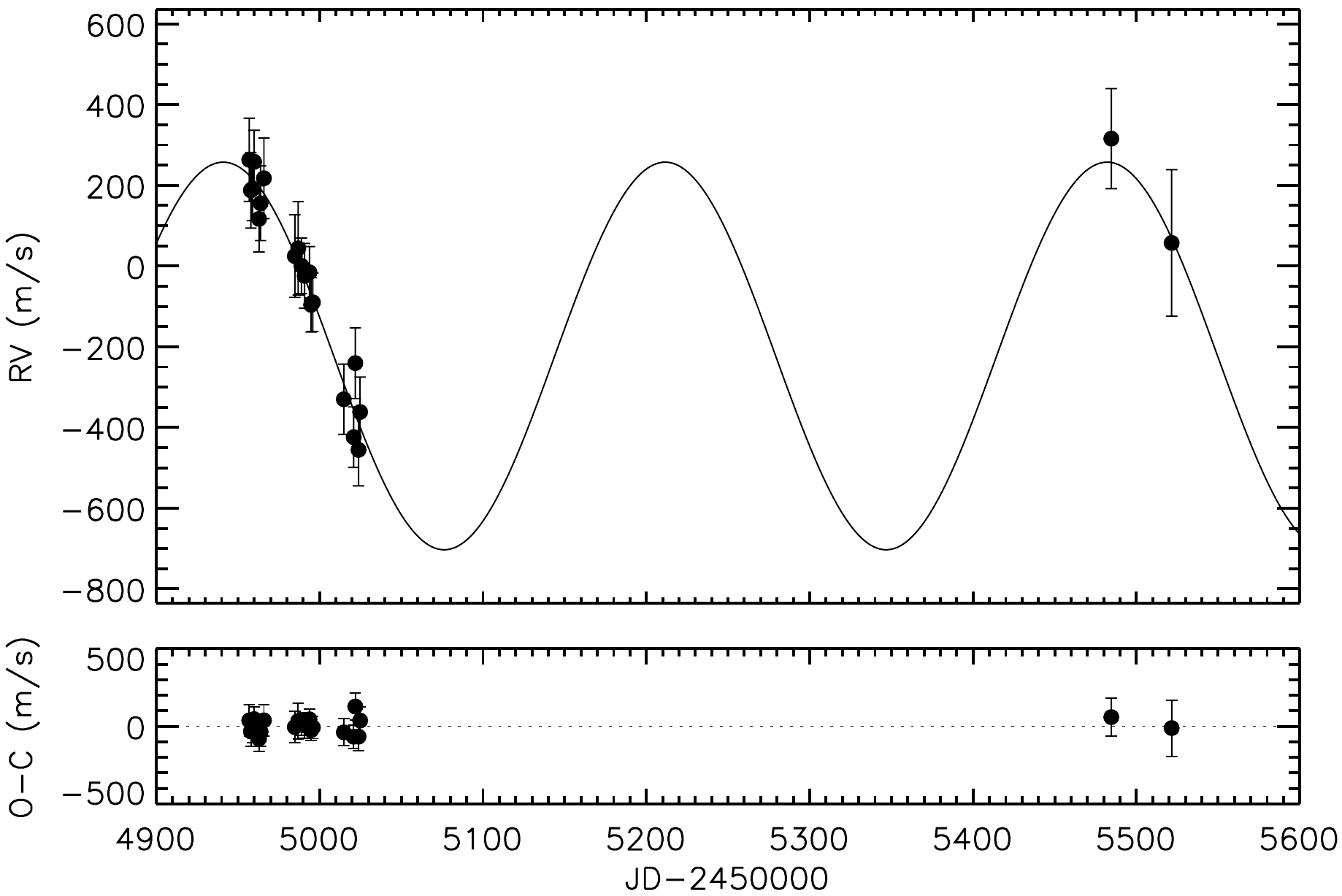}
	\hfill
	\centering
	\includegraphics[width=0.45\linewidth]{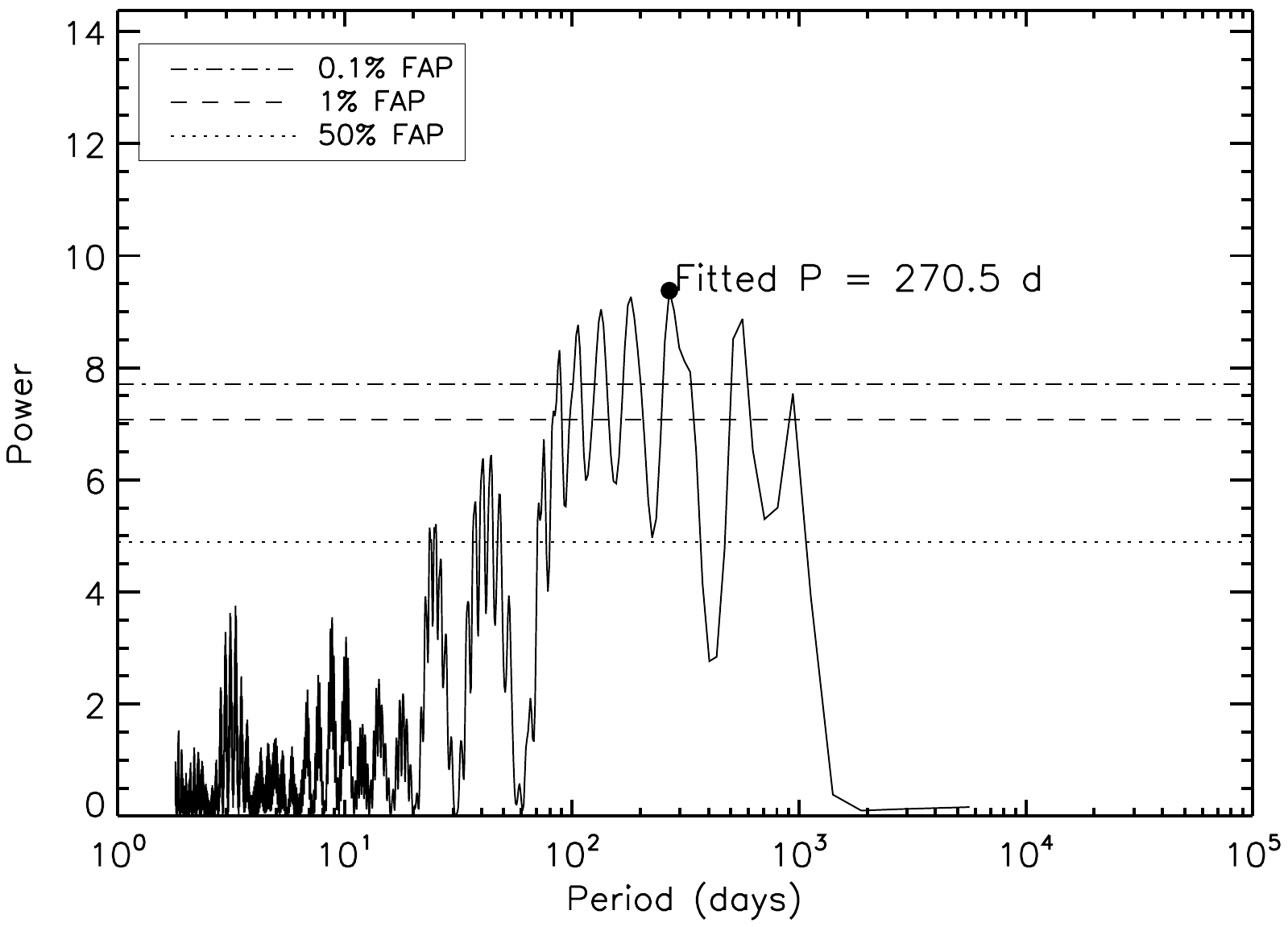}
	}
\caption{These three stars were all observed on the same MARVELS SDSS-III plate. Each plate has the same observation times and may suffer from similar systematic errors. Given the similarity of both the RV variation and periodograms of these three targets we remove them as potential BD candidates as their signals may be due to systematic errors associated with this specific plate.}
\label{fig:keplate}
\end{figure}

Finally we analyze these remaining 23 candidates by looking at both their RV measurements and LS periodograms. We find that our period fit for TYC 1717-02014-1 sits just above the 0.1\% FAP level, and when looking at the RV, a stellar companion with a long period seems just as likely as the low-mass fit our vetting procedure produces (figure \ref{fig:tyc1717}). Thus, we remove TYC 1717-02014-1 from our potential BD candidates. We also find that three of our potential BD candidates all have similar RV fits and LS periodogram peaks (figure \ref{fig:keplate}). These three candidates are all on the same SDSS-III MARVELS plate (each plate contains fiber holes for the 60 stars that were observed at the same time) and may suffer from the same systematic errors associated with the measurements of this plate. We remove TYC 3545-02661-1, TYC 3549-00411-1, and TYC 3545-00813-1 from our list of potential BD companions as these RV variations may be due to systematic errors rather than real low-mass companions. This final periodogram analysis leaves us with 19 potential BD candidates.

\subsection{Stellar Contamination}

\begin{figure}
	\centering
	 \subfloat[TYC 3010-01494-1 \citep{Mack2013}]{
	 \includegraphics[width=0.99\columnwidth]{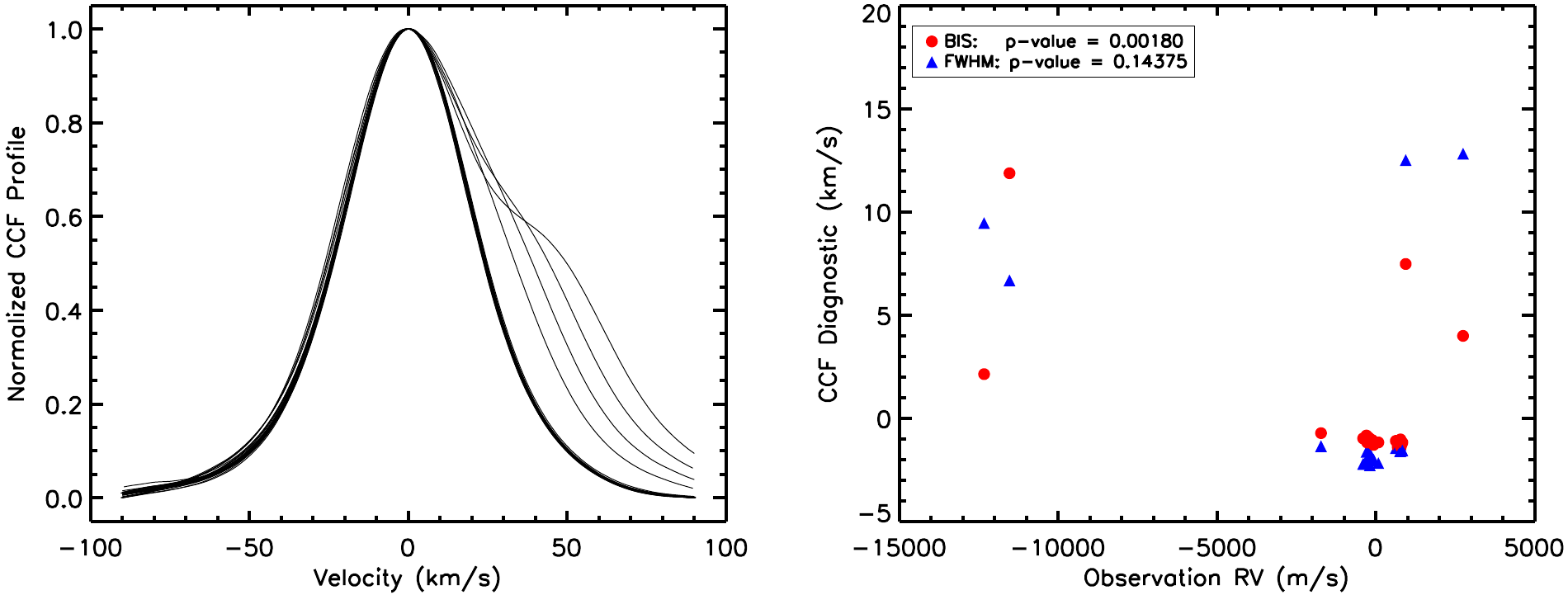}\label{fig:ccf_prev_a}
	 }
	\hfill
	 \centering
	 \subfloat[TYC 1240-00945-1 \citep{Lee2011,Wright2013}]{
	 \includegraphics[width=0.99\columnwidth]{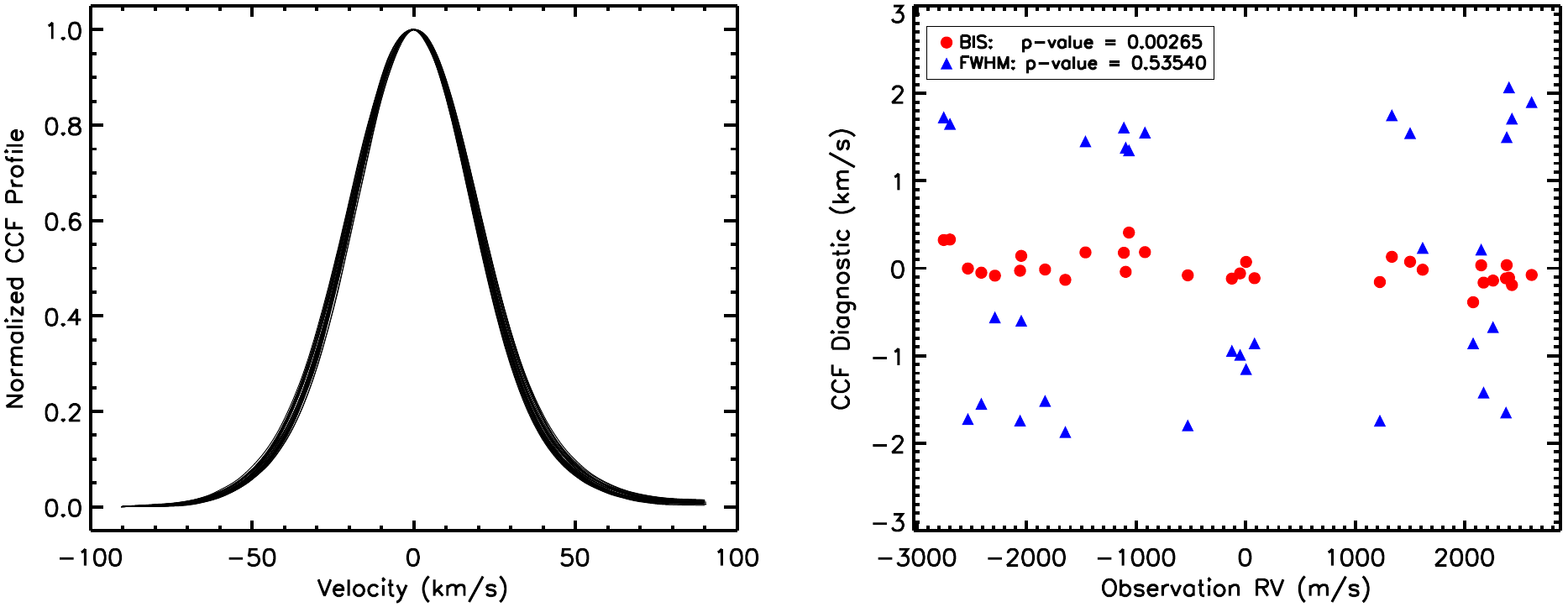}\label{fig:ccf_prev_b}
	 }	 
    	\caption{\textit{Left}: CCF line profile plots (each CCF line profile is calculated from the average of both CCF line profiles of the two spectral beams for each star). The top shows a clear case of stellar contamination, while the bottom is a known binary star but has relatively smooth line profiles. \textit{Right}: Bisector velocity span (BIS) and FWHM line-profile asymmetry diagnoses with their corresponding observation's RV measurement. Both BIS and FWHM values have the mean value of the star subtracted before plotting. The Kendall rank correlation p-values for both the BIS and RV and FWHM and RV are displayed in the legend.}
	\label{fig:ccf_prev}
\end{figure}

RV signals containing amplitudes compatible with substellar companions are not necessarily planets or BDs; the RV signal might be mimicked by stellar activity \citep[e.g.][]{Queloz2001}. \citet{Wright2013} and \citet{Mack2013} displayed this type of contamination in previous MARVELS companion candidates. Many methods to reduce this type of contamination employ the analysis of a cross-correlation function (CCF), which can find a star's RV by measuring the Doppler shift of spectral lines. The CCF technique averages hundreds to thousands of spectral lines to create a spectral line profile that can be approximated by a Gaussian function. The CCF has a much higher SNR compared with individual lines and can be characterized by a given location in the RV space, a contrast, and a FWHM. An asymmetry in the observed line profile can be produced by the convective blue shift, pulsations, stellar spots, and contaminating stars \citep{Santerne2015}.  

We use the bisector velocity span \citep[BIS;][]{Queloz2001} to measure line-profile asymmetry in the CCFs of our potential BD candidates. The BIS consists of computing the velocity span between the average of the top portion of the bisector and the bottom portion of the bisector, where the bisector is defined as the mid-points of horizontal line segments bounded by the sides of the line profile. We use the definition and limits described in \citet{Santerne2015} for our BIS calculations. We also use the FWHM of the CCF line profiles to measure CCF line-profile asymmetry. The FWHM may present a clear correlation with RV when other line-profile asymmetry diagnoses do not; therefore, the FWHM is a good diagnosis to use as a complement to other line-asymmetry diagnoses \citep{Santerne2015}. For each BD companion candidate we calculated the Kendall rank correlation coefficient between their RV and BIS and RV and FWHM values. However, as pointed out by \citet{Mack2013}, MARVELS low resolution makes line bisector analysis less reliable than other studies which use high resolution spectra for this method. Thus, we analyzed both the CCF line profiles and kendall rank correlation p-values for these candidates. For our CCF line profiles we take the average CCF line profile of both spectral beams of the star for each observation. The CCF line profile may contain an obvious second component, as seen in figure \ref{fig:ccf_prev_a}, which shows the CCF line profiles and line-asymmetry diagnoses plots for the double-lined spectroscopic binary TYC 3010-01494-1 \citep{Mack2013}.

The 19 BD candidates from the section above includes TYC 1240-00945-1, the face-on binary previously reported to be a BD \citep{Lee2011,Wright2013}. Figure \ref{fig:ccf_prev_b} shows the RV and line-asymmetry diagnoses plots with the associated p-values and the CCF line profiles for this star. We see that there is not a visible second component in the CCF line profile but this star has a low BIS p-value of $\sim$0.003. In order to set a significance level, we calculate p-values for the $\sim$2,340 MARVELS stars with $\ge$ 10 RV observations, time baselines longer than 300 days, and RPM$_{J}$ dwarf/subgiant designation. This sample has a mean RV and BIS correlation p-value of $\sim$0.31 and a mean RV and FWHM correlation p-value of $\sim$0.30. The BIS p-value of 0.00265 for the known binary TYC 1240-00945-1 is greater than $\sim$16\% of stars in this sample, while the relatively high FWHM p-value of $\sim$0.54 for this star is greater than $\sim$75\% of stars in this sample. We set our p-value significance levels at the bottom 20\% of this sample, which corresponds to a BIS p-value of $\sim$0.008 or lower and a FWHM p-value of $\sim$0.01 or lower.

\begin{figure}
	\centering
	 \includegraphics[width=0.99\columnwidth]{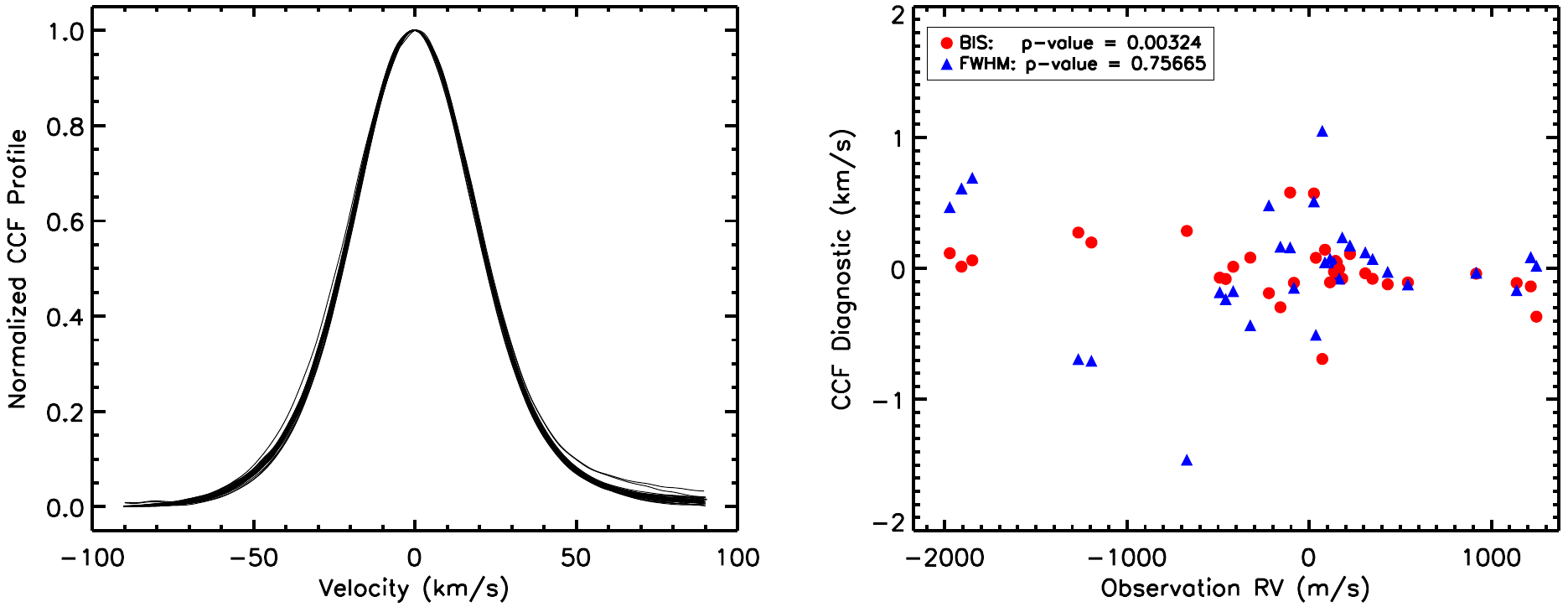}
    	\caption{CCF line profile and line-profile asymmetry diagnoses plots as in figure \ref{fig:ccf_prev}, now for TYC 3005-00460-1. The RV fit for this star does not contain a linear trend to account for the small BIS p-value and CCF line profile variation, thus we remove this star as a BD candidate and label it as a likely double-lined spectroscopic binary.}
	\label{fig:ccf_hd89}
\end{figure}

One other star besides TYC 1240-00945-1 has a BIS p-value below 0.008: TYC 3005-00460-1. Considering the CCF line profile variation and low p-value seen in figure \ref{fig:ccf_hd89}, as well as no associated linear trend in the RV fit that may account for such a variation, we remove TYC 3005-00460-1 as a BD candidate. Two stars have a FWHM p-value below 0.01: TYC 3413-02367-1 (MCKGS1-94 from \citet{Ghezzi2014}) and TYC 5634-00322-1. We also remove these stars as BD candidates. Figures \ref{fig:marv-8} -  \ref{fig:marv-21} show the CCF line profiles and line-asymmetry diagnoses plots for all of the new MARVELS BD candidates. TYC 0173-02410-1 (MARVELS-9b, figure \ref{fig:marv-9}) shows relatively large CCF line profile variation, but our RV fit contains a linear component that may account for the CCF asymmetry.

\begin{table*}
    \caption{MARVELS Host Star Parameters}
     \label{tab:starpar}
    \begin{center}
    \begin{tabular}{lcccccc}
    \hline
   \hline
   	Starname & T$_{\text{eff}}$ 	& [Fe/H]	& log \textit{g} 	& M$_{\star}$ 	& R$_{\star}$ & V$_{\text{mag}}$ \\
			& (K)				&  (dex)	& (cgs)		&(M$_{\odot}$) & (R$_{\odot}$) & (GSC 2.3) \\
	\hline
TYC 0123-00549-1$^{\dagger}$ (MCKGS1-50)	&	5540	$\pm$	39	&	0.20	$\pm$	0.07	&	4.48	$\pm$	0.23	&	1.02	$\pm$	0.08	&	0.95	$\pm$	0.27	&	10.3	\\
TYC 0173-02410-1	&	5470	$\pm$	112	&	0.18	$\pm$	0.09	&	4.25	$\pm$	0.18	&	1.06	$\pm$	0.10	&	1.26	$\pm$	0.30	&	10.5	\\
TYC 4955-00369-1	&	6190	$\pm$	143	&	-0.43	$\pm$	0.09	&	4.55	$\pm$	0.24	&	1.03	$\pm$	0.09	&	0.90	$\pm$	0.26	&	10.2	\\
TYC 1451-01054-1 	&	5980	$\pm$	140	&	-0.34	$\pm$	0.09	&	4.60	$\pm$	0.24	&	0.99	$\pm$	0.08	&	0.83	$\pm$	0.24	&	9.9	\\
TYC 3469-00492-1	&	5285	$\pm$	82	&	0.36	$\pm$	0.06	&	3.78	$\pm$	0.13	&	1.29	$\pm$	0.12	&	2.39	$\pm$	0.46	&	11.3	\\
GSC 03467-00030$^{\dagger}$  (MCKGS1-135)	&	5525	$\pm$	71	&	-0.17	$\pm$	0.08	&	4.39	$\pm$	0.23	&	0.96	$\pm$	0.09	&	1.03	$\pm$	0.30	&	12.2	\\
TYC 3091-00716-1$^{\dagger}$ (MCKGS1-153)	&	5614	$\pm$	29	&	-0.09	$\pm$	0.06	&	4.61	$\pm$	0.09	&	0.95	$\pm$	0.06	&	0.80	$\pm$	0.09	&	10.0	\\
TYC 3097-00398-1	&	5340	$\pm$	98	&	0.16	$\pm$	0.07	&	3.85	$\pm$	0.15	&	1.21	$\pm$	0.12	&	2.14	$\pm$	0.47	&	11.3	\\
TYC 3130-00160-1$^{\dagger}$ (MCUF1-11)	&	5315	$\pm$	44	&	0.33	$\pm$	0.06	&	4.23	$\pm$	0.19	&	1.06	$\pm$	0.10	&	1.28	$\pm$	0.33	&	11.0	\\
TYC 3547-01007-1	&	6090	$\pm$	190	&	0.06	$\pm$	0.11	&	4.57	$\pm$	0.27	&	1.12	$\pm$	0.10	&	0.91	$\pm$	0.30	&	10.5	\\
TYC 3556-03602-1	&	6295	$\pm$	200	&	0.11	$\pm$	0.09	&	4.35	$\pm$	0.30	&	1.25	$\pm$	0.14	&	1.23	$\pm$	0.48	&	10.6	\\
TYC 3148-02071-1	&	6270	$\pm$	271	&	0.09	$\pm$	0.14	&	4.13	$\pm$	0.27	&	1.33	$\pm$	0.18	&	1.63	$\pm$	0.62	&	9.5	\\
	\hline
       \end{tabular}    \\
       	       \footnotesize{$^{\dagger}$ T$_{\text{eff}}$, [Fe/H], and log \textit{g} obtained from \citet{Ghezzi2014}'s precise measurements with high resolution spectra. Star name in \citet{Ghezzi2014} is in parentheses. For all other stars these parameters were obtained using \citet{Ghezzi2014}'s spectral indices method.}

\end{center}
\end {table*}    

    \begin{table*}
    \caption{MARVELS Companion Parameters}
    \label{tab:compar}
    \begin{center}
    \begin{tabular}{llcccc}
    \hline
   \hline
   	Star & MARVELS &	M sin$\textit{i}$ 	&	Period & 	Ecc & 	K   \\
		& designation	& 	(M$_{\text{Jup}}$)  & (days) & 		& 	(m s$^{-1}$)  \\
	\hline
TYC 0123-00549-1 (MCKGS1-50)	&	MARVELS-8b	&	37.7	$\pm$	0.4	&	36.045	$\pm$	0.005	&	0.101	$\pm$	0.004	&	2241	$\pm$	8	\\
TYC 0173-02410-1	&				MARVELS-9b	&	76.0	$\pm$	5.1	&	216.5	$\pm$	1.8	&	0.37	$\pm$	0.05	&	2537	$\pm$	210	\\
TYC 4955-00369-1	&				MARVELS-10b	&	14.8	$\pm$	2.2	&	217.3	$\pm$	1.1	&	0.53	$\pm$	0.15	&	574	$\pm$	128	\\
TYC 1451-01054-1	 &						MARVELS-11b	&	52.1	$\pm$	0.6	&	11.6121	$\pm$	0.0003	&	0.000	$\pm$	0.002	&	4551	$\pm$	12	\\
TYC 3469-00492-1	&				MARVELS-12b	&	32.8	$\pm$	1.1	&	252.5	$\pm$	0.1	&	0.60	$\pm$	0.01	&	1093	$\pm$	35	\\
GSC 03467-00030  (MCKGS1-135)	&	MARVELS-13b	&	41.8	$\pm$	2.9	&	147.6	$\pm$	0.3	&	0.50	$\pm$	0.04	&	1868	$\pm$	167	\\
TYC 3091-00716-1 (MCKGS1-153)	&	MARVELS-14b	&	100.2	$\pm$	1.4	&	269.1	$\pm$	0.3	&	0.417	$\pm$	0.003	&	3359	$\pm$	14	\\
TYC 3097-00398-1	&				MARVELS-15b	&	97.6	$\pm$	1.8	&	95.54	$\pm$	0.04	&	0.223	$\pm$	0.004	&	3730	$\pm$	17	\\
TYC 3130-00160-1 (MCUF1-11)	&	MARVELS-16b	&	42.7	$\pm$	0.7	&	31.656	$\pm$	0.007	&	0.690	$\pm$	0.006	&	3552	$\pm$	42	\\
TYC 3547-01007-1	&				MARVELS-17b	&	13.1	$\pm$	1.6	&	103.4	$\pm$	0.9	&	0.24	$\pm$	0.08	&	537	$\pm$	64	\\
TYC 3556-03602-1	&				MARVELS-18b	&	24.4	$\pm$	2.4	&	117.6	$\pm$	1.1	&	0.46	$\pm$	0.09	&	968	$\pm$	83	\\
TYC 3148-02071-1	&				MARVELS-19b	&	17.3	$\pm$	2.4	&	122.2	$\pm$	1.7	&	0.53	$\pm$	0.10	&	683	$\pm$	82	\\
\hline
\hline
\end{tabular}    
\begin{tabular}{lccccccc}
MARVELS & a & T$_{\text{p}}$ & $\omega$ & $\gamma$ &  $\frac{dv}{dt}$ & fit RMS &  	$\chi_{\text{red}}^{2}$ \\
designation & (AU) & (JD - 2450000)  & (deg) & (m s$^{-1})$ & (m s$^{-1}$day$^{-1}$) & (m s$^{-1}$) & \\
	\hline
MARVELS-8b	&	0.215	$\pm$	0.006	&	4864.3	$\pm$	0.3	&	146.8	$\pm$	2.6	&	-130.1	$\pm$	7.3	&	0	$\pm$	0	&	62.1	&	0.37	\\
MARVELS-9b	&	0.72	$\pm$	0.02	&	4911.1	$\pm$	6.1	&	251.2	$\pm$	8.0	&	46203.0	$\pm$	3095.1	&	5.27	$\pm$	0.35	&	432.4	&	10.00	\\
MARVELS-10b	&	0.71	$\pm$	0.02	&	5598.2	$\pm$	4.0	&	0.1	$\pm$	27.3	&	3685.9	$\pm$	804.9	&	0.46	$\pm$	0.10	&	77.5	&	0.71	\\
MARVELS-11b	&	0.100	$\pm$	0.003	&	5581.0	$\pm$	1.6	&	66.2	$\pm$	44.4	&	132.4	$\pm$	9.1	&	0	$\pm$	0	&	48.0	&	0.71	\\
MARVELS-12b	&	0.85	$\pm$	0.03	&	4926.9	$\pm$	0.0	&	326.8	$\pm$	2.7	&	-283.5	$\pm$	30.4	&	0	$\pm$	0	&	110.6	&	1.34	\\
MARVELS-13b	&	0.54	$\pm$	0.02	&	5007.4	$\pm$	0.9	&	196.7	$\pm$	2.7	&	-145.2	$\pm$	34.8	&	0	$\pm$	0	&	116.2	&	0.63	\\
MARVELS-14b	&	0.80	$\pm$	0.02	&	5046.1	$\pm$	0.5	&	124.3	$\pm$	0.7	&	-89.6	$\pm$	9.0	&	0	$\pm$	0	&	75.7	&	0.46	\\
MARVELS-15b	&	0.44	$\pm$	0.01	&	5012.7	$\pm$	0.4	&	4.7	$\pm$	1.7	&	64.5	$\pm$	13.8	&	0	$\pm$	0	&	57.5	&	0.29	\\
MARVELS-16b	&	0.200	$\pm$	0.006	&	4945.2	$\pm$	0.1	&	308.7	$\pm$	0.9	&	288.1	$\pm$	17.1	&	0	$\pm$	0	&	102.5	&	1.18	\\
MARVELS-17b	&	0.45	$\pm$	0.01	&	4994.9	$\pm$	6.2	&	316.0	$\pm$	21.2	&	16584.2	$\pm$	2394.8	&	1.91	$\pm$	0.28	&	102.2	&	1.93	\\
MARVELS-18b	&	0.51	$\pm$	0.02	&	5025.1	$\pm$	3.1	&	176.2	$\pm$	10.8	&	-115.2	$\pm$	59.1	&	0	$\pm$	0	&	185.9	&	1.44	\\
MARVELS-19b	&	0.52	$\pm$	0.03	&	4903.0	$\pm$	2.8	&	218.2	$\pm$	12.4	&	29.2	$\pm$	77.8	&	0	$\pm$	0	&	124.1	&	0.85	\\
	\hline
	       \end{tabular}    
	       \footnotesize{Companion parameters correspond to the best Keplerian orbital fits found with RVLIN, which are displayed in figures \ref{fig:marv-8} -  \ref{fig:marv-21}.} 
\end{center}
\end {table*}    

\subsection{Previous Candidates \& Vetting Results}

We removed TYC 1240-945-1 \citep[MARVELS-1b,][]{Lee2011,Wright2013} as well as TYC 3005-00460-1, TYC 3413-02367-1 (MCKGS1-94), and TYC 5634-00322-1 in the previous section, leaving 15 BD candidates left in our vetting procedure. However, three of these 15 candidates were previously published: the $\sim$32 M$_{\text{Jup}}$ BD around GSC 03546-01452 \citep[MARVELS-6b,][]{DeLee2013}, the $\sim$65 M$_{\text{Jup}}$ BD around HIP 67526 \citep[MARVELS-5b,][]{Jiang2013}, and the $\sim$98 M$_{\text{Jup}}$ stellar companion around TYC 4110-01037-1 \citep[MARVELS-3b,][]{Wisniewski2012}. We find similar companion parameters as previous studies for these three stars using the new UF2D data. With the exclusion of these three companions, our full vetting process produces 12 new BD candidates. 

The other MARVELS stars with published companions or double-lined spectroscopic binaries (SB2s) did not make it through the vetting procedure for various reasons. We did not find a companion around TYC 2087-00255-1 \citep[MARVELS-4b,][]{Ma2013} due to three outliers in the new UF2D data, which prevent the vetting procedure to converge on the best fit initially. After removing these three outliers, we recover the companion reported in \citet{Ma2013}. We did not find any companions around TYC 2930-00872-1 \citep[MARVELS-2b,][]{Fleming2012}; the initial fit was poor as the star likely requires a multiple companion fit. The vetting criteria also eliminated HD 87646 \citep[MARVELS-7b,c,][]{Ma2016}, again likely due to the multiple companion fitting requirements. The UF2D pipeline produces large outliers for the SB2 TYC 3010-01494-1 \citep{Mack2013}, preventing a good-fit and possible false-positive that could have made it through all of our initial criteria. The SB2 TYC 1240-00945-1 \citep[MARVELS-1b,][]{Lee2011,Wright2013} has a companion fit that makes it through all 11 initial criteria, but is excluded in the stellar contamination section (see figure \ref{fig:ccf_prev_b}).

\section{Host Star Parameters}\label{sec:host}

\citet{Ghezzi2014} determine precise atmospheric parameters (effective temperature T$_{\text{eff}}$, metallicity [Fe/H], and surface gravity log \textit{g}) using a method based on spectral indices. This method is necessary because blending of the atomic lines is severe at the moderate resolution of the MARVELS spectra, precluding individual line analysis through classical spectroscopic methods (e.g., excitation and ionization equilibria). The spectral indices method does not rely on the accuracy of single atomic lines that can be blended together in moderate and low resolution spectra. \citet{Ghezzi2014} test the spectral indices method using 30 MARVELS stars that also have reliable atmospheric parameters derived from high-resolution spectra and spectroscopic analysis based on the excitation and ionization equilibria method. Using MARVELS spectra and the spectral indices method, \citet{Ghezzi2014} recover the parameters within 80 K for T$_{\text{eff}}$, 0.05 dex for [Fe/H], and 0.15 dex for log \textit{g}. In a separate test using 138 stars from the ELODIE stellar library, they recover the literature atmospheric parameters within 125 K for  T$_{\text{eff}}$, 0.10 dex for [Fe/H], and 0.29 dex for log \textit{g}.

Of the 12 new companions, four were previously considered to be MARVELS candidates, for which \citet{Ghezzi2014} obtained high-resolution spectra and analysis. We use \citet{Ghezzi2014}'s high resolution values for these four stars, while the other stellar parameters are found with the spectral indices method using the same pipeline developed by \citet{Ghezzi2014}. However, we use input spectra that were pre-processed using the newest UF1D/2D pipeline \citep{Thomas2016, Thomas2015UFphd}, while \citet{Ghezzi2014} used input spectra that were pre-processed using the older CCF+DFDI MARVELS pipeline released in the SDSS DR11. We tested the same 30 MARVELS stars as \citet{Ghezzi2014} using the newest UF1D input spectra and recovered parameters within 85 K for T$_{\text{eff}}$, 0.06 dex for [Fe/H], and 0.17 dex for log \textit{g} compared to the high-resolution analysis results. Atmospheric parameters and their associated uncertainties for the indices method are found with a reduced $\chi^{2}$ minimization process, which is detailed in Section 4.4 of \citet{Ghezzi2014}. The T$_{\text{eff}}$, [Fe/H], and log \textit{g} values for the companion host stars are presented in Table \ref{tab:starpar}.

We determine the mass and radius of each host star from T$_{\text{eff}}$, [Fe/H], and log \textit{g} using the empirical polynomial relations of \citet{Torres2010}, which were derived from a sample of eclipsing binaries with precisely measured masses and radii. We estimate the uncertainties in M$_{\star}$ and R$_{\star}$  by propagating the uncertainties in T$_{\text{eff}}$, [Fe/H], and log \textit{g} using the covariance matrices of the \citet{Torres2010} relations (kindly provided by G. Torres). The mass and radius of each host star is also presented in Table \ref{tab:starpar}.

\section{RESULTS \& Conclusions}\label{sec:res}

\subsection{Brown Dwarf Parameters}
Our candidate filtering process produced 12 new companions within $\sim$10 - 110 M$_{\text{Jup}}$. We run these stars through RVLIN a final time to obtain an optimal fit. Figures \ref{fig:marv-8} -  \ref{fig:marv-21} show the best fit to the RV data for each of the new MARVELS companion candidates. The orbital parameters for our final fits of the MARVELS BD candidates introduced in this paper are presented in Table \ref{tab:compar}. Values for M sin\textit{i} are determined using the host star mass derived in the previous section. Uncertainties in all parameters are calculated through bootstrapping (with 1,000 bootstrap replicates) using the publicly available BOOTTRAN package, which is detailed in \citet{Wang2012x}. Our final parameters produce two low-mass stellar companions (M sin$\textit{i}$ $\sim$97.6 and 100.2 M$_{\text{Jup}}$), and 10 new BD candidates in the mass range 13 M$_{\text{Jup}}$ $<$ M sin$\textit{i}$ $<$ 80 M$_{\text{Jup}}$. TYC 3547-01007-1 (MARVELS-17b, figure \ref{fig:marv-19}) resides just above the giant planet regime with M sin$\textit{i}$ = 13.1 $\pm$ 1.6 M$_{\text{Jup}}$. Given the errors of M sin$\textit{i}$ on MARVELS-17b it may be a giant planet if its orbit is edge-on. 

\subsection{Caveats}

We designed our vetting procedure to only extract substellar companion candidates with a low likelihood of being false positives. However, the new substellar companions presented in this work are still only candidates. Follow-up observations are required to confirm them as true BDs. Some follow-up observations have already begun using the EXtremely high Precision Extrasolar Planet Tracker \citep[EXPERT;][]{Ge2010} instrument at the Kitt Peak National Observatory. Additional CCF line profile analysis with high resolution spectra would help confirm that these candidates are not double-lined spectroscopic binaries (SB2s). 

Some candidates pass all of our selection criteria but have caveats which make them more prone to being false positives. These caveats consist of weak phase coverage, multiple periodogram peaks, and aliases that make other RV fits just as likely as those presented. More observations are required to fully constrain some of our new candidate parameters. 

Specifically we note the poor phase coverage of MARVELS-12b (figure \ref{fig:marv-13}). Given its high eccentricity, this candidate could have failed to pass criterion \ref{cri9}; however, this star had at least one fit pass all 11 initial criteria. When fitting some stars a final time with RVLIN we may find a better fit with higher eccentricity, which also passes the periodogram FAP test. More phase coverage is needed to validate this candidate, but given that at least one fit can pass all 11 criteria and it passes both our stellar contamination and periodogram tests we put a high likelihood on this star having some type of low-mass companion. 

\subsection{Statistical Results \& Conclusions}

\begin{figure}
\center
\includegraphics[width=\columnwidth]{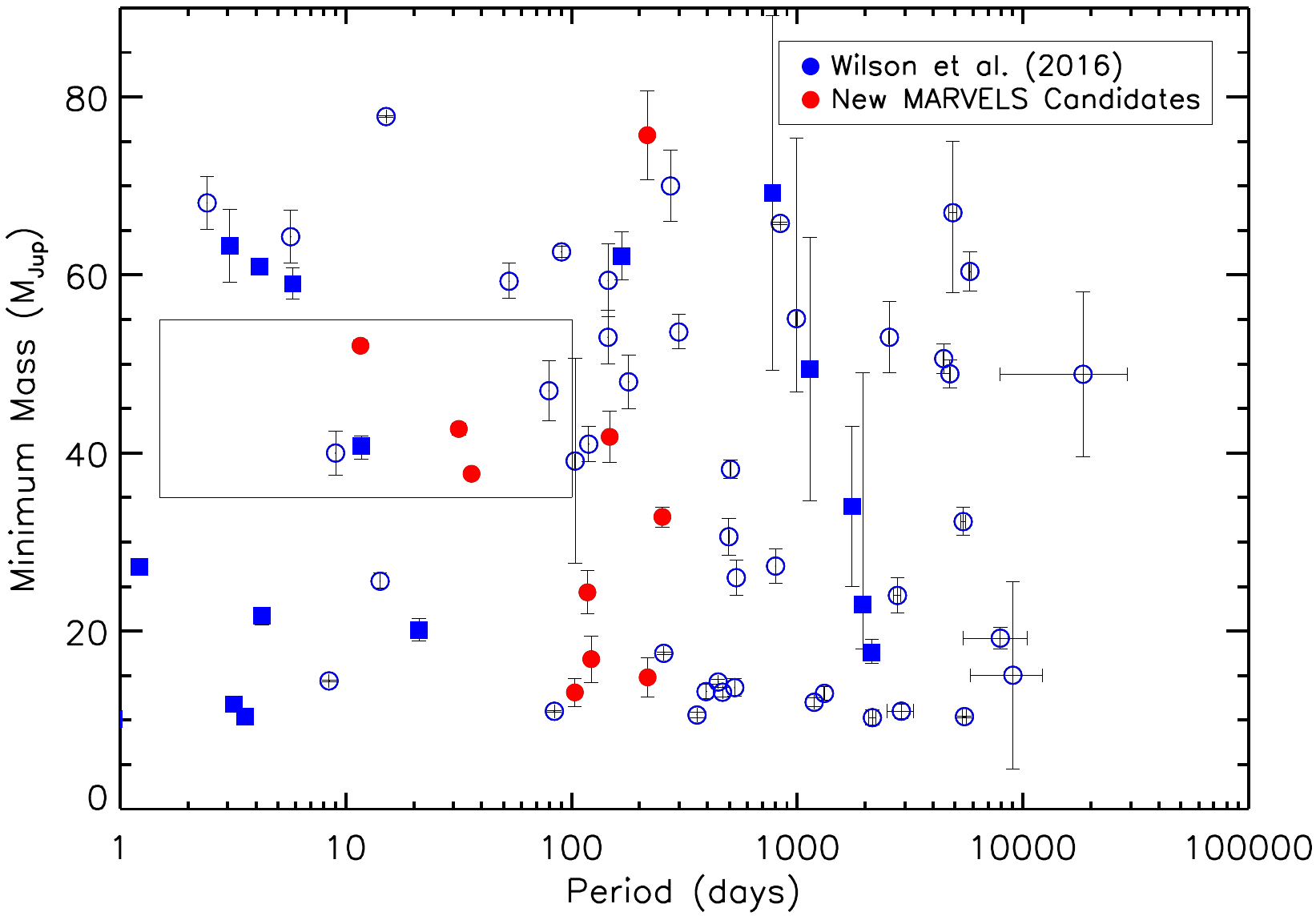}
\caption{Minimum mass vs period for current BD candidates with projected masses of 10 M$_{\text{Jup}}$ $\le$ M sin$\textit{i}$ $\le$ 80 M$_{\text{Jup}}$. Blue symbols are BD candidates presented in Table 7 of \citet{Wilson2016}. Objects from \citet{Wilson2016} with known masses are indicated using blue filled squares and objects from \citet{Wilson2016} with only minimum mass measurements are shown as blue open circles. Red filled circles represent the new MARVELS BD candidates presented in this paper. The rectangular area represents the `driest' part of the brown dwarf desert introduced by \citet{MaGe2014}.}
\label{fig:period_msini}
\end{figure}

\citet{MaGe2014} collected data from literature about currently known BD candidate companions around FGK-type stars. While examining the period-mass distribution of BDs they found a statistically significant gap in the short period (P $<$ 100 d) and medium-mass region (35 M$_{\text{Jup}}$ $<$ M$_{\text{BD}}$ $<$ 55 M$_{\text{Jup}}$), which they deemed the `driest' part of the brown dwarf desert and likely to be real as it is unlikely caused by detection sensitivity (RV precision) or observation biases (due to survey incompleteness). Three of our new BD candidates, however, are within this driest part of the desert. Figure \ref{fig:period_msini} shows the M sin$\textit{i}$ vs period plot of \citet{MaGe2014}, but now with candidates from Table 7 of \citet{Wilson2016} combined with our new MARVELS candidates. 

\begin{figure}
\center
\includegraphics[width=\columnwidth]{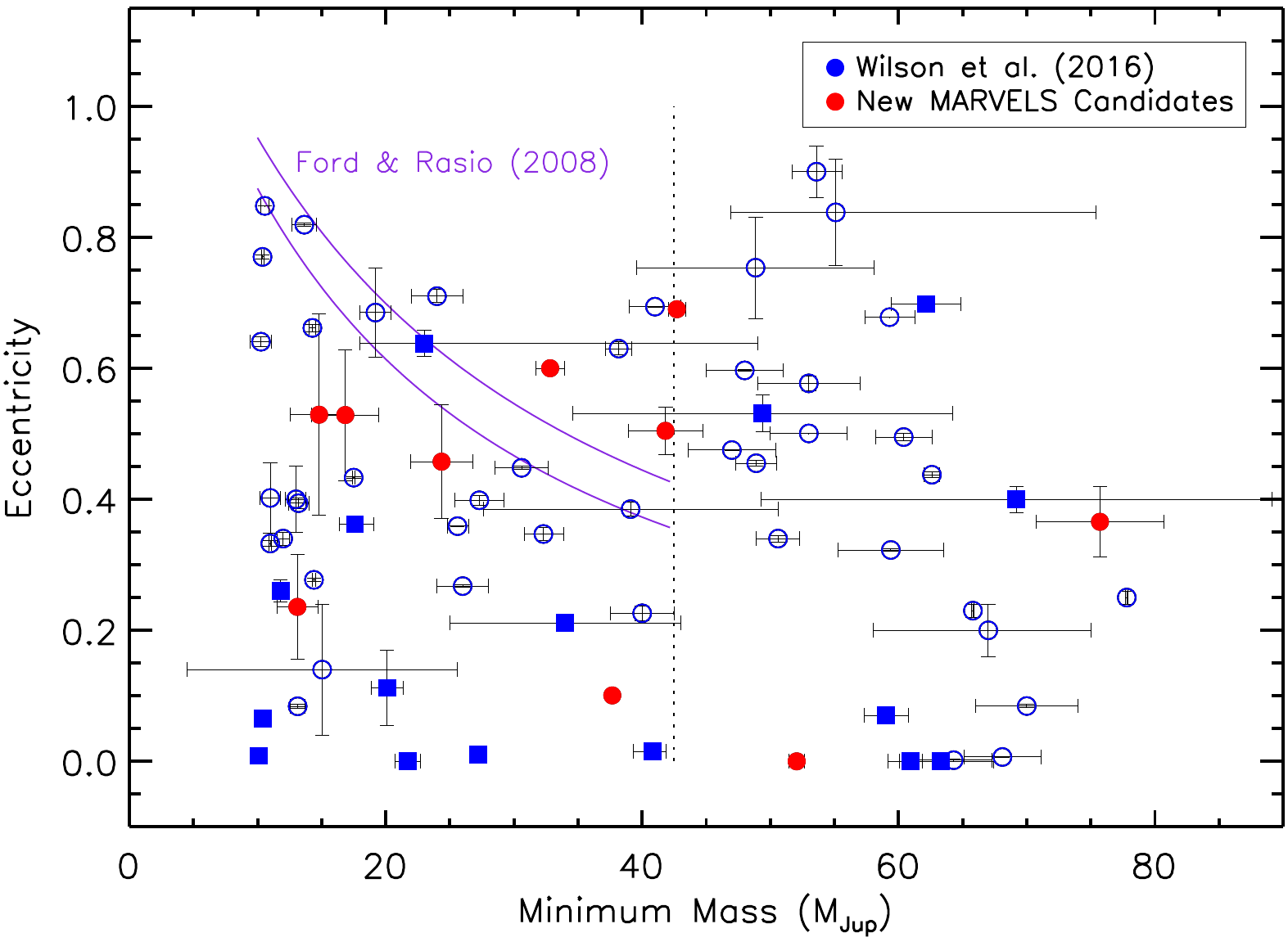}
\caption{Eccentricity vs minimum mass for current BD candidates with projected masses of 10 M$_{\text{Jup}}$ $\le$ M sin$\textit{i}$ $\le$ 80 M$_{\text{Jup}}$. As in figure \ref{fig:period_msini}, blue symbols are BD candidates presented in Table 7 of \citet{Wilson2016} and red filled circles represent the new MARVELS BD candidates presented in this paper. Two solid curves show predictions of BD eccentricity after scattering with another object of 20 or 25 M$_{\text{Jup}}$, respectively from bottom to top. These curves are calculated using a fitting formula of a planet-planet scattering model from \citet{FordRasio2008}. The vertical dotted line shows M sin$\textit{i}$ = 42.5 M$_{\text{Jup}}$.}
\label{fig:msini_ecc}
\end{figure}

As previously discussed, \citet{MaGe2014} concluded that the orbital eccentricities for BDs with masses lower than 42.5 M$_{\text{Jup}}$ and those higher than 42.5 M$_{\text{Jup}}$ have noticeable differences in their orbital eccentricities. Their study showed that BDs below $\sim$42.5 M$_{\text{Jup}}$ trend such that more massive BDs tend to have lower maximum eccentricities, similar to giant planets, while BDs above this mass threshold do not show such a trend, instead showing more diversity in their eccentricities. This trend is consistent with less massive BD formation occurring in protoplanetary discs and subsequential scattering with other objects in the disc pumping these BDs to higher eccentricities. In the scattering model \citep{Chatterjee2008,FordRasio2008}, it is harder to pump a higher mass BD to higher eccentricity. Figure \ref{fig:msini_ecc} shows the eccentricity vs M sin$\textit{i}$ plot from \citet{MaGe2014}, now with \citet{Wilson2016} BDs and new MARVELS BD candidates as in figure \ref{fig:period_msini}. The two violet curves show eccentricities if BDs were scattered by another object of 20 or 25 M$_{\text{Jup}}$. We see a few outliers from \citet{Wilson2016} as well as one new MARVELS BD candidate in the high eccentricity region above the \citet{FordRasio2008} curves; however, in general our new BD candidates still support a two population trend.

Two of our candidates have host stars that are classified as subgiants according to \citet{Troup2016} criteria: T$_{\text{eff}}$ $>$ 4800 K, log \textit{g} $\ge$ 3.7 + 0.1[Fe/H] if T$_{\text{eff}}$ < 5500 K, and  log \textit{g} $\le$ 4 - (7 x 10$^{-5}$)(T$_{\text{eff}}$ - 8000 K); the third criterion was derived by mapping the main-sequence turnoff in Dartmoth isochrones. This includes TYC 3097-00398-1 (MARVELS-15, log \textit{g} $\sim$ 3.85 dex) which hosts the $\sim$98 M$_{\text{Jup}}$ stellar companion (figure \ref{fig:marv-17}) and TYC 3469-00492-1 (MARVELS-12, log \textit{g} $\sim$ 3.78 dex) which hosts the $\sim$33 M$_{\text{Jup}}$ BD candidate with a $\sim$253 day period (figure \ref{fig:marv-13}). 

The host stars of our 10 new BD candidates (M sin$\textit{i}$ $\le$ 80 M$_{\text{Jup}}$) have a mean metallicity of [Fe/H] = 0.04 $\pm$ 0.07 dex. When including MARVELS-4b, MARVELS-5b, MARVELS-6b, and MARVELS-7c with our new candidates we find a mean metallicity of [Fe/H] = 0.03 $\pm$ 0.08 dex. This result agrees with \citet{MataSanchez2014} who study chemical abundances of stars with BD companions. \citet{MataSanchez2014} find that stars with BD companions have $\alpha$-elements and iron peak elements at about solar abundance, whereas stars with low-mass and high-mass planets have [$X_{\alpha}$/H] and [$X_{\text{Fe}}$/H] peak abundaces of $\sim$0.1 dex and $\sim$0.15 dex, respectively. Three of the 10 new BD candidates have host stars with subsolar metallicity. Core accretion, the giant planet formation mechanism dependent on high metallicity \citep{IdaLin2004}, is less likely than other mechanisms for these BD candidates. 

In this work we sifted through MARVELS data to create a sample of $\sim$2,340 dwarf/subgiant stars with reasonable RV coverage ($\ge$10 observations) and time baselines ($\ge$ 300 days) and found 10 new BD candidates. Thus, including the three previously found MARVELS BDs with periods less than 300 days (MARVELS-4b, \citet{Ma2013}; MARVELS-5b, \citet{Jiang2013}; MARVELS-6b, \citet{DeLee2013}), we estimate the BD occurrence rate for solar-type stars up to $\sim$300 days (approximately half the typical baseline of our stellar sample) to be $\sim$0.56\%. Unlike \citet{Troup2016}, our results do not show a truncated brown dwarf desert, but one that extends up to at least 1 AU. However, \citet{Troup2016}'s APOGEE sample is dominated by evolved stars, whereas our stellar sample was designed to exclude giant stars. Our work gives further support for the brown dwarf desert around solar-type stars.

\section*{Acknowledgements}

Funding for the MARVELS multi-object Doppler instrument was provided by the W. M. Keck Foundation and NSF with grant AST-0705139. The MARVELS survey was partially funded by the SDSS-III consortium, NSF grant AST-0705139, NASA with grant NNX07AP14G, and the University of Florida. Funding for SDSS-III has been provided by the Alfred P. Sloan Foundation, the Participating Institutions, the National Science Foundation, and the U.S. Department of Energy Office of Science. L.G. would like to thank the financial support provided by CAPES/PNPD Fellowship. G.F.P.M. acknowledges the CNPq research grant 474972/2009-7.

\balance


\onecolumn
\appendix

\section{Vetting Criteria Dependencies} \label{sec:cridep}
\begin{figure}
\center
\includegraphics[width=0.63\columnwidth]{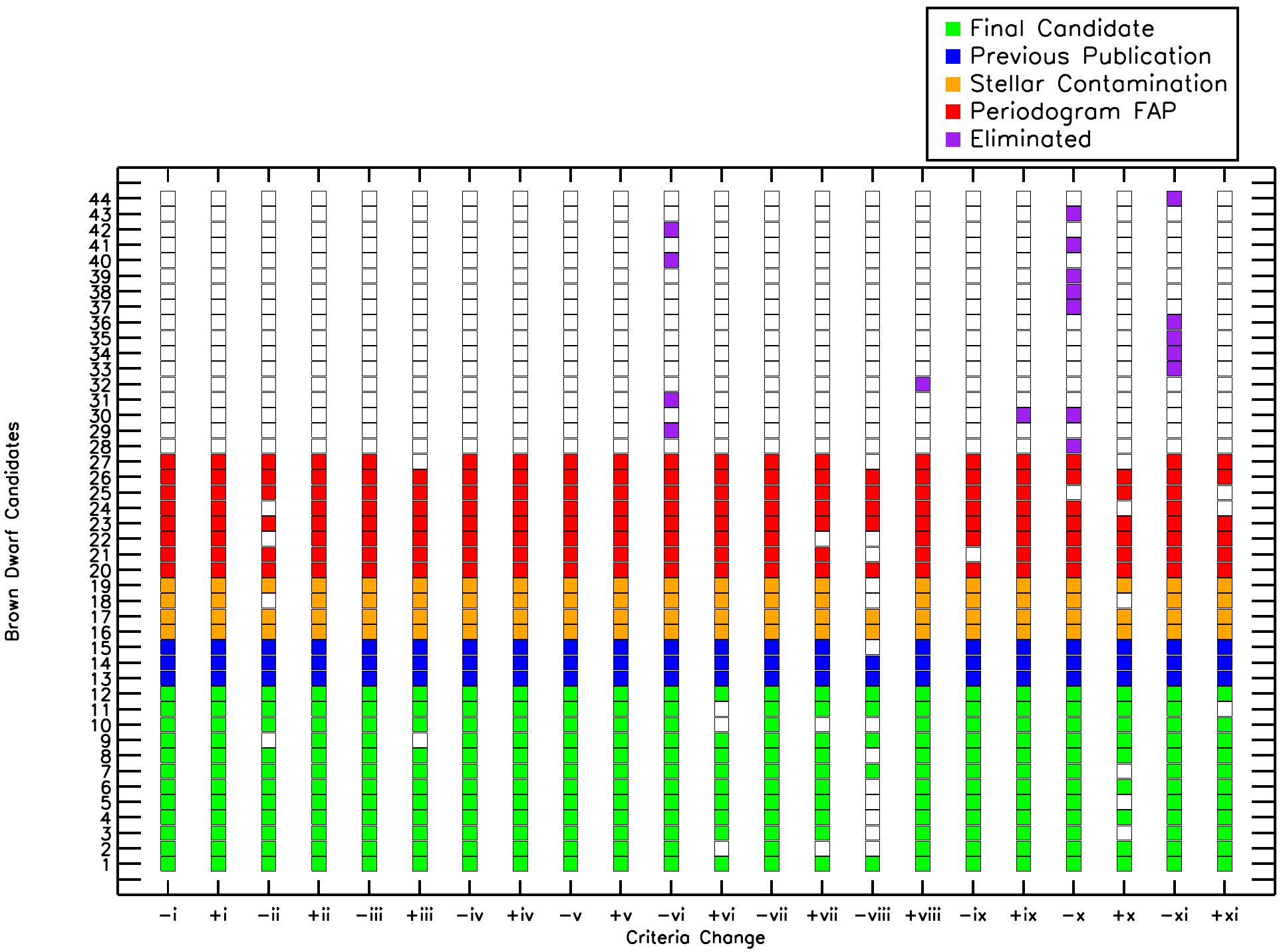}
\caption{Vetting results from varying each criterion -1$\sigma$ and +1$\sigma$, as shown in table \ref{tab:cridep}. The x-axis shows the criterion variation, e.g. -i represents changing the criterion \ref{cri1} dependence by -1$\sigma$ and keeping all other criteria dependencies the same. Colored squares represent a star/candidate that passed all criteria with that variation, while empty squares represent when a candidate did not pass. BD candidates 1 - 12 on the y-axis of the plot, shown with green squares (when all criteria are passed), represent the final 12 BD candidates, MARVELS-8b - MARVELS-19b, respectively. BD candidates 13, 14, 15, shown with blue squares, represent the previously published BD candidates that made it through all vetting processes: MARVELS-3b \citep{Wisniewski2012}, MARVELS-5b \citep{Jiang2013}, and MARVELS-6b \citep{DeLee2013}, respectively. BD candidates 16-19, shown with orange squares, represent the four candidates removed after our stellar contamination tests. BD candidates 20-27, shown as red squares, represent the eight candidates removed after our periodogram testing. BD candidates 28-44, shown as purple squares, are candidates that did not pass the original 11 criteria, but will occasionally pass all 11 criteria when a certain criterion is loosened.}
\label{fig:cri_comp}
\end{figure}

Here we determine rough estimates for the dependence of our results on each vetting criterion defined in $\S$ \ref{sec:initcomsec}. For each criterion we add and subtract 1$\sigma$ from the main dependence and determine the number of stars that have companion fits that pass all 11 criteria, labeled as \# pass in table \ref{tab:cridep}, without changing any of the other 10 criteria. Out of these stars with passing companion fits we then find the number of companions that are within the mass range we consider for BD candidates (M sin$\textit{i}$ values within 10 M$_{\text{Jup}}$ - 110 M$_{\text{Jup}}$ assuming a 1 M$_{\odot}$ host star), labeled as \#BD pass in table \ref{tab:cridep}. The original criteria have 84 stars with companion fits that pass all 11 criteria and 27 of them are within our mass range for BD candidate consideration. Figure \ref{fig:cri_comp} further illustrates criteria dependencies and shows the results of each BD candidate as we vary each criterion -1$\sigma$ and +1$\sigma$. As illustrated in both figure \ref{fig:cri_comp} and table \ref{tab:cridep}, the final number of candidates is most sensitive to the reduction of criterion \ref{cri8}. This is expected because making this criterion more strict will remove several candidates with relatively high semi-amplitudes compared to their velocity variations.

    \begin{table} 
    \center
    \caption{Vetting Criteria Dependencies}
    \label{tab:cridep}
   \begin{tabular}{cccccccc}
    \hline
   \hline
Criteria	& 	\multicolumn{1}{c}{original}		&	\multicolumn{3}{c}{ minus 1$\sigma$}  					&	\multicolumn{3}{c}{ plus 1$\sigma$}	 	\\
\cmidrule(lr){2-2} \cmidrule(lr){3-5}  \cmidrule(lr){6-8}
		&	dependence	&	dependence		&  \# pass	& \#BD pass	&	dependence	& \# pass 	& \#BD pass	\\
\hline
\ref{cri1} 	&	5\%	&	3.30\%	&	84	&	27	&	6.71\%	&	84	&	27	\\
\ref{cri2} 	&	$e > 0.934$	&	$e > 0.62$	&	76	&	23	&	$e > 0.99$	&	86	&	27	\\
\ref{cri3} 	&	$U_{N}V_{N} < 0.5 $	&	$U_{N}V_{N} < 0.33 $	&	84	&	27	&	$U_{N}V_{N} < 0.67 $	&	79	&	25	\\
\ref{cri4} 	&	$3 + 3(1 - V_{N})e$	&	$1.98 + 1.98(1 - V_{N})e$	&	85	&	27	&	$4.02 + 4.02(1 - V_{N})e$	&	80	&	27	\\
\ref{cri5} 	&	$3 + 3(1 - V_{N})e$	&	$1.98 + 1.98(1 - V_{N})e$	&	84	&	27	&	$4.02 + 4.02(1 - V_{N})e$	&	84	&	27	\\
\ref{cri6} 	&	$3 + 3(1 - V_{N})e$	&	$1.98 + 1.98(1 - V_{N})e$	&	95	&	31	&	$4.02 + 4.02(1 - V_{N})e$	&	77	&	24	\\
\ref{cri7} 	&	$3 + 3(1 - V_{N})e$	&	$1.98 + 1.98(1 - V_{N})e$	&	84	&	27	&	$4.02 + 4.02(1 - V_{N})e$	&	80	&	24	\\
\ref{cri8} 	&	$< 4$	&	$< 2.64$	&	43	&	14	&	$< 5.36$	&	85	&	28	\\
\ref{cri9} 	&	$e$ $>$ 0.4	&	$e$ $>$ 0.26	&	78	&	26	&	$e$ $>$ 0.54	&	86	&	28	\\
\ref{cri10} 	&	$\Delta T \ge 2P$	&	$\Delta T \ge 1.32P$	&	94	&	33	&	$\Delta T \ge 2.68P$	&	76	&	21	\\
\ref{cri11} 	&	$\ge$ 1/3	&	$\ge$ 0.22	&	89	&	32	&	$\ge$ 0.45	&	76	&	24	\\														
	\hline
	 \end{tabular}    
\end {table}    

\section{New MARVELS Substellar Companions}
\begin{figure*}
      	\centering
	\includegraphics[width=0.49\linewidth]{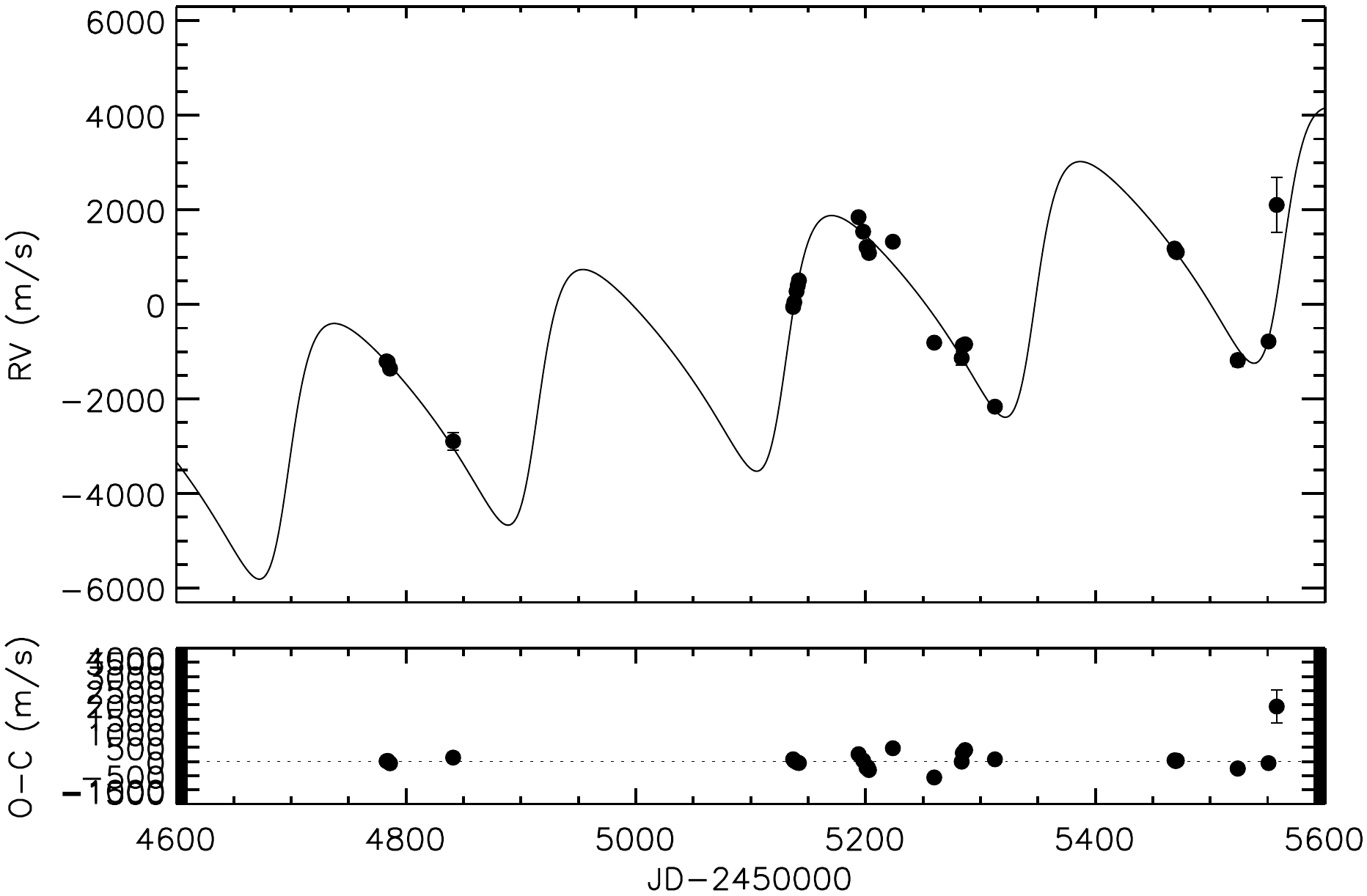}
	\hfill
	\centering
	\includegraphics[width=0.47\linewidth]{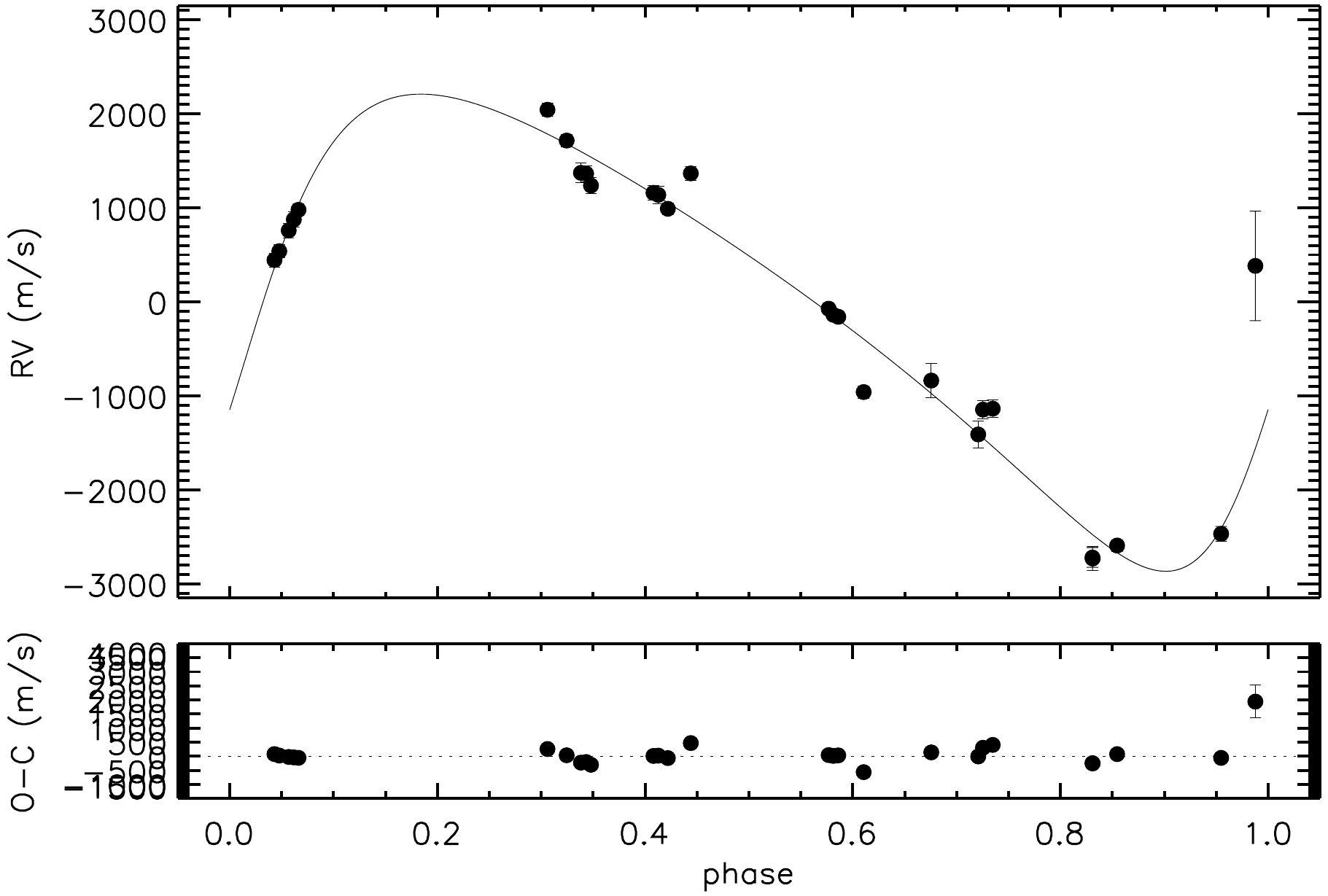}
	\hfill
	\centering
	\includegraphics[width=0.32\linewidth]{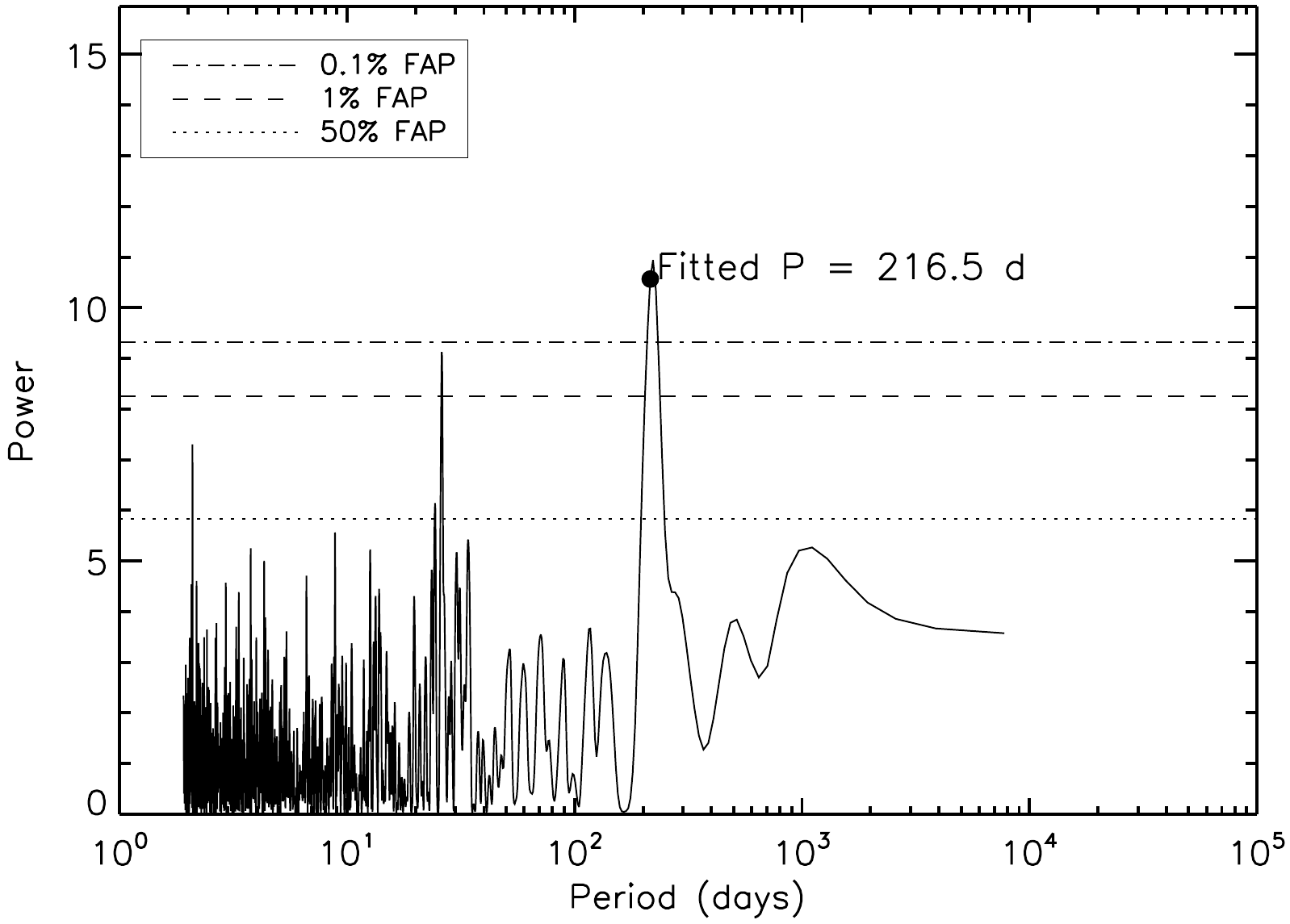}
	\hfill
	\centering
	\includegraphics[width=0.6\linewidth]{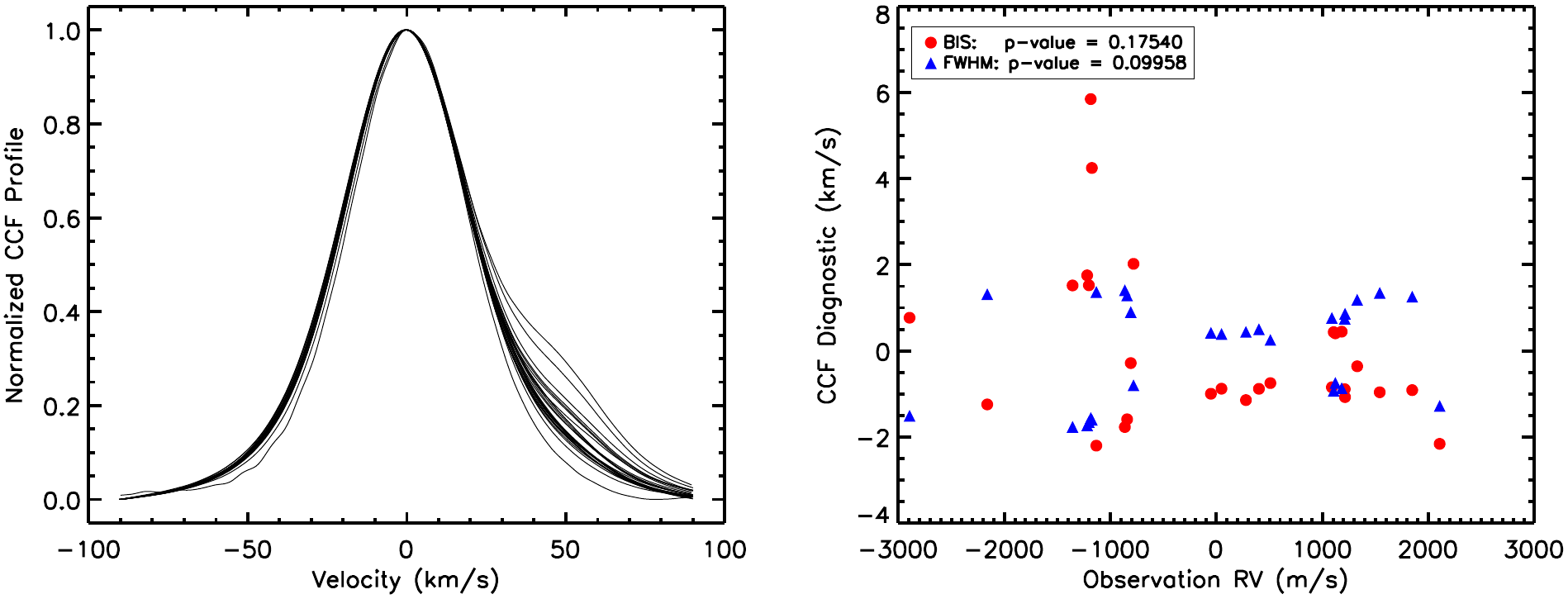}
	\hfill
\caption{\textbf{TYC 0173-02410-1, MARVELS-9b}: M sin$\textit{i}$ $\sim$76.0 M$_{\text{Jup}}$, Period $\sim$216.5 days, ecc $\sim$0.37.}
\label{fig:marv-9}
\end{figure*}

\begin{figure*}
      	\centering
	\includegraphics[width=0.49\linewidth]{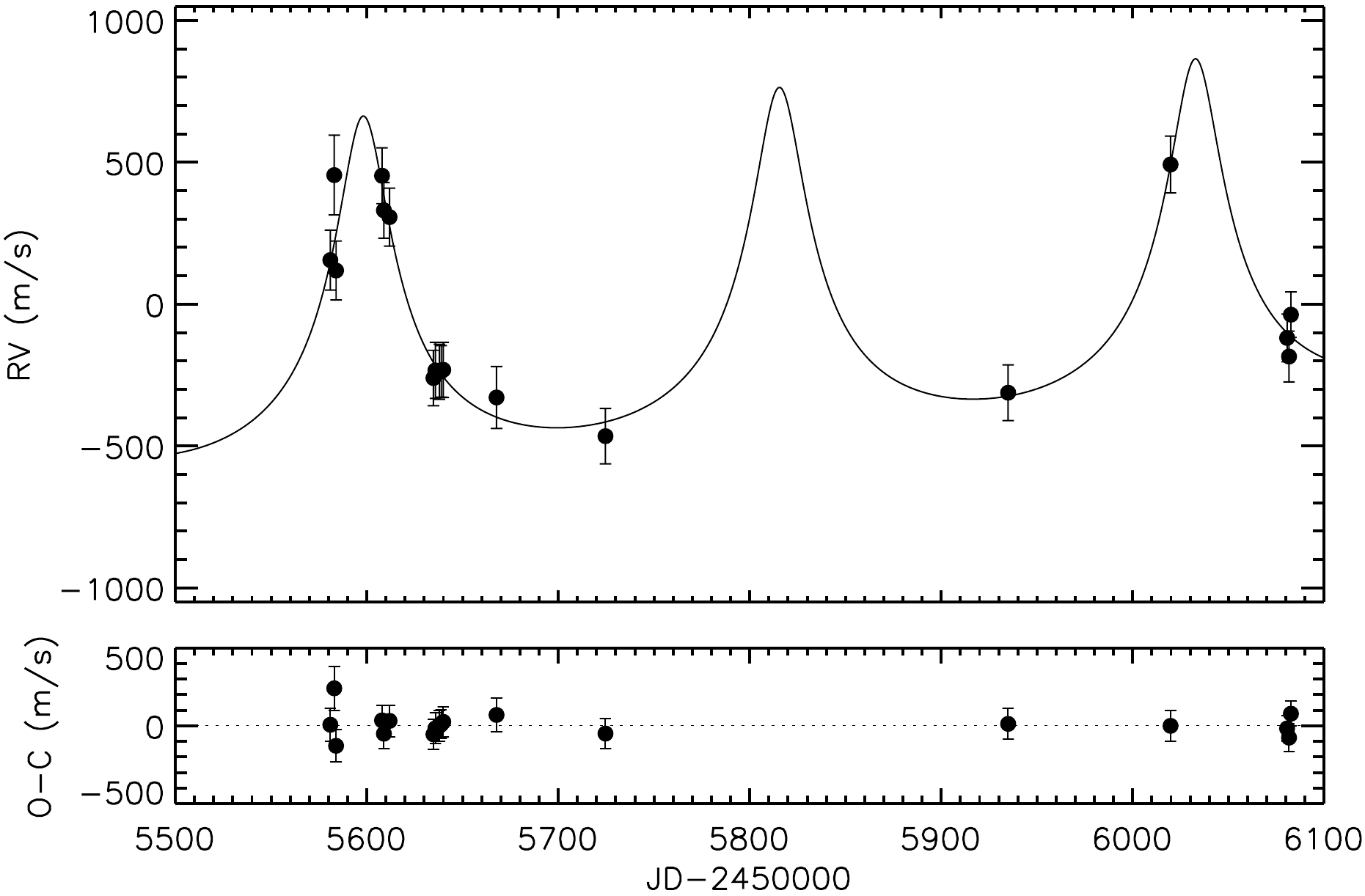}
	\hfill
	\centering
	\includegraphics[width=0.47\linewidth]{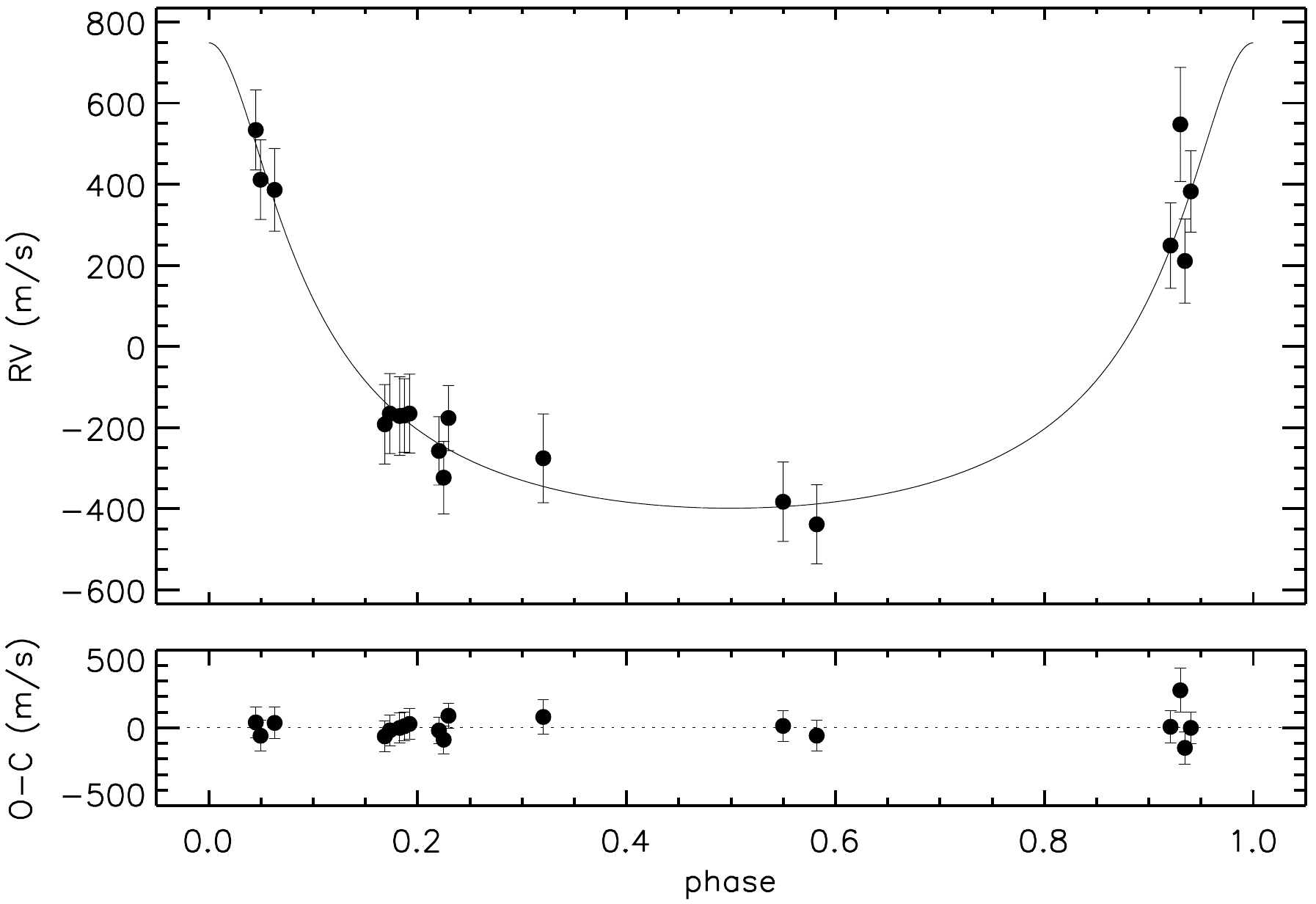}
	\hfill
	\centering
	\includegraphics[width=0.32\linewidth]{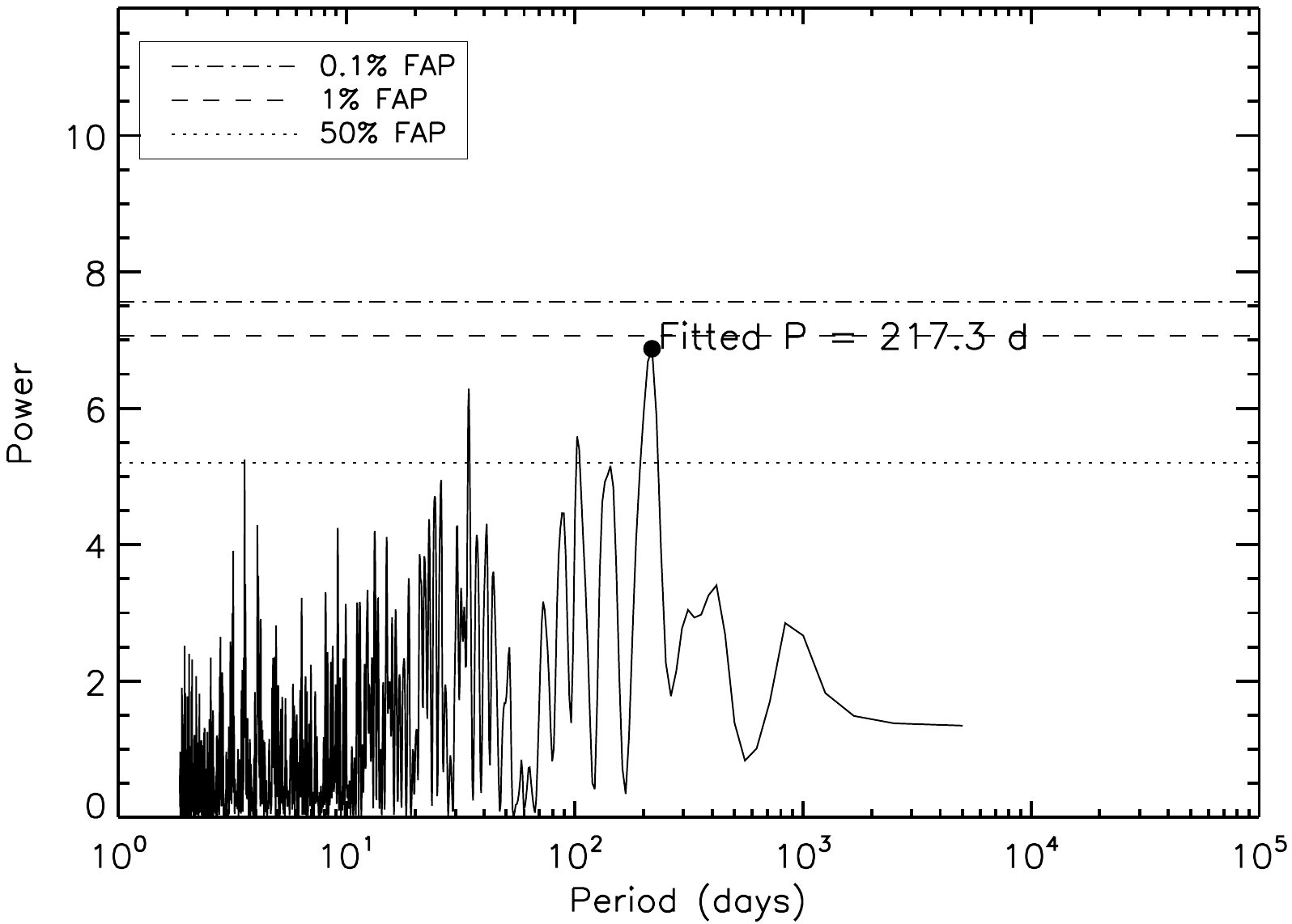}
	\hfill
	\centering
	\includegraphics[width=0.6\linewidth]{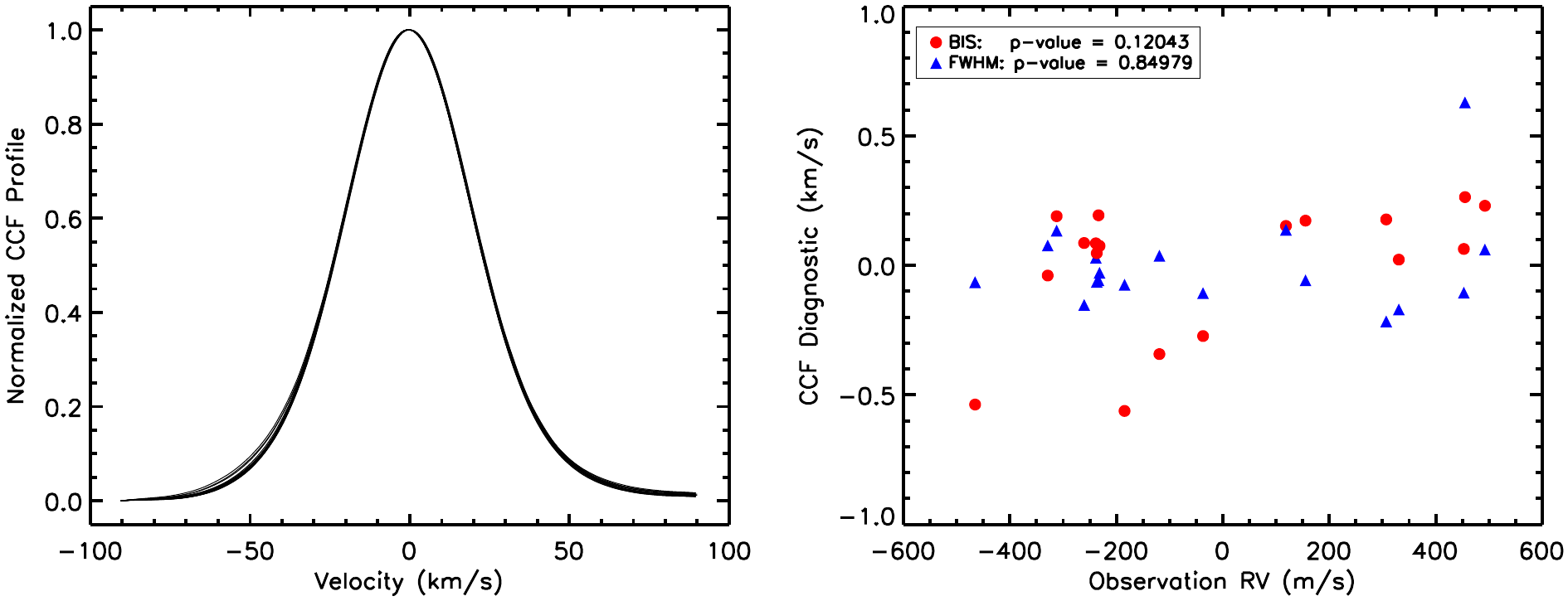}
	\hfill
\caption{\textbf{TYC 4955-00369-1, MARVELS-10b}: M sin$\textit{i}$ $\sim$14.8 M$_{\text{Jup}}$, Period $\sim$217.3 days, ecc $\sim$0.53.}
\label{fig:marv-11}
\end{figure*}

\begin{figure*}
      	\centering
	\includegraphics[width=0.49\linewidth]{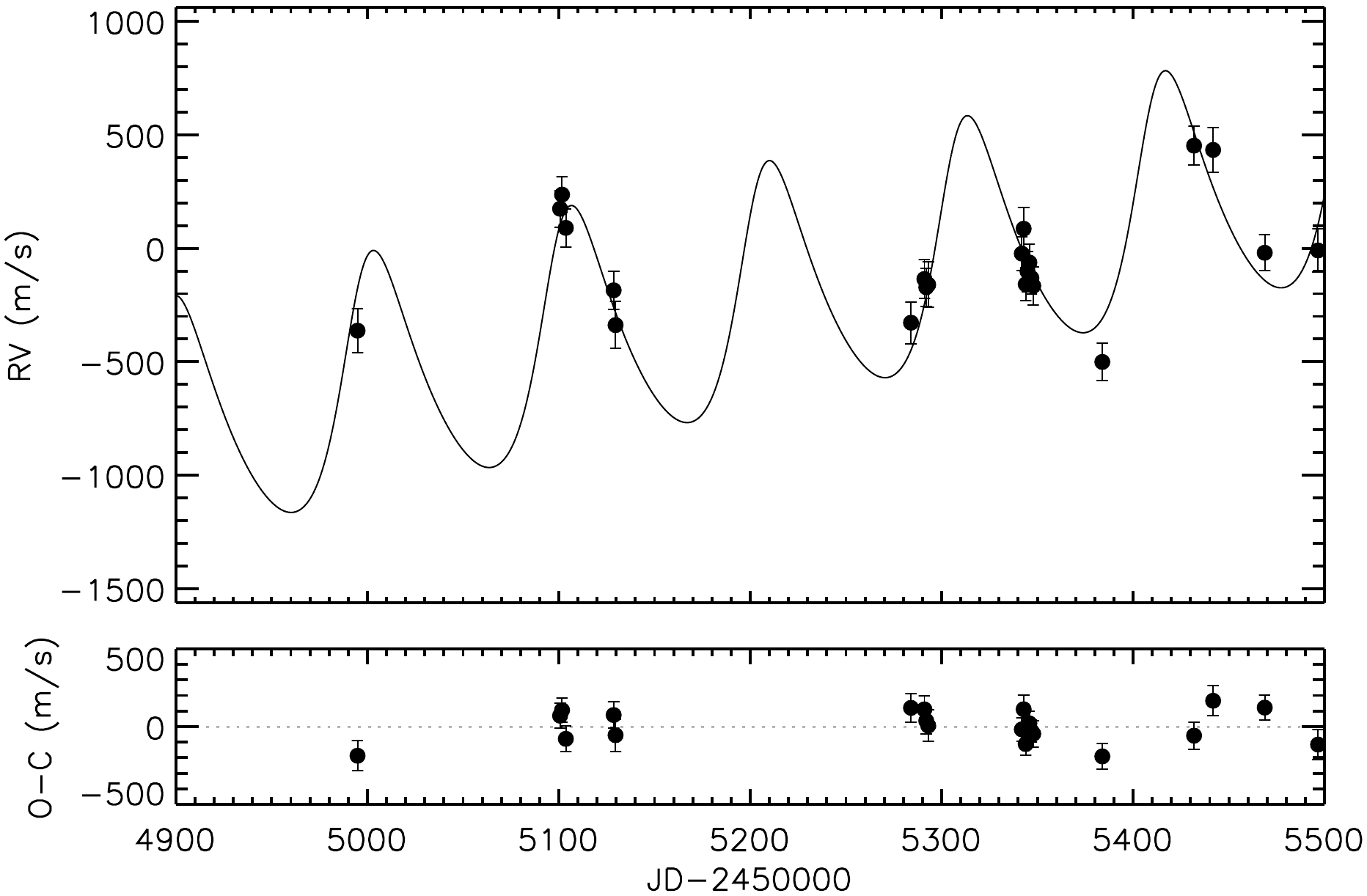}
	\hfill
	\centering
	\includegraphics[width=0.47\linewidth]{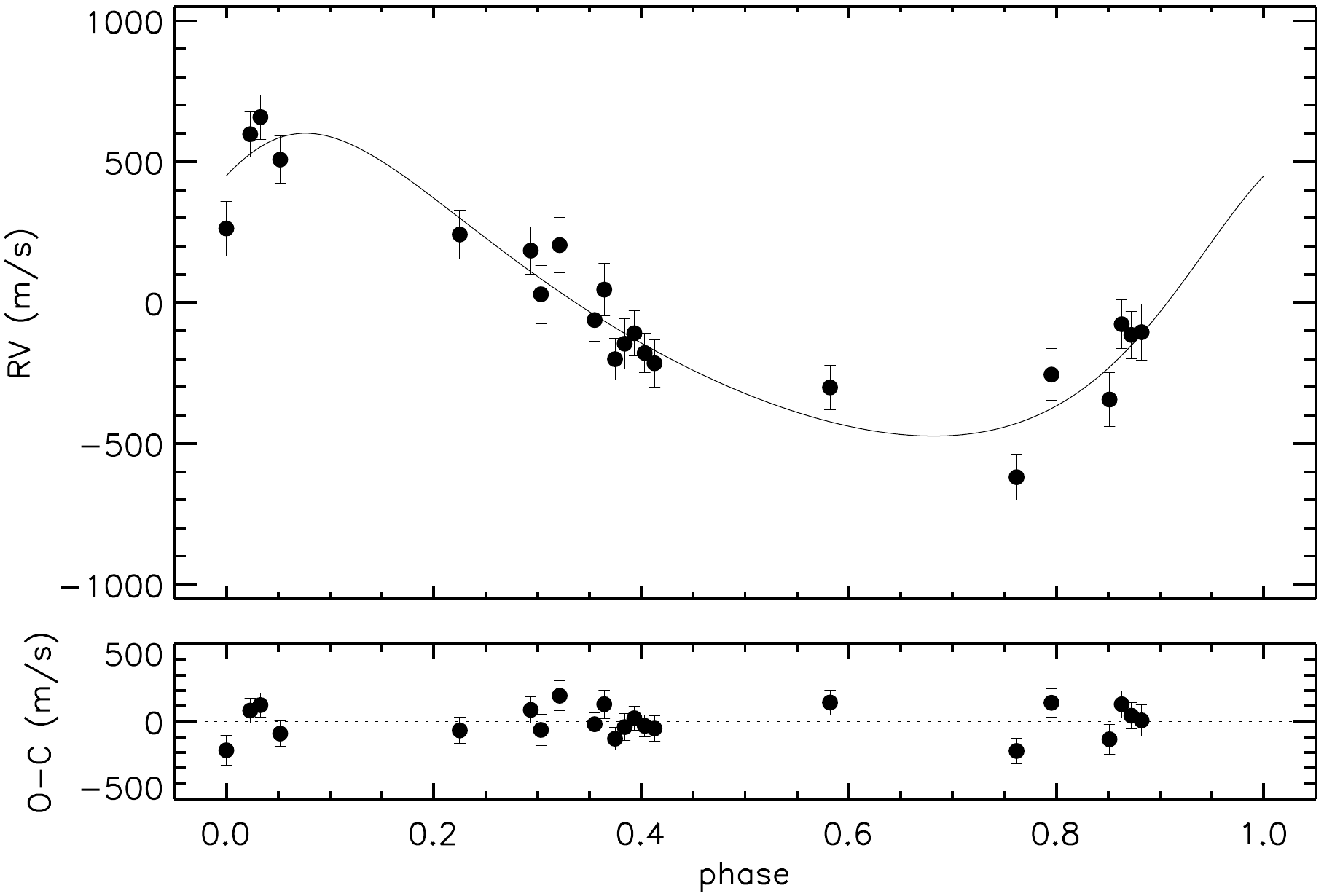}
	\hfill
	\centering
	\includegraphics[width=0.32\linewidth]{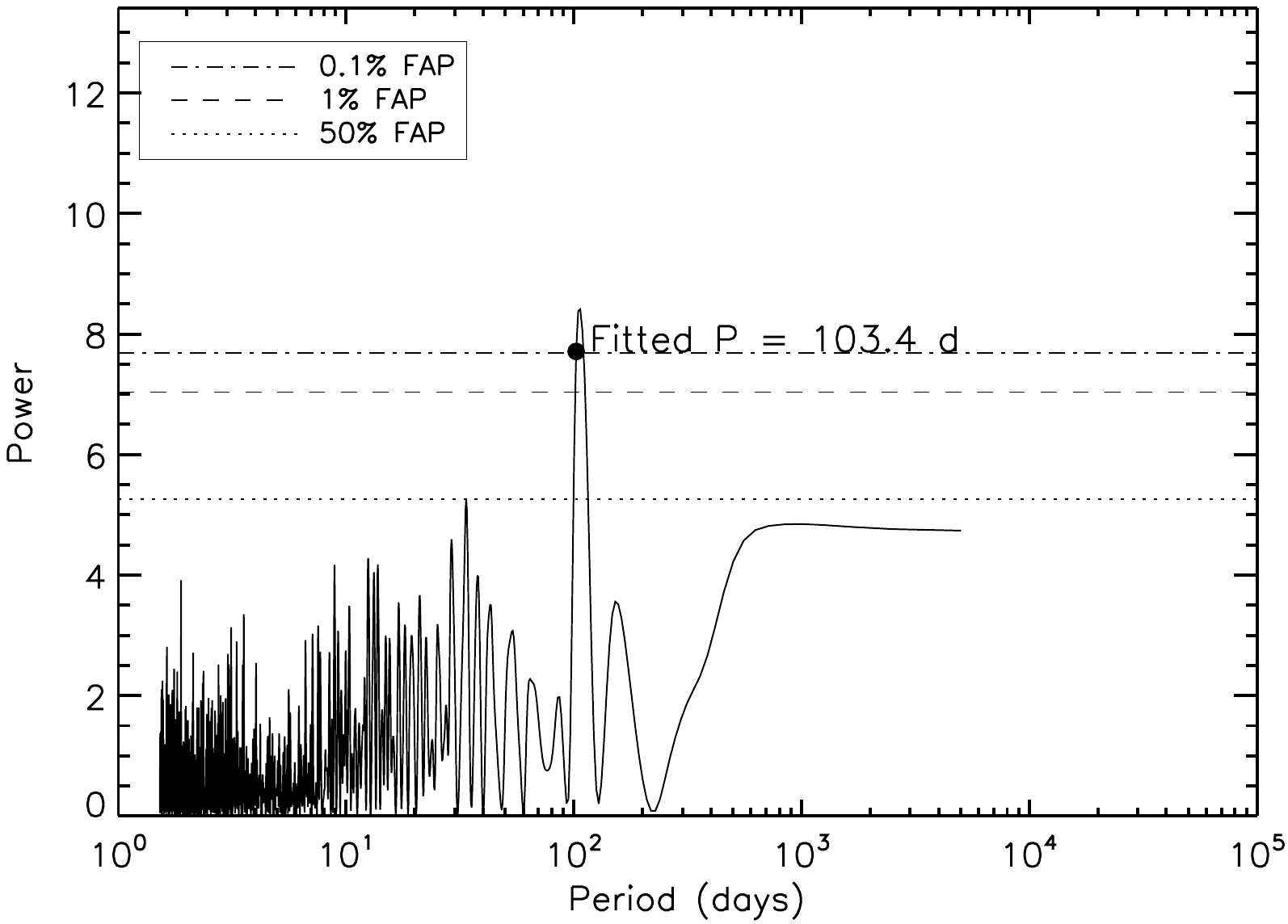}
	\hfill
	\centering
	\includegraphics[width=0.6\linewidth]{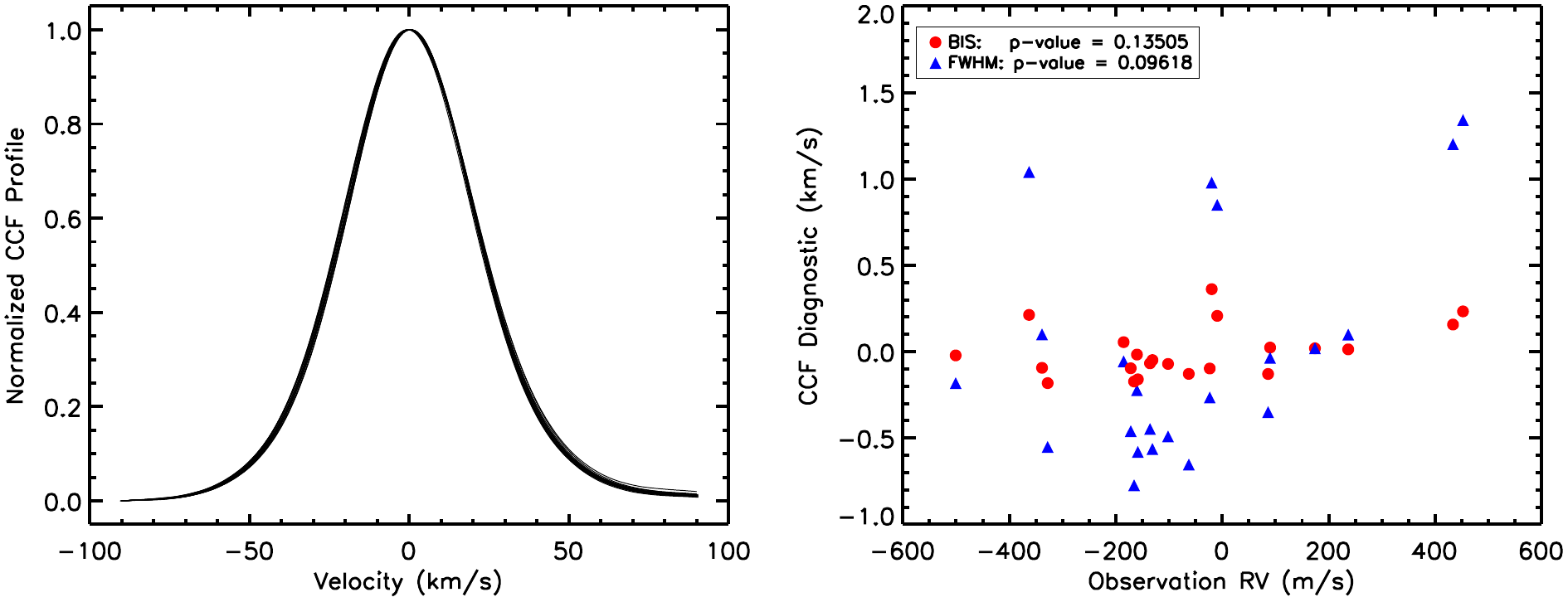}
	\hfill
\caption{\textbf{TYC 3547-01007-1, MARVELS-17b}: M sin$\textit{i}$ $\sim$13.1 M$_{\text{Jup}}$, Period $\sim$103.4 days, ecc $\sim$0.24.}
\label{fig:marv-19}
\end{figure*}

Here we include both the full-length (\textit{top left}) and phased (\textit{top right}) radial velocity plots, Lomb-Scargle periodograms (\textit{bottom left}), CCF line profiles (\textit{bottom middle}), and BIS and FWHM vs RV plots (\textit{bottom right}) for each new MARVELS companion candidate. Radial velocity measurements are obtained from MARVELS and processed with the UF2D pipeline. The top panel for each radial velocity plot shows the best fit to the data, and the bottom panel shows the best-fit residuals. Lomb-Scargle periodograms show False Alarm Probabilities (FAPs) that are determined via a randomized bootstrapping method with 10,000 iterations. The dotted lines represent 50\% FAP, dashed lines represent 1\% FAP, and the dashed and dotted lines represent 0.1\% FAP. Filled circles represent the best-fit period of each companion. The best-fit period must be above the 0.1\% FAP, unless companion's fit has an eccentricity $>$ 0.5, then the period's peak must only be within 90\% of the max peak of the bootstrapped data. MARVELS-9b (figure \ref{fig:marv-9}), MARVELS-10b (figure \ref{fig:marv-11}), and MARVELS-17b (figure \ref{fig:marv-19}) had the best-fit linear trend removed from the RV data before running the LS periodograms. The BIS and FWHM vs RV plots show the bisector velocity span (BIS) and the FWHM of the CCF line profiles as a function of radial velocity for each of the MARVELS candidate host stars. Both BIS and FWHM values have the mean value of the star subtracted before plotting. We show the p-value of the Kendall rank correlation between the BIS and RV and FWHM and RV values in each plot.

\begin{figure*}
      	\centering
	\includegraphics[width=0.49\linewidth]{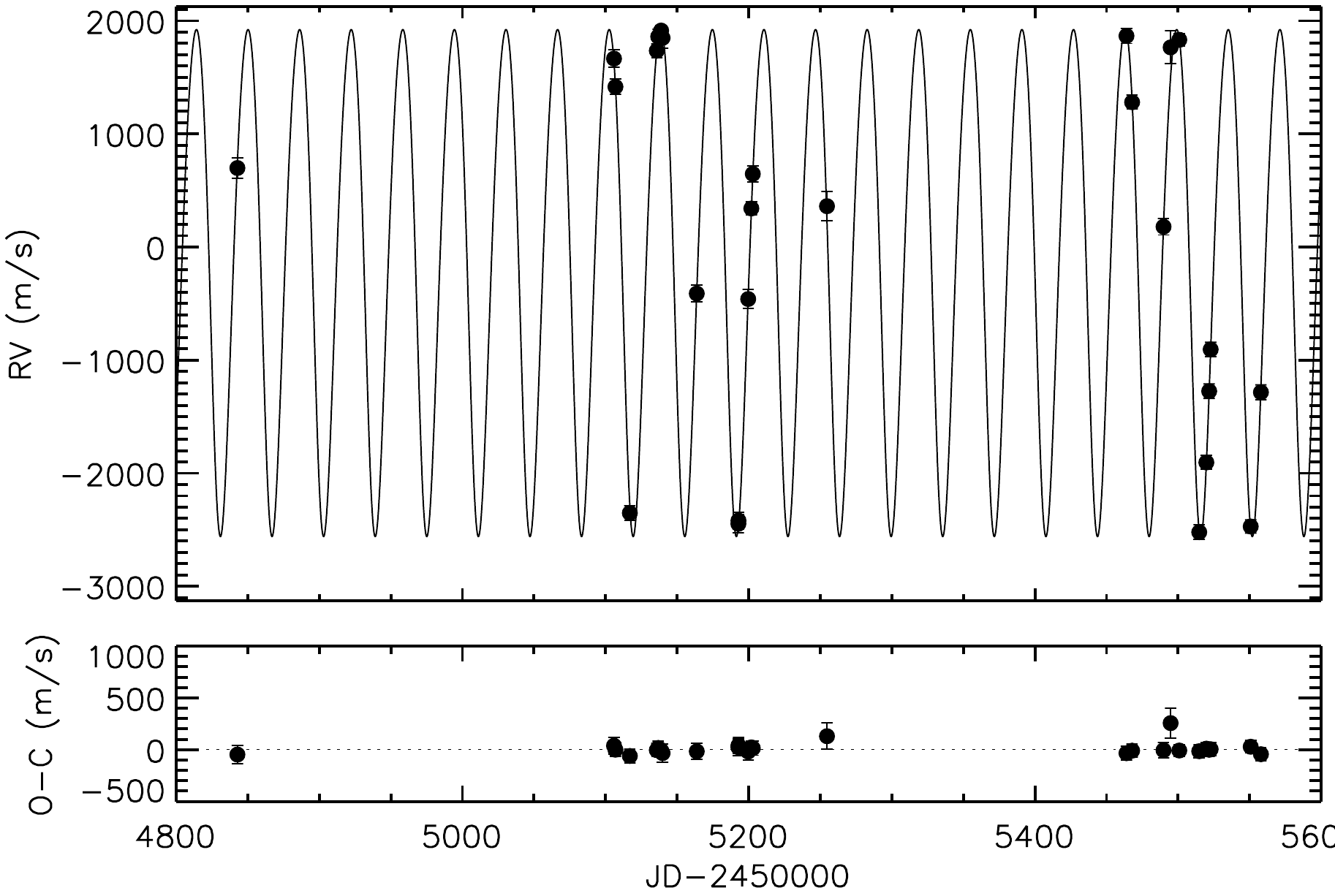}
	\hfill
	\centering
	\includegraphics[width=0.47\linewidth]{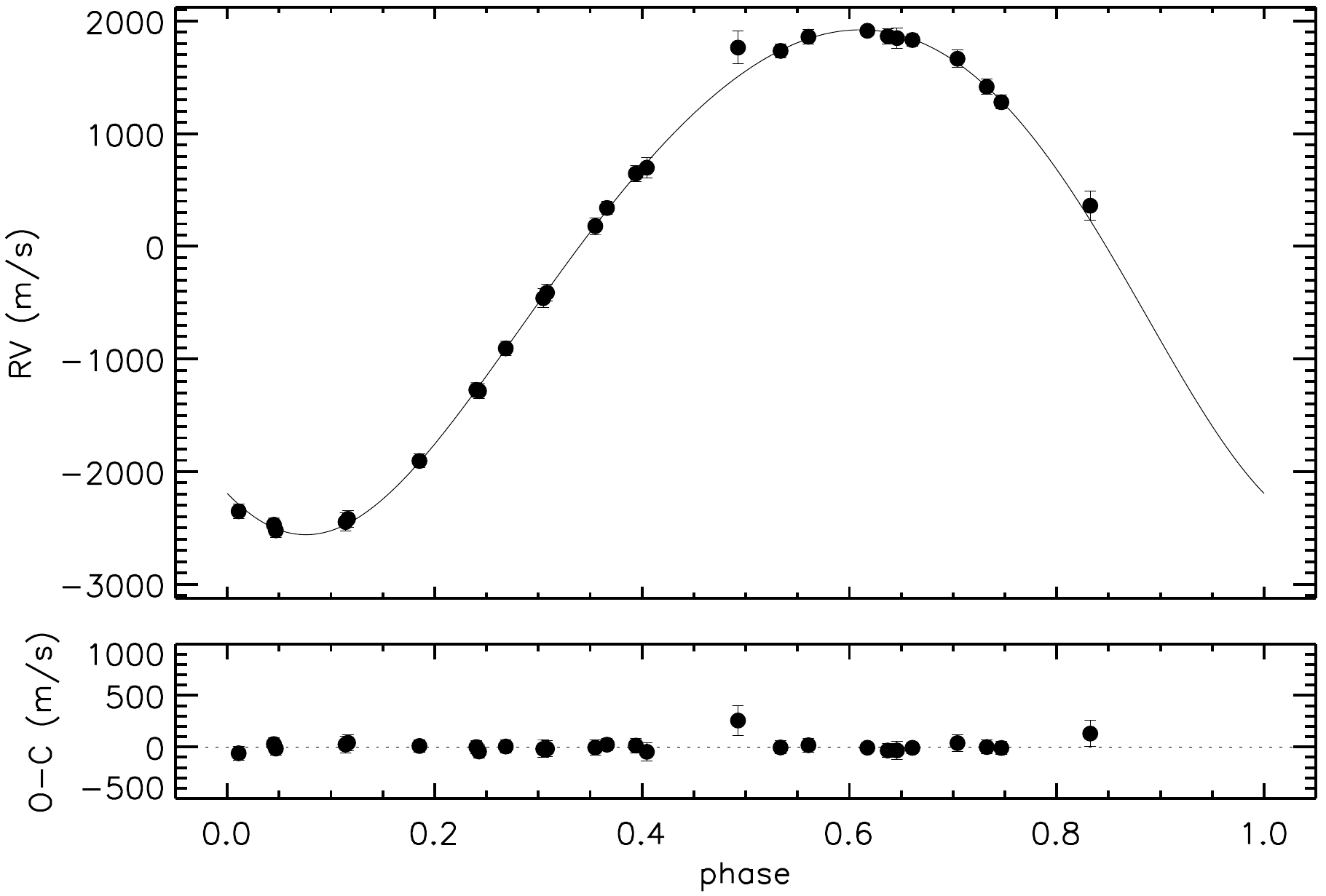}
	\hfill
	\centering
	\includegraphics[width=0.32\linewidth]{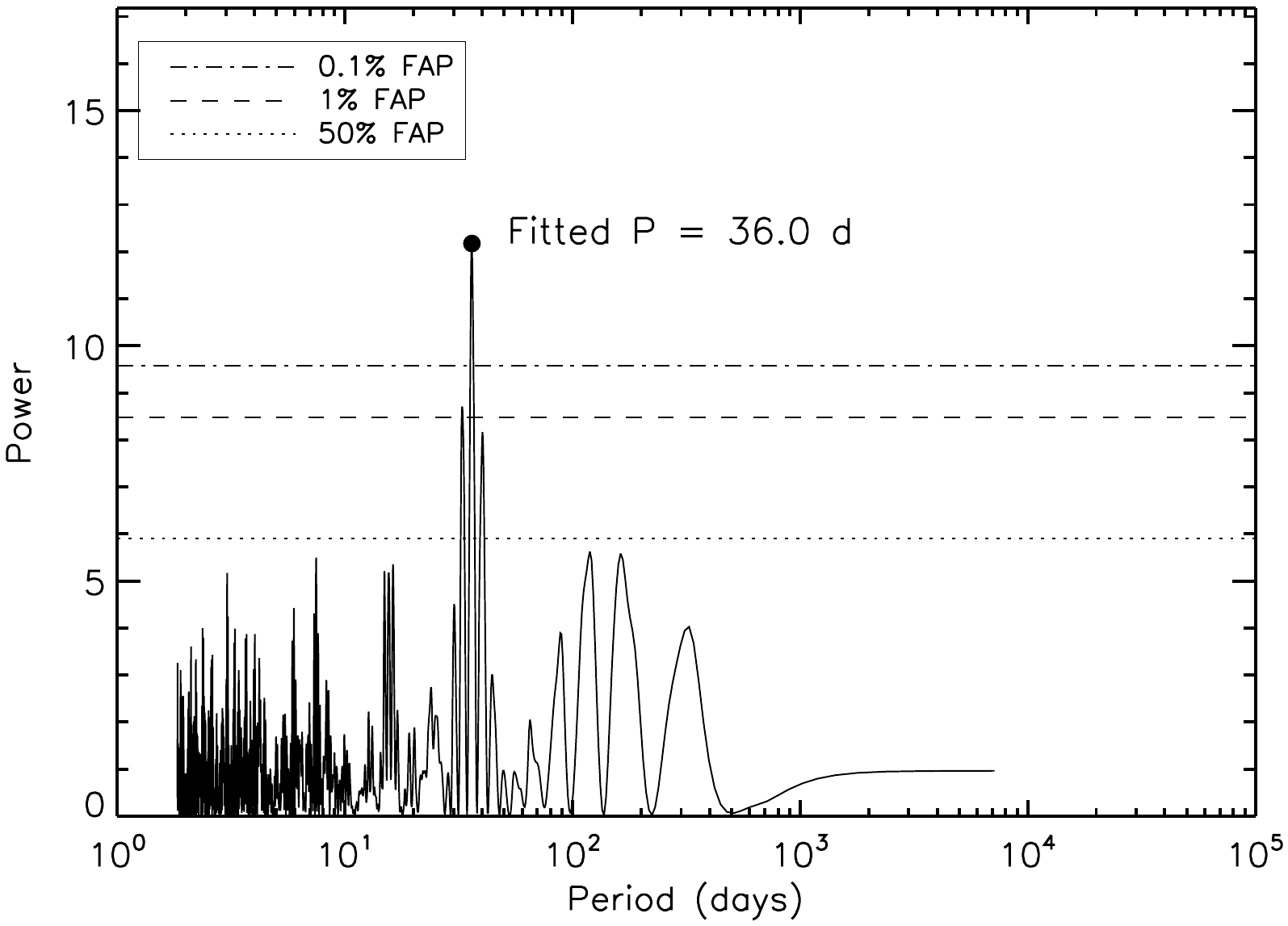}
	\hfill
	\centering
	\includegraphics[width=0.6\linewidth]{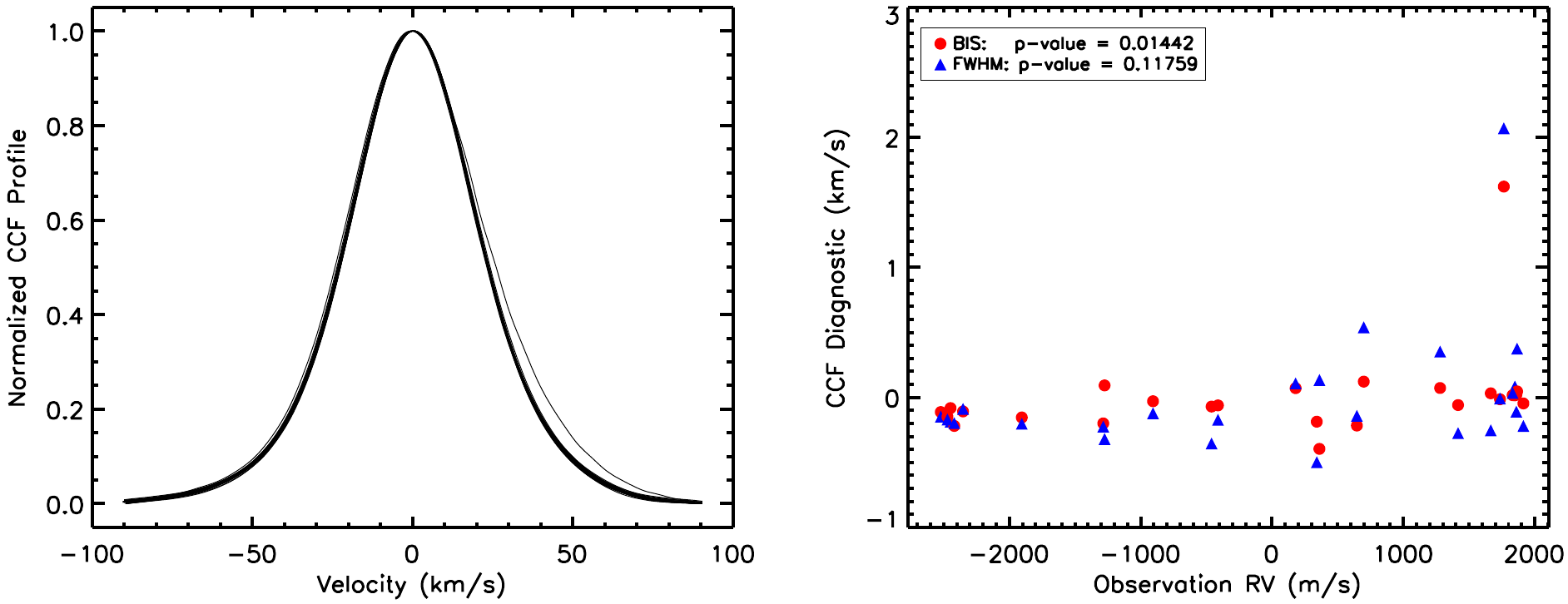}
	\hfill
\caption{\textbf{TYC 0123-00549-1 (MCKGS1-50), MARVELS-8b}: M sin$\textit{i}$ $\sim$37.7 M$_{\text{Jup}}$, Period $\sim$36.045 days, ecc $\sim$0.101.}
\label{fig:marv-8}
\end{figure*}
\begin{figure*}
      	\centering
	\includegraphics[width=0.49\linewidth]{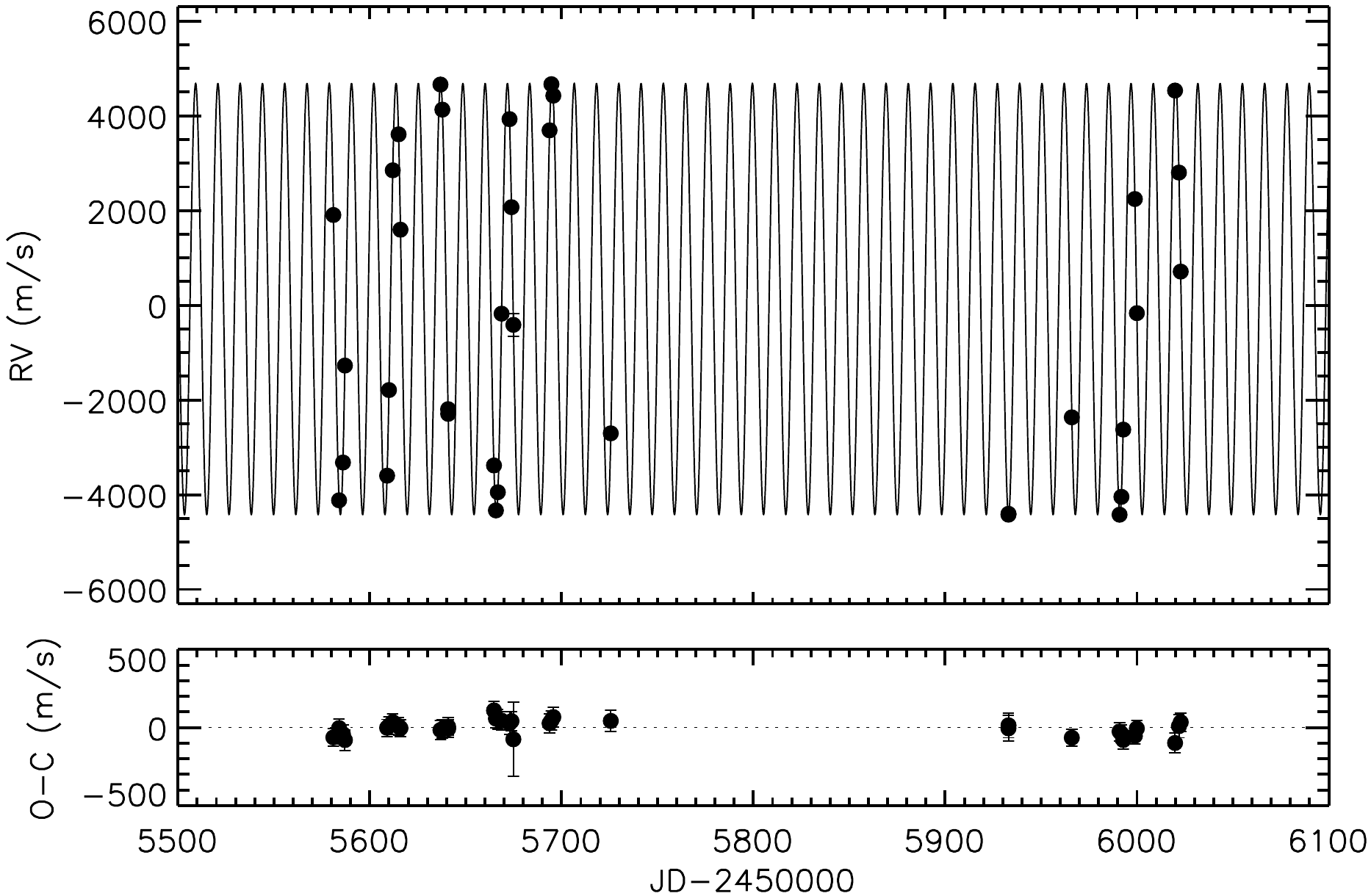}
	\hfill
	\centering
	\includegraphics[width=0.47\linewidth]{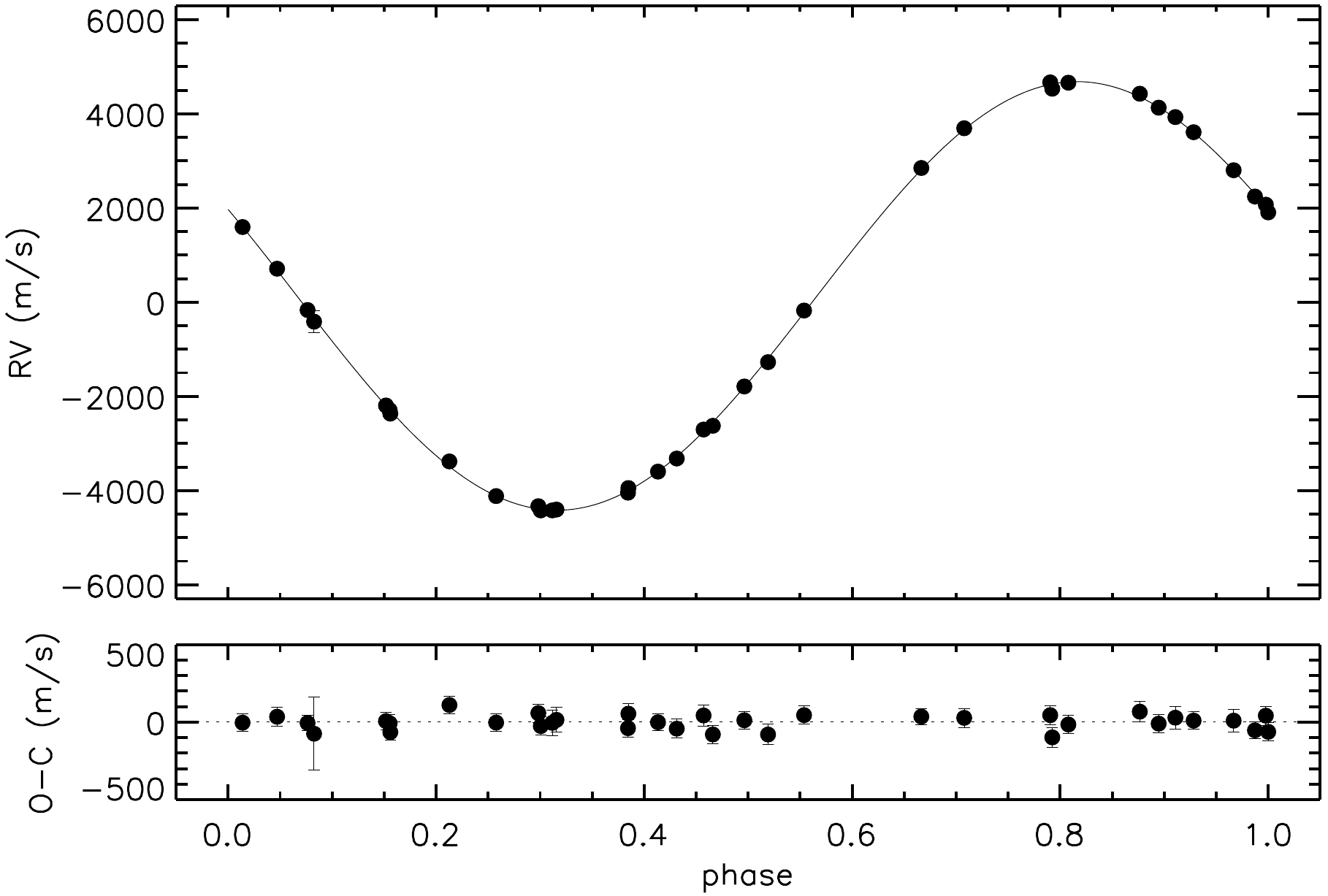}
	\hfill
	\centering
	\includegraphics[width=0.32\linewidth]{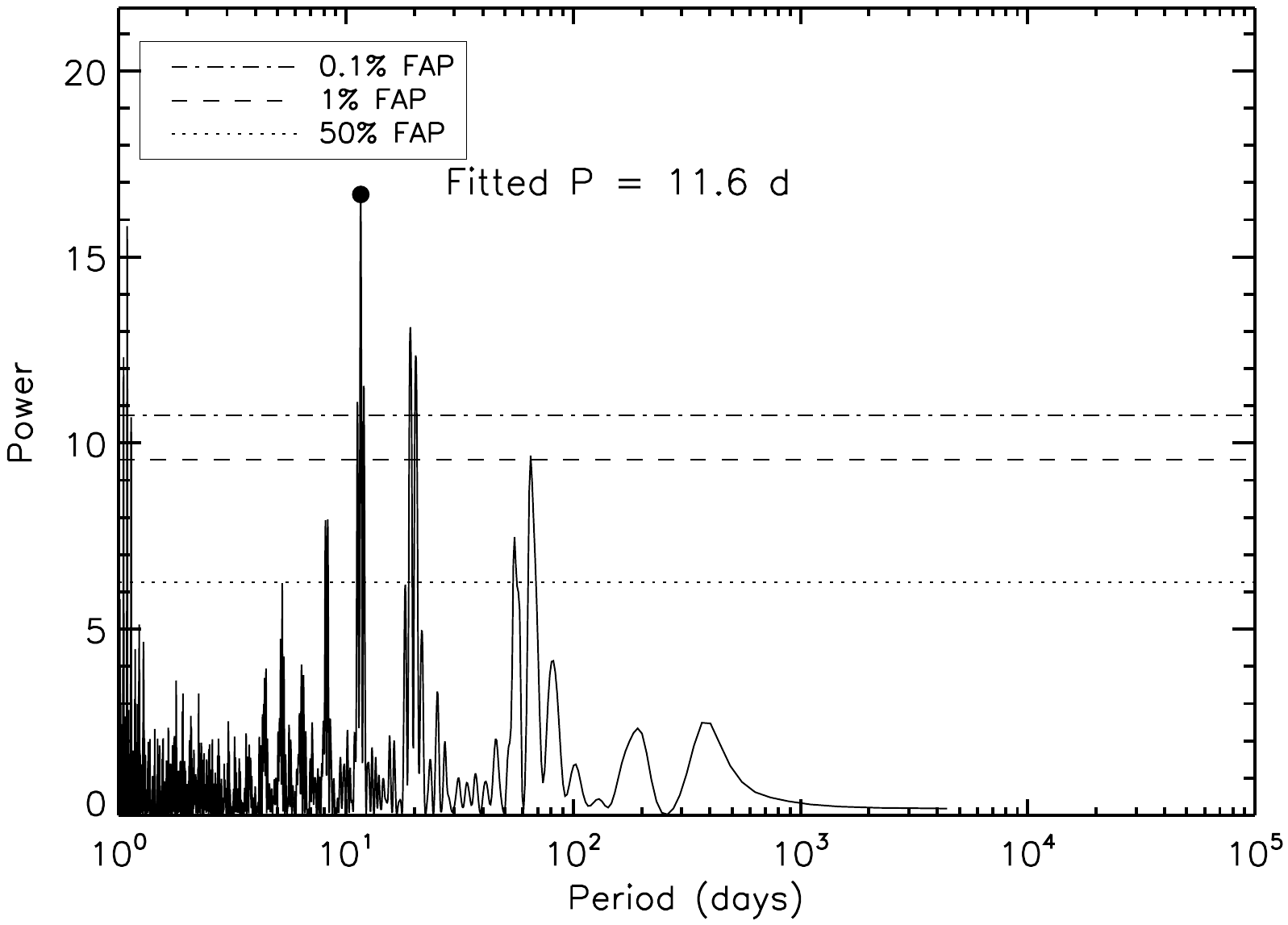}
	\hfill
	\centering
	\includegraphics[width=0.6\linewidth]{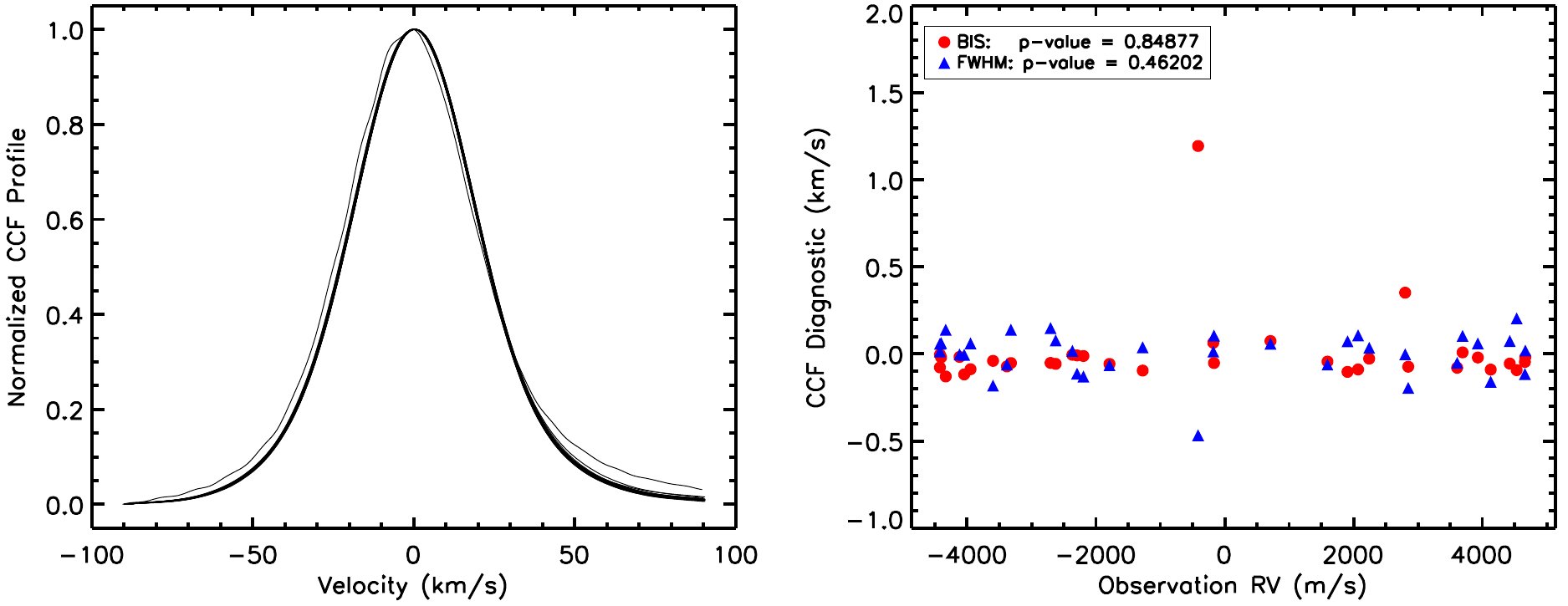}
	\hfill
\caption{\textbf{TYC 1451-01054-1, MARVELS-11b}: M sin$\textit{i}$ $\sim$52.1 M$_{\text{Jup}}$, Period $\sim$11.6121 days, ecc $\sim$0.000.}
\label{fig:marv-12}
\end{figure*}
\begin{figure*}
      	\centering
	\includegraphics[width=0.49\linewidth]{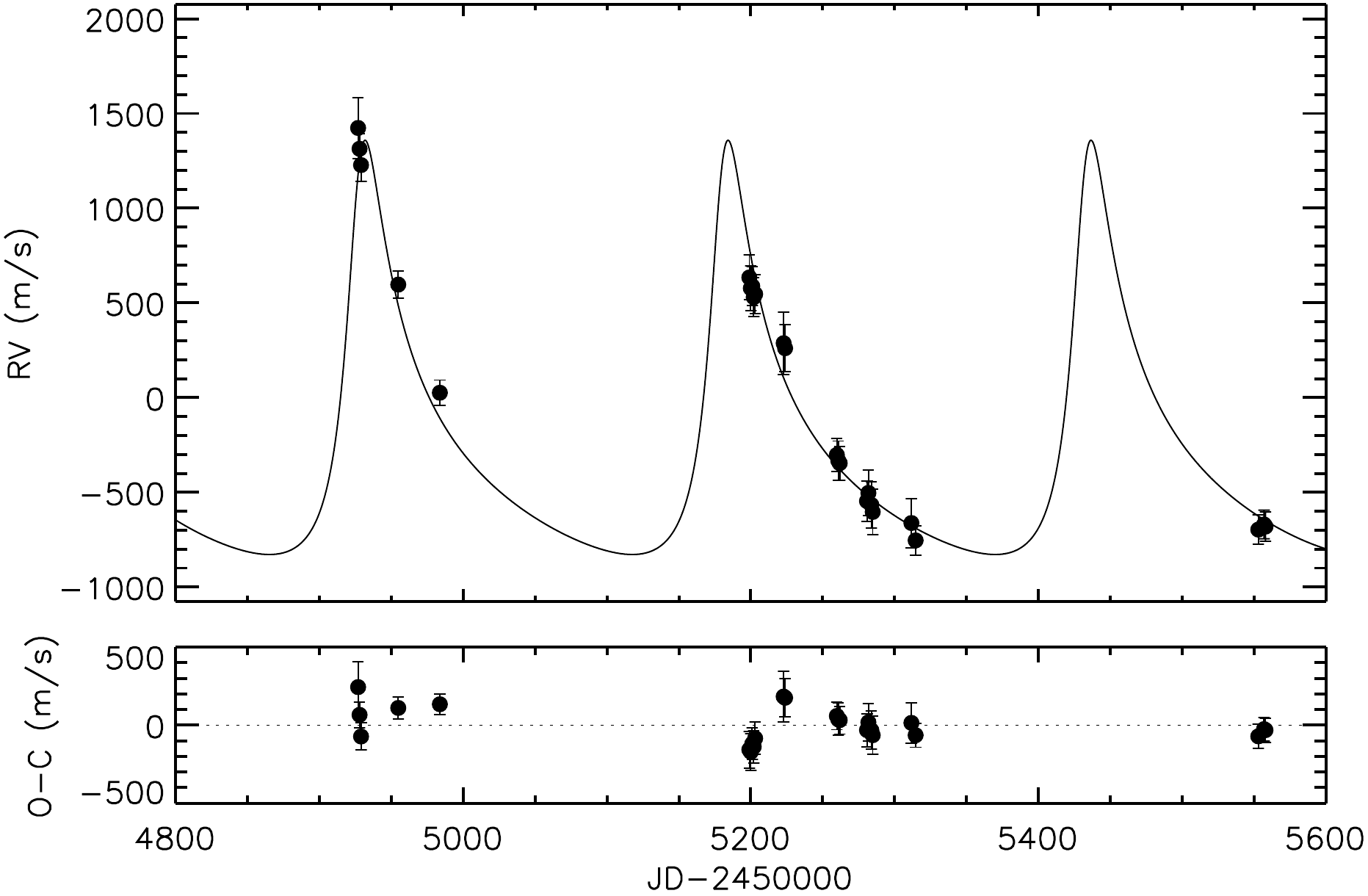}
	\hfill
	\centering
	\includegraphics[width=0.47\linewidth]{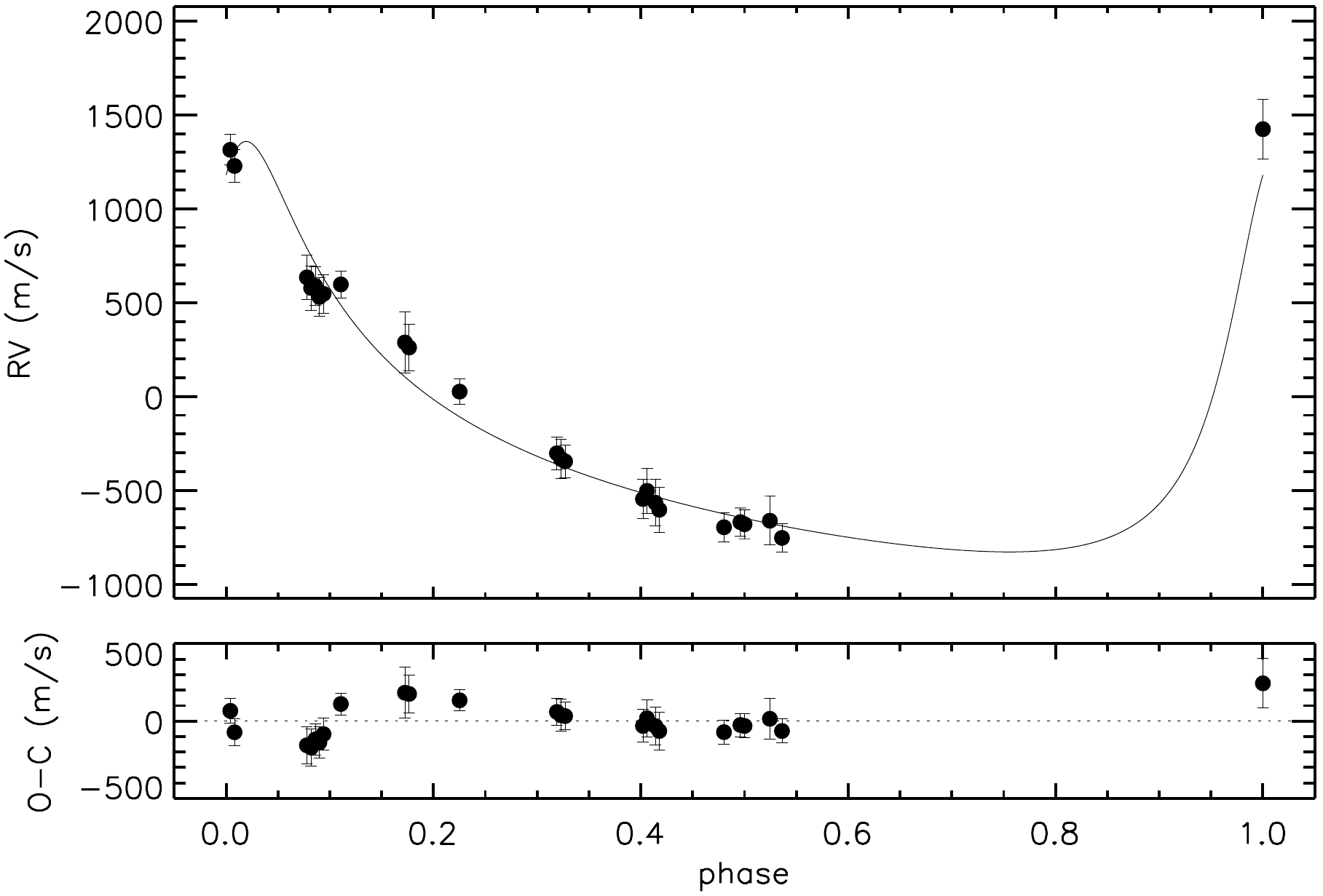}
	\hfill
	\centering
	\includegraphics[width=0.32\linewidth]{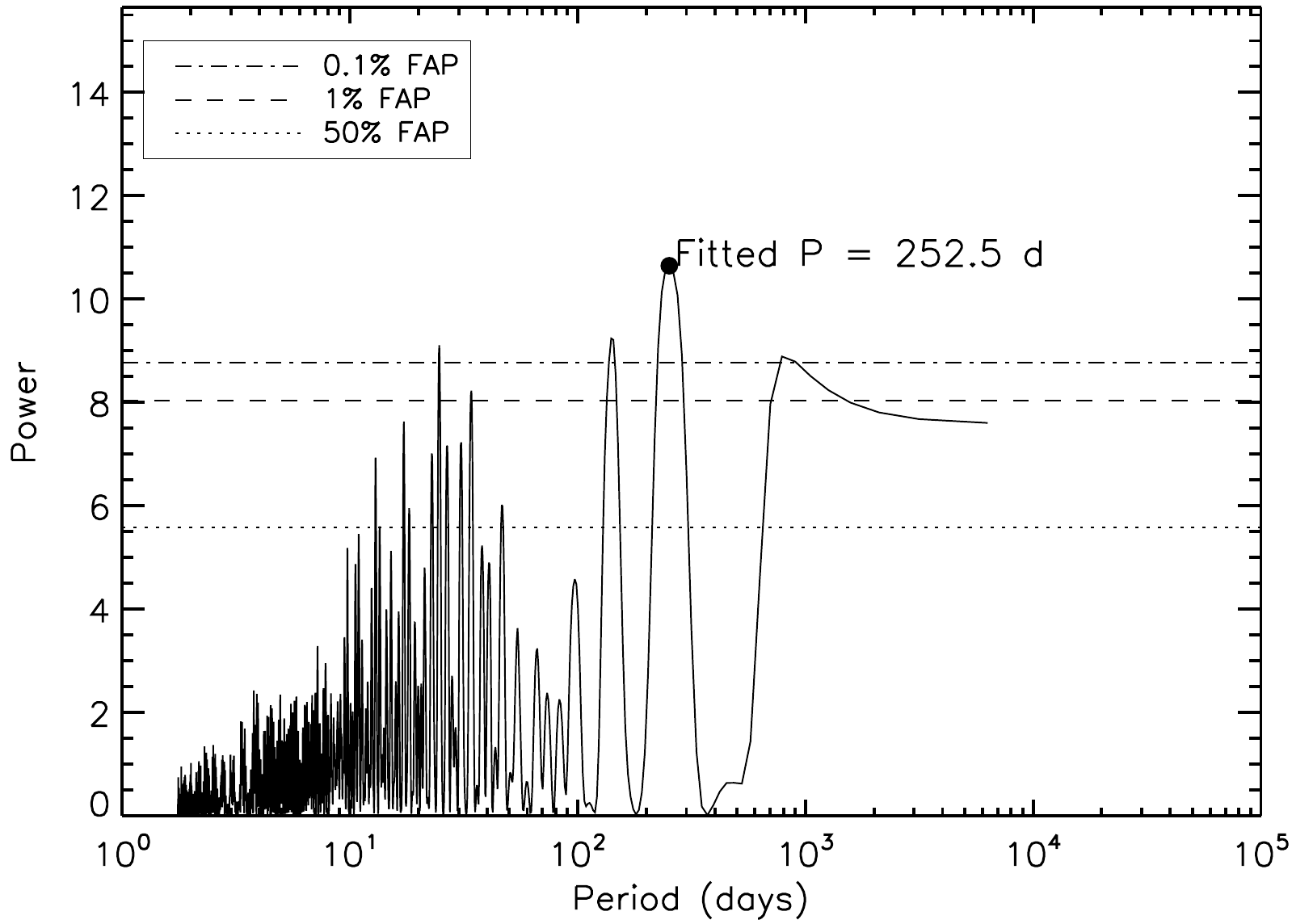}
	\hfill
	\centering
	\includegraphics[width=0.6\linewidth]{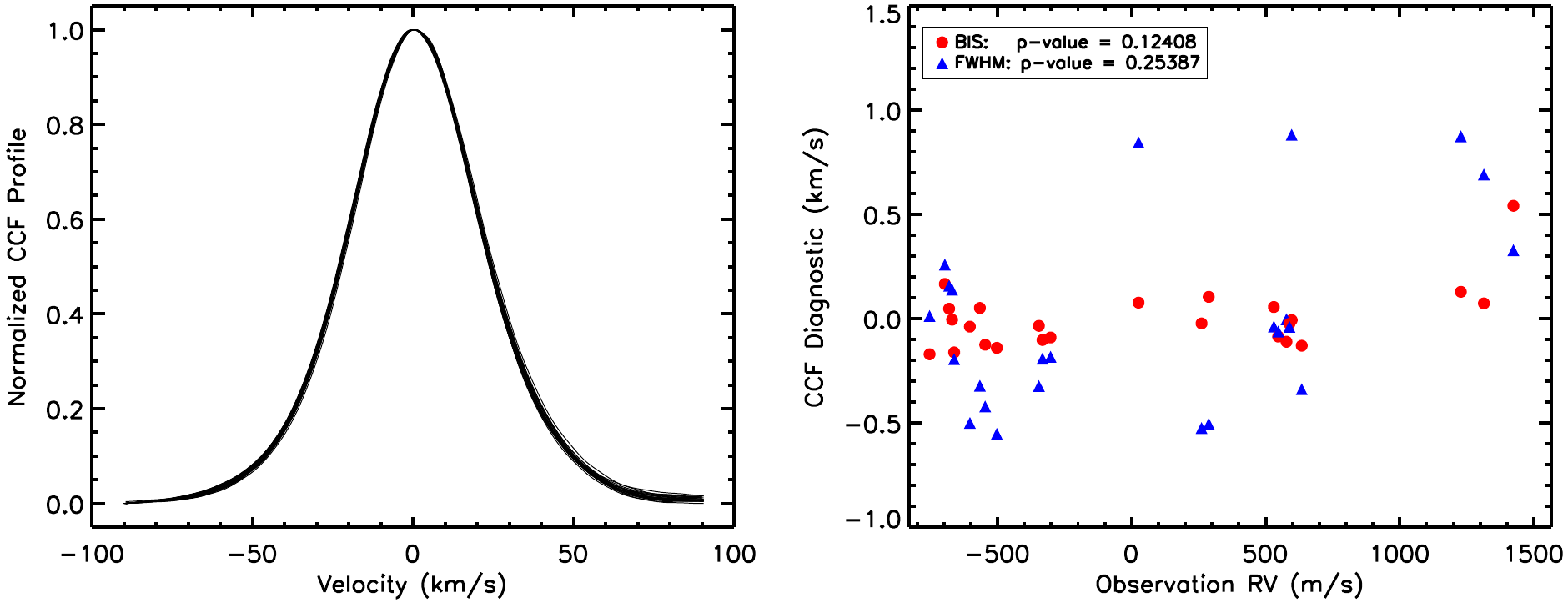}
	\hfill
\caption{\textbf{TYC 3469-00492-1, MARVELS-12b}: M sin$\textit{i}$ $\sim$32.8 M$_{\text{Jup}}$, Period $\sim$252.5 days, ecc $\sim$0.60.}
\label{fig:marv-13}
\end{figure*}
\begin{figure*}
      	\centering
	\includegraphics[width=0.49\linewidth]{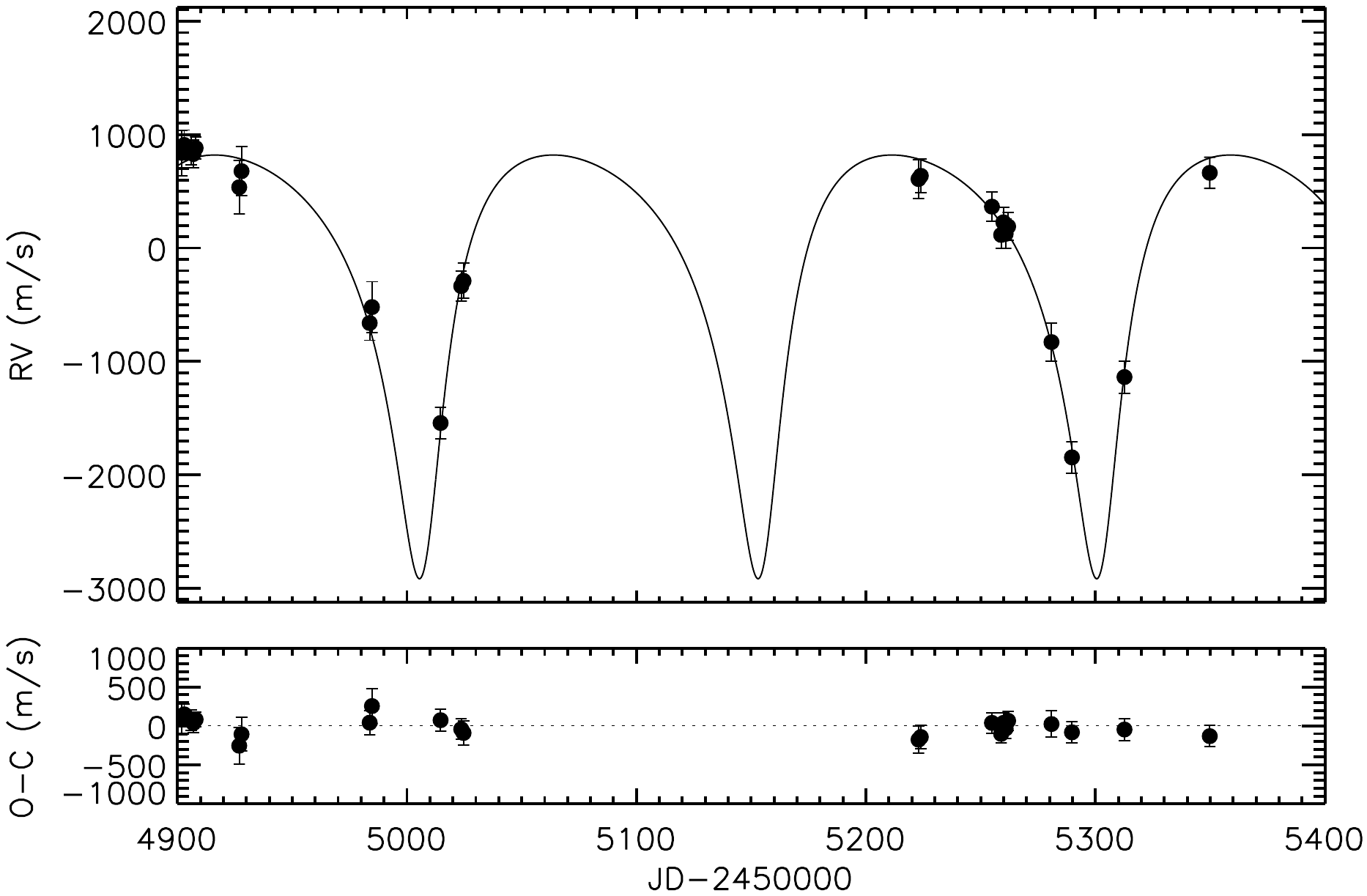}
	\hfill
	\centering
	\includegraphics[width=0.47\linewidth]{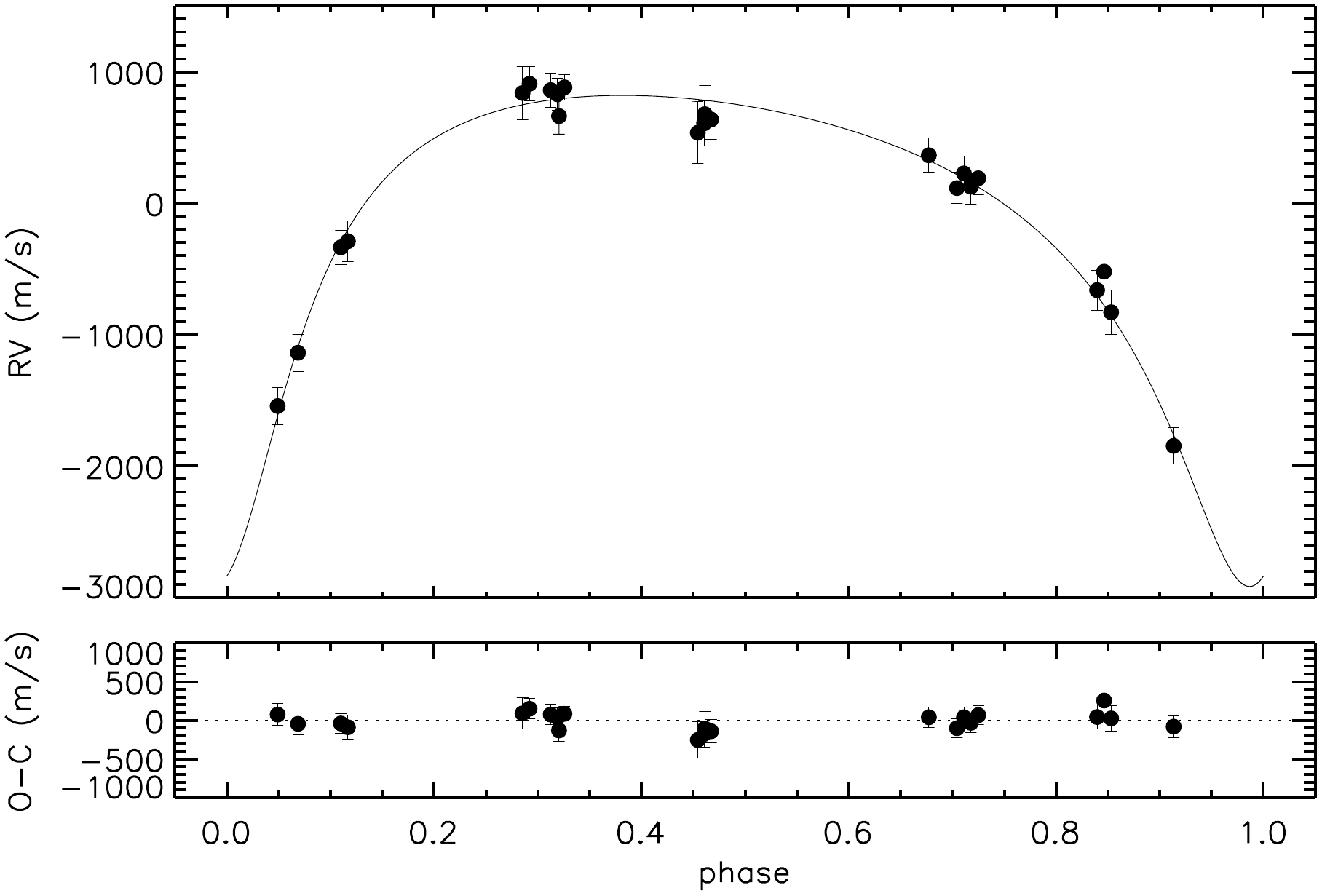}
	\hfill
	\centering
	\includegraphics[width=0.32\linewidth]{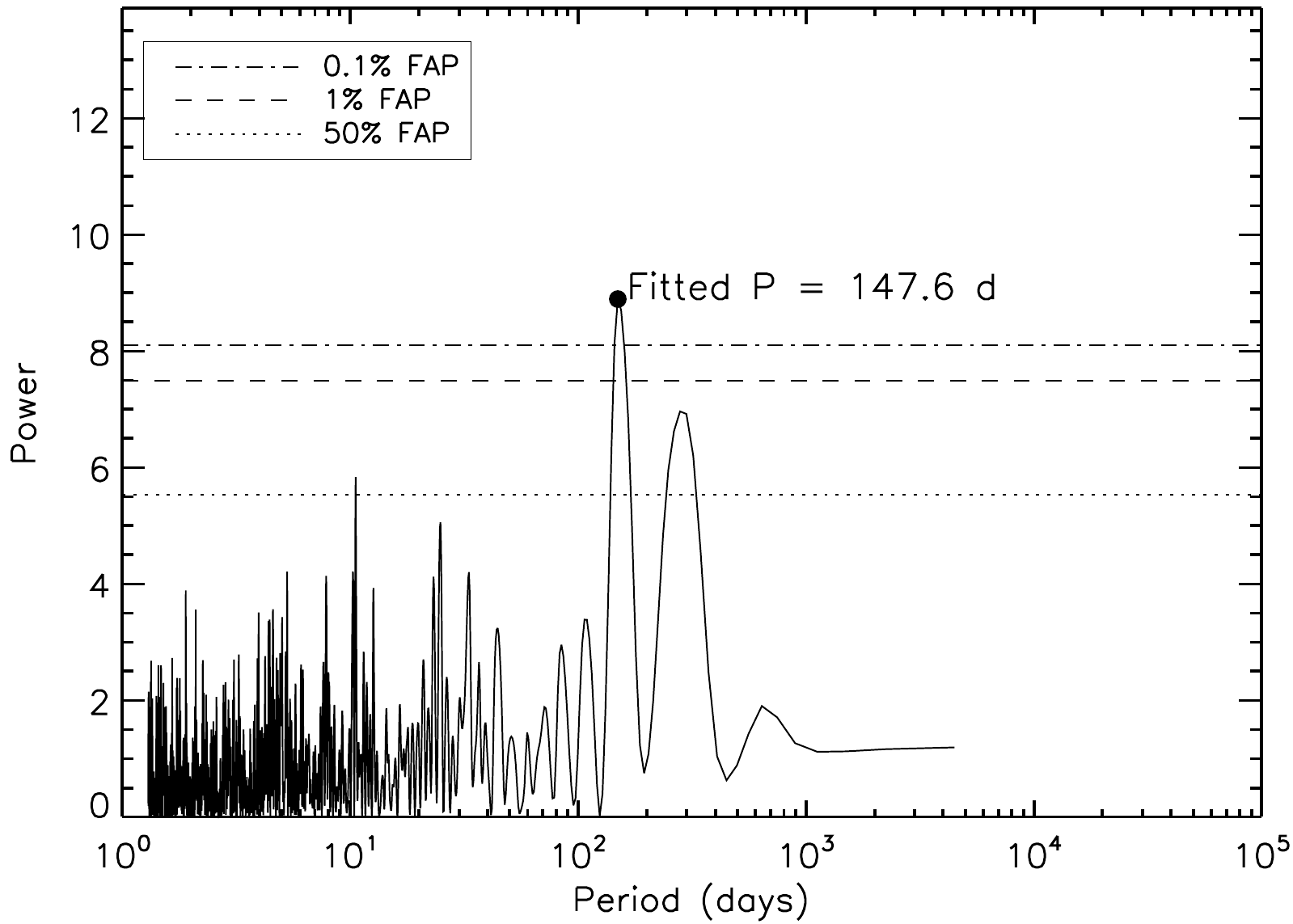}
	\hfill
	\centering
	\includegraphics[width=0.6\linewidth]{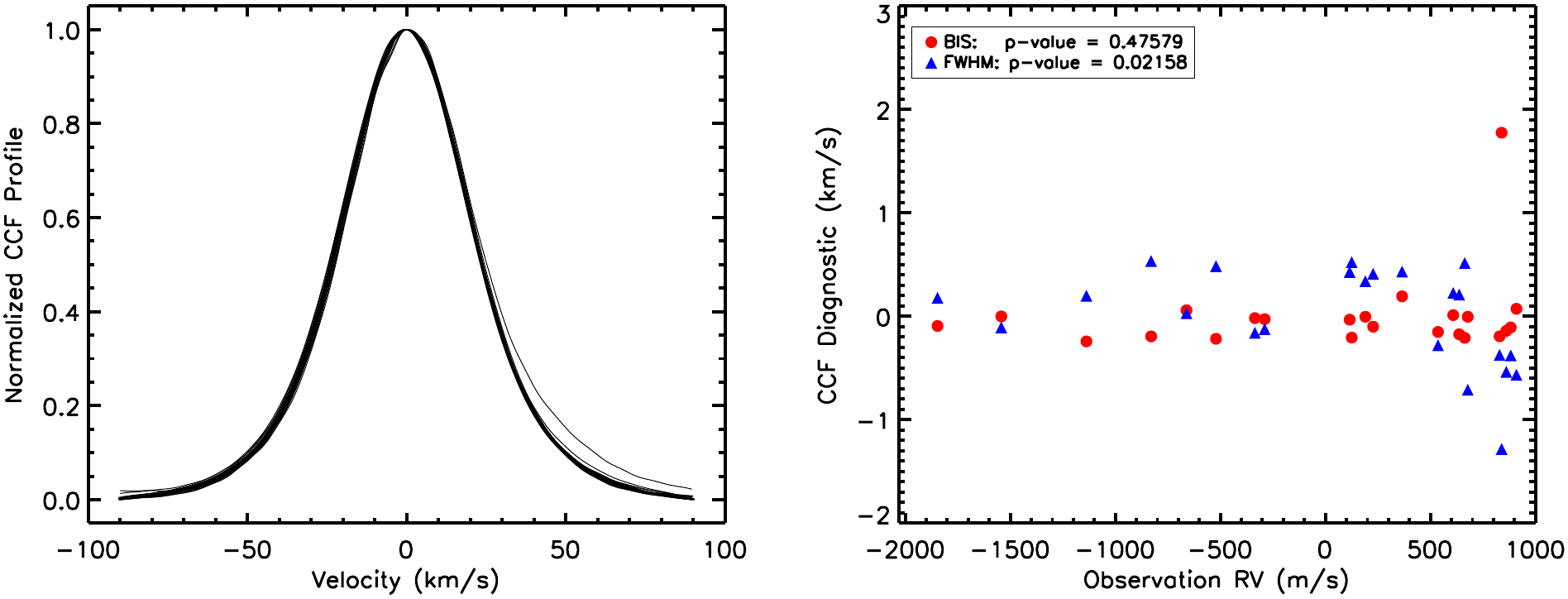}
	\hfill
\caption{\textbf{GSC 03467-00030 (MCKGS1-135), MARVELS-13b}: M sin$\textit{i}$ $\sim$41.8 M$_{\text{Jup}}$, Period $\sim$147.6 days, ecc $\sim$0.50.}
\label{fig:marv-14}
\end{figure*}
\begin{figure*}
      	\centering
	\includegraphics[width=0.49\linewidth]{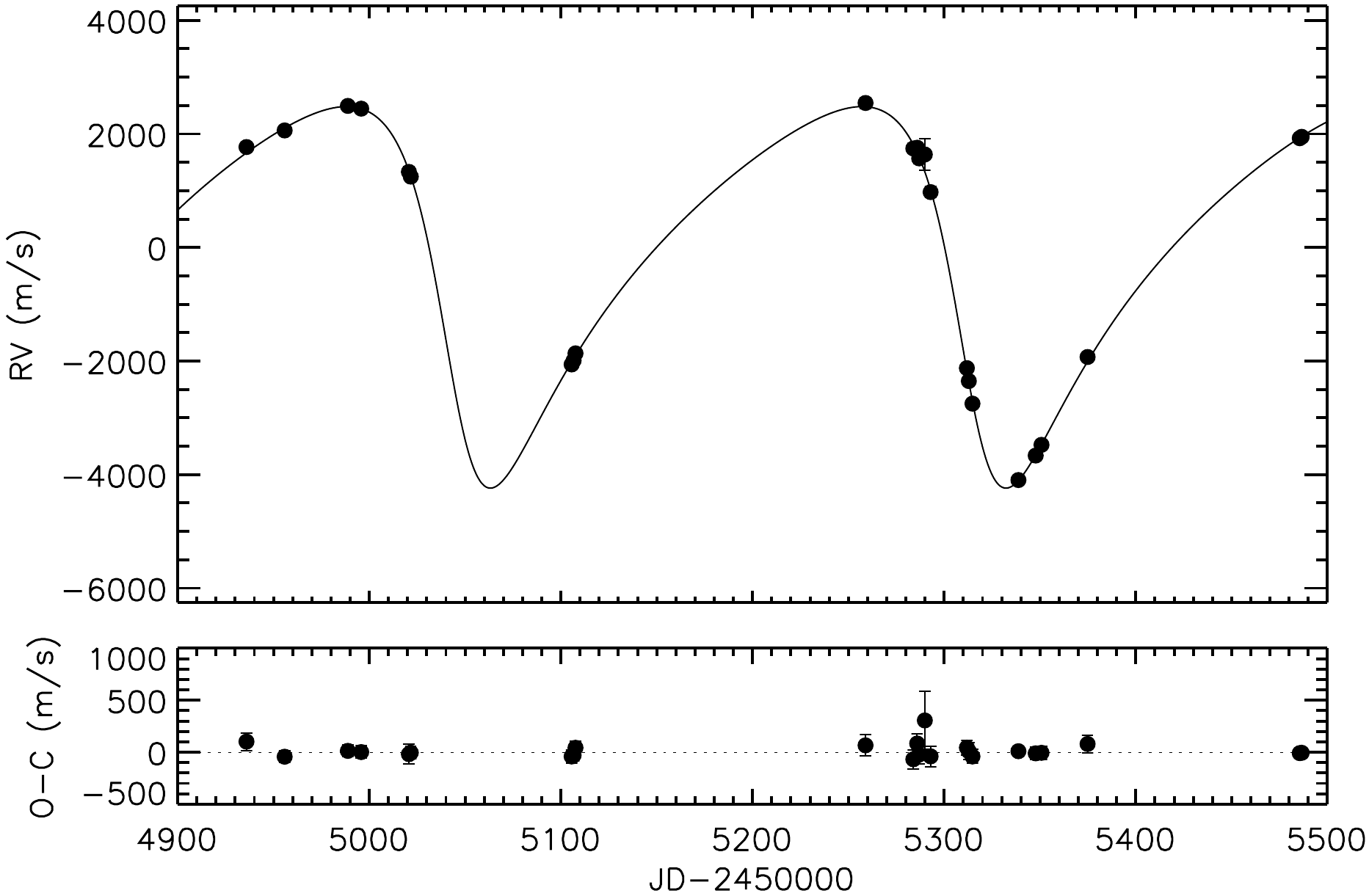}
	\hfill
	\centering
	\includegraphics[width=0.47\linewidth]{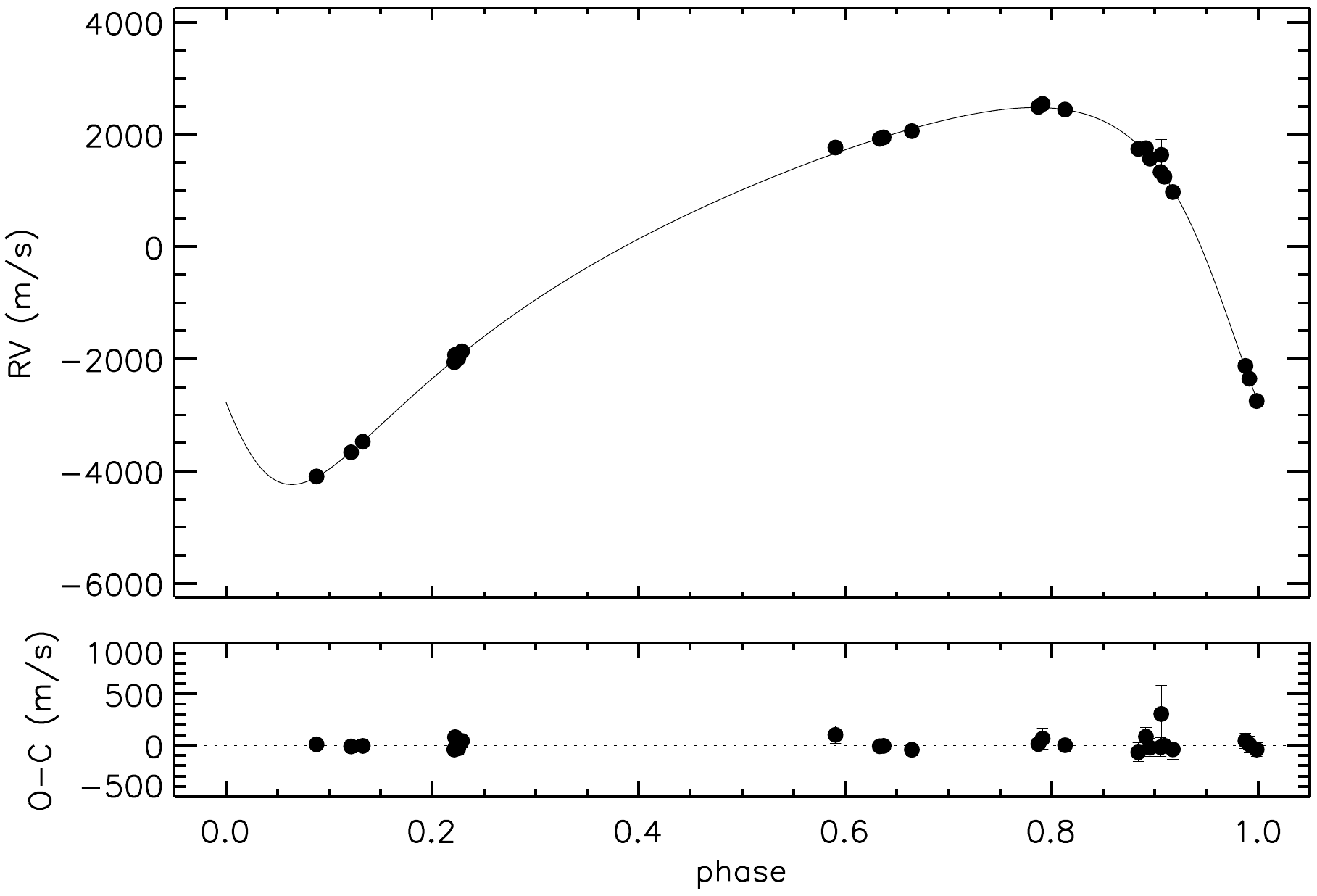}
	\hfill
	\centering
	\includegraphics[width=0.32\linewidth]{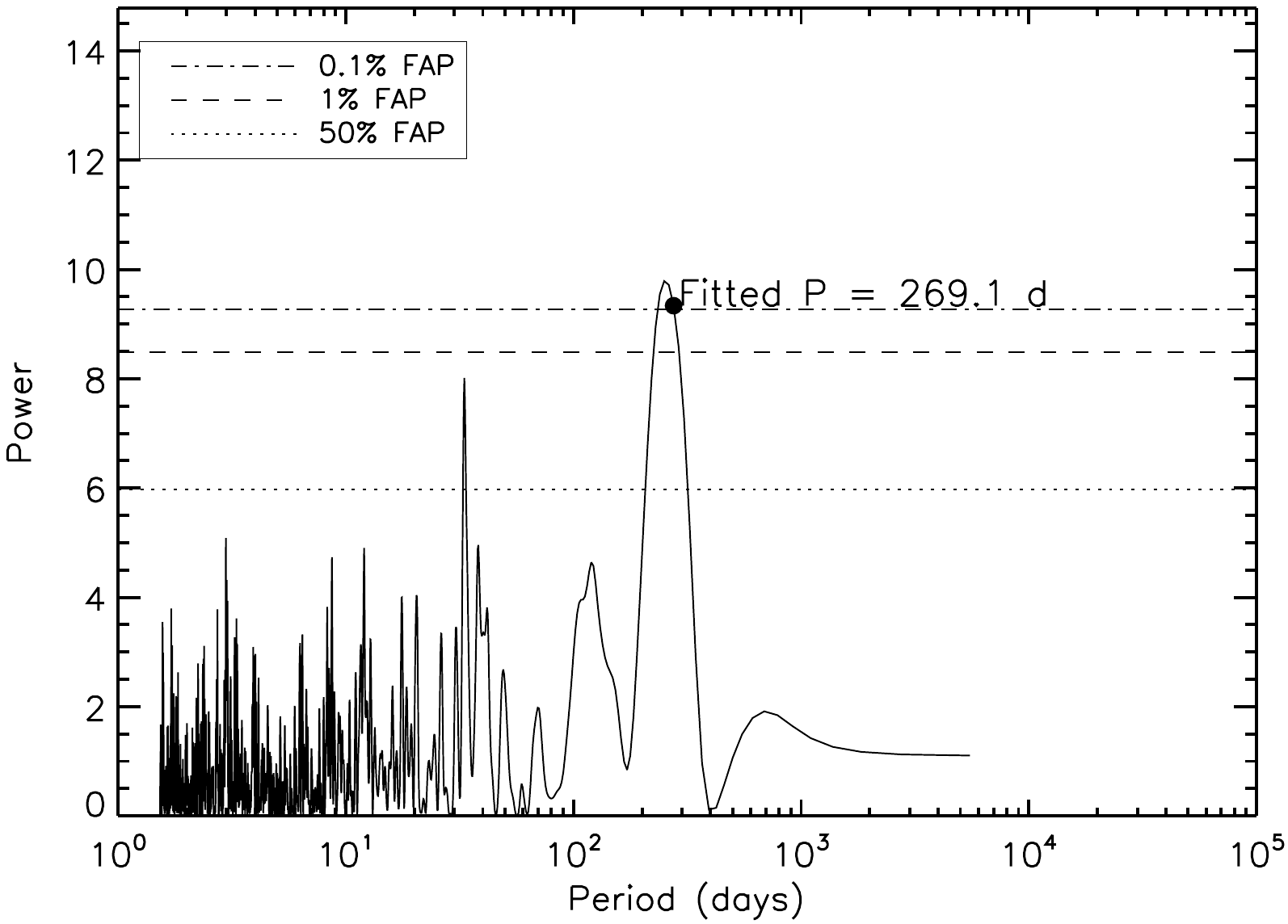}
	\hfill
	\centering
	\includegraphics[width=0.6\linewidth]{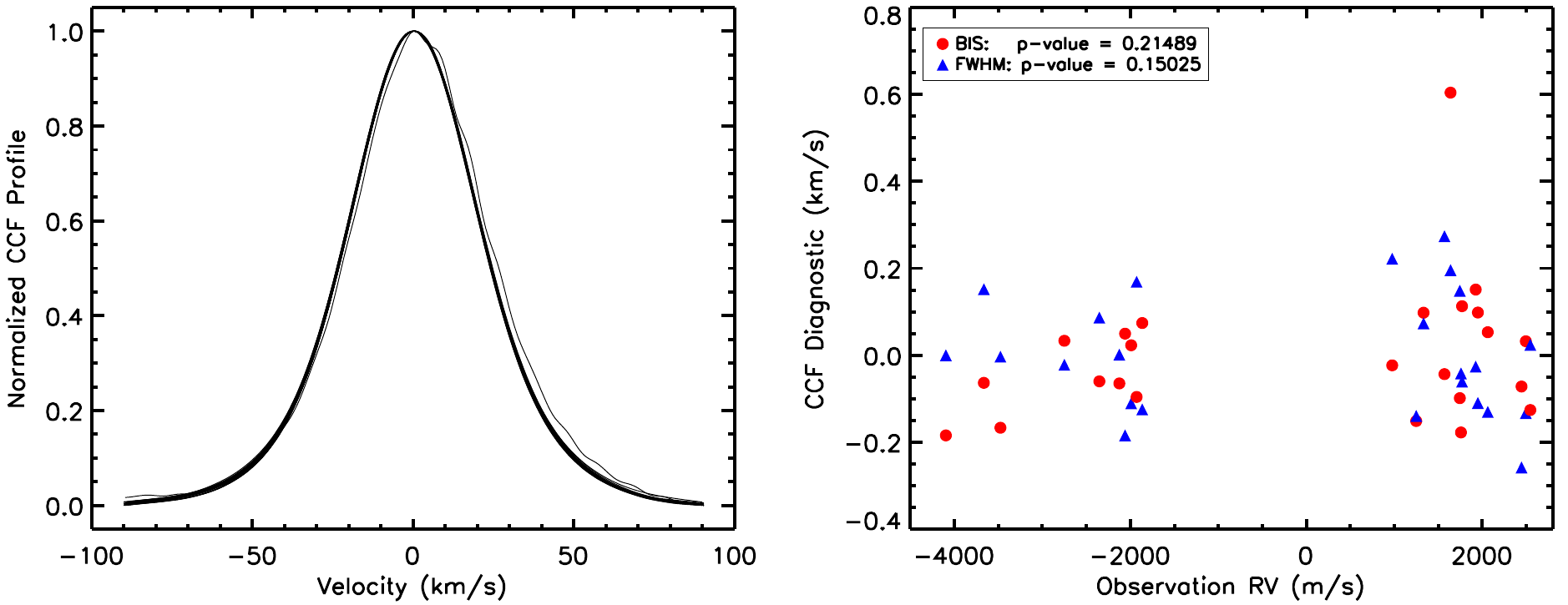}
	\hfill
\caption{\textbf{TYC 3091-00716-1 (MCKGS1-153), MARVELS-14b}: M sin$\textit{i}$ $\sim$100.2 M$_{\text{Jup}}$, Period $\sim$269.1 days, ecc $\sim$0.417.}
\label{fig:marv-16}
\end{figure*}
\begin{figure*}
      	\centering
	\includegraphics[width=0.49\linewidth]{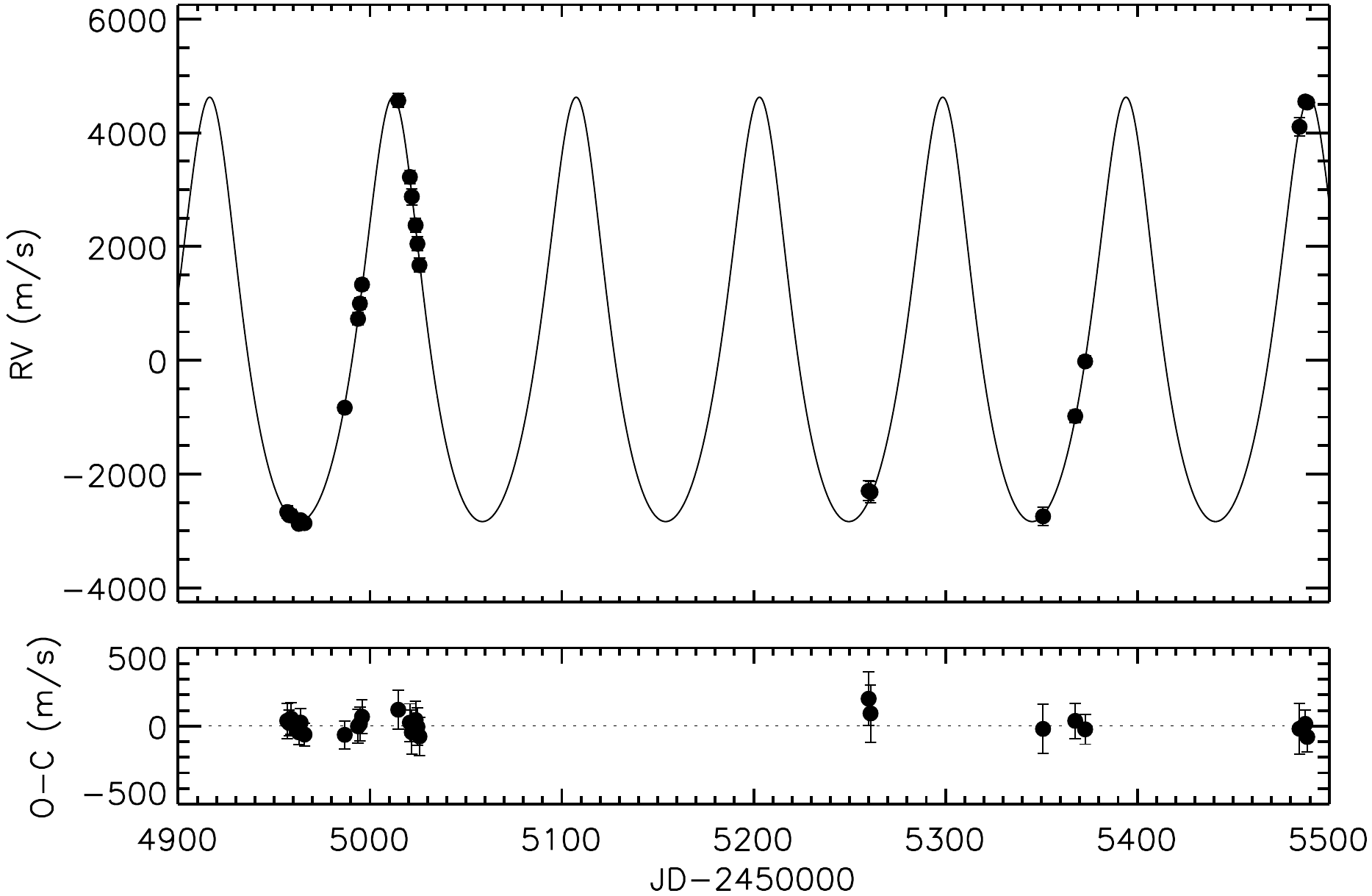}
	\hfill
	\centering
	\includegraphics[width=0.47\linewidth]{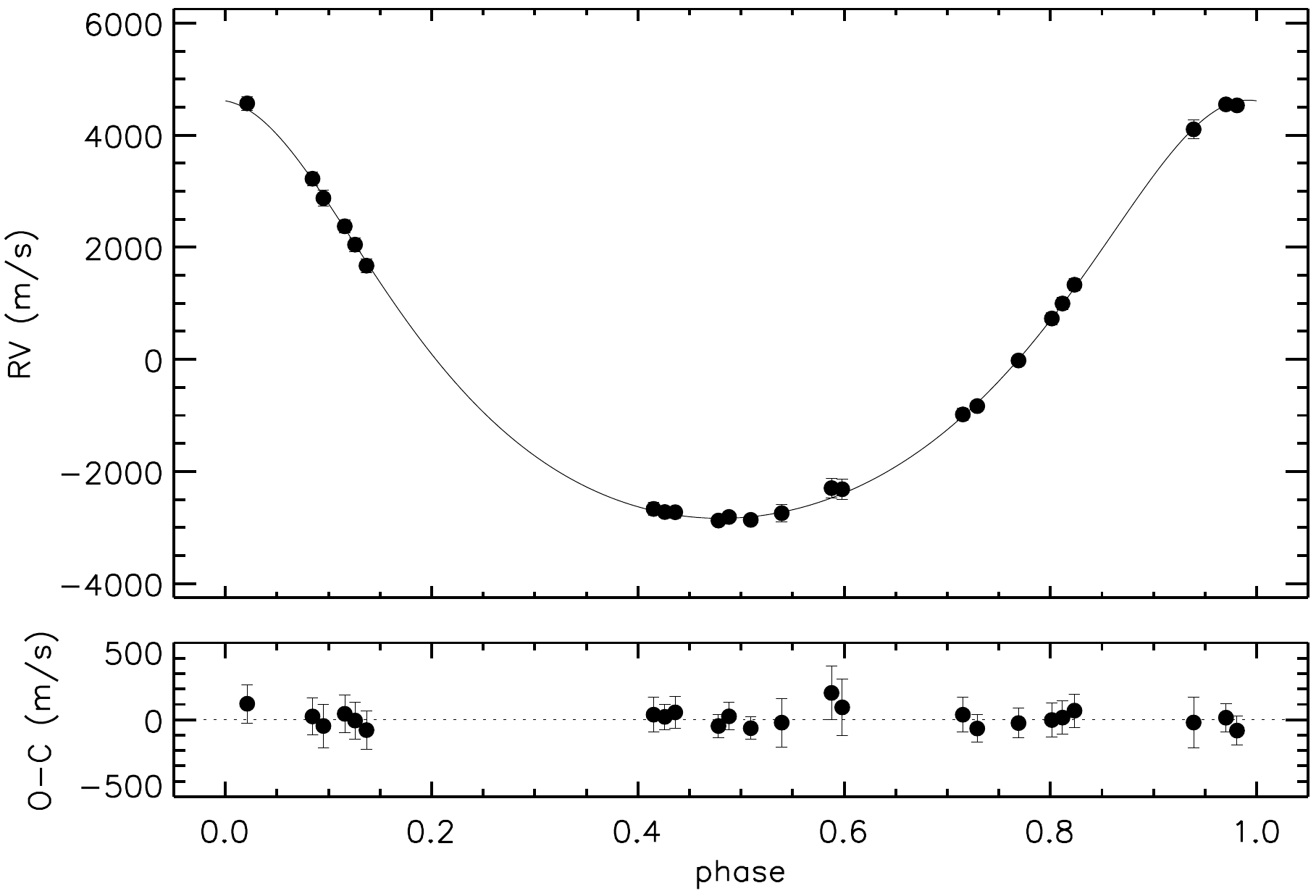}
	\hfill
	\centering
	\includegraphics[width=0.32\linewidth]{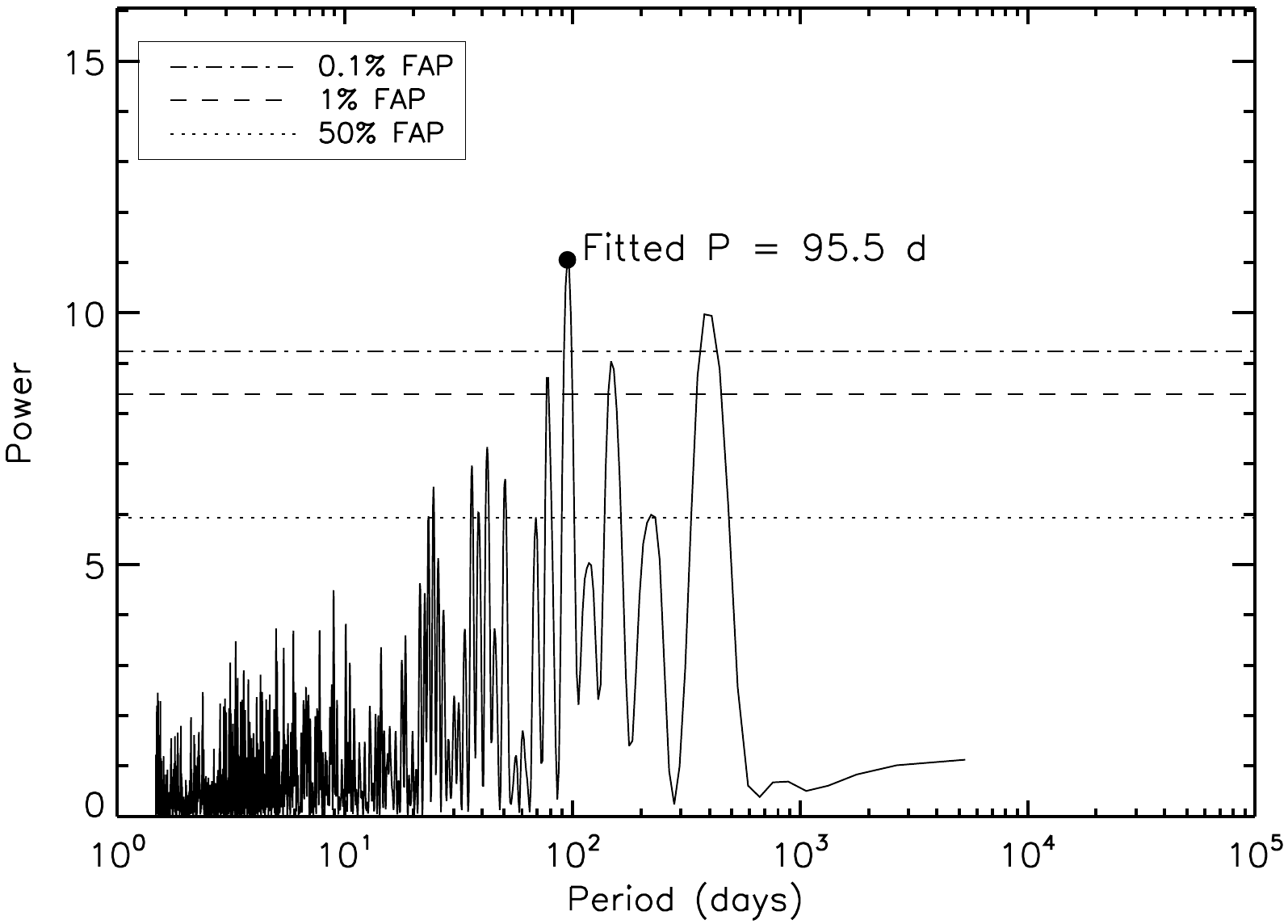}
	\hfill
	\centering
	\includegraphics[width=0.6\linewidth]{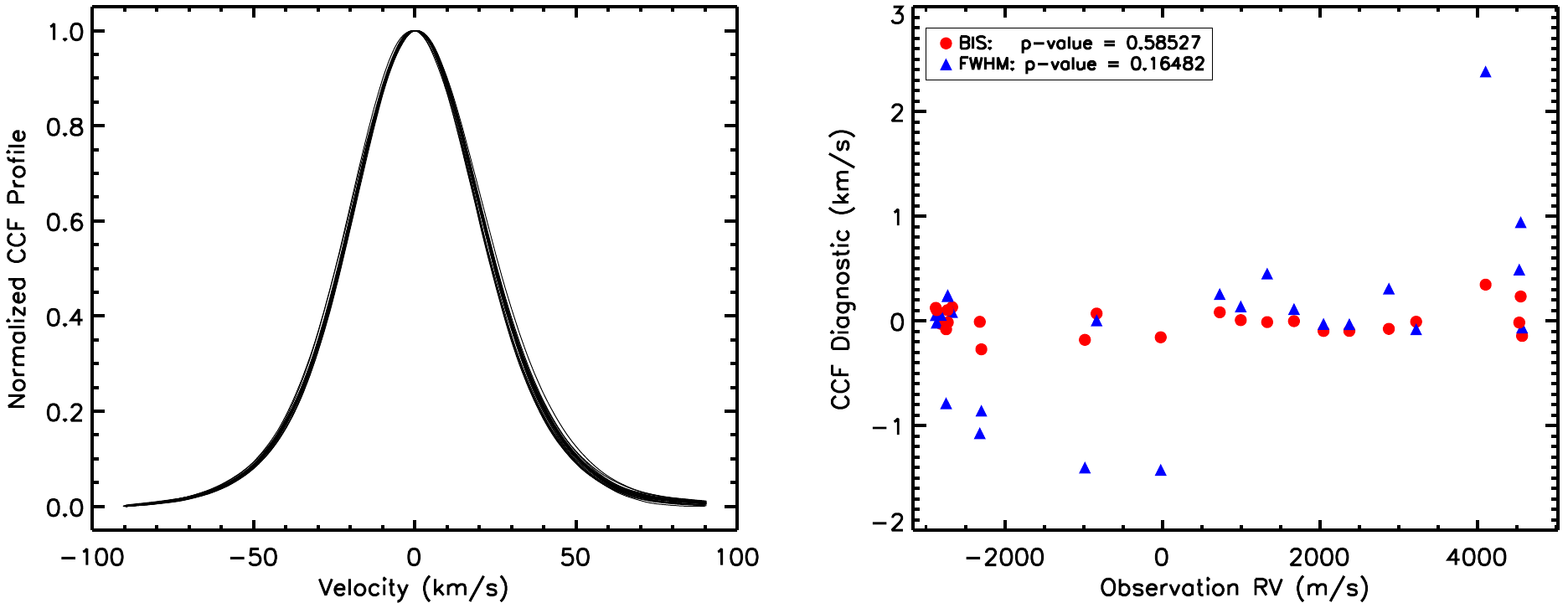}
	\hfill
\caption{\textbf{TYC 3097-00398-1, MARVELS-15b}: M sin$\textit{i}$ $\sim$97.6 M$_{\text{Jup}}$, Period $\sim$95.54 days, ecc $\sim$0.223.}
\label{fig:marv-17}
\end{figure*}
\begin{figure*}
      	\centering
	\includegraphics[width=0.49\linewidth]{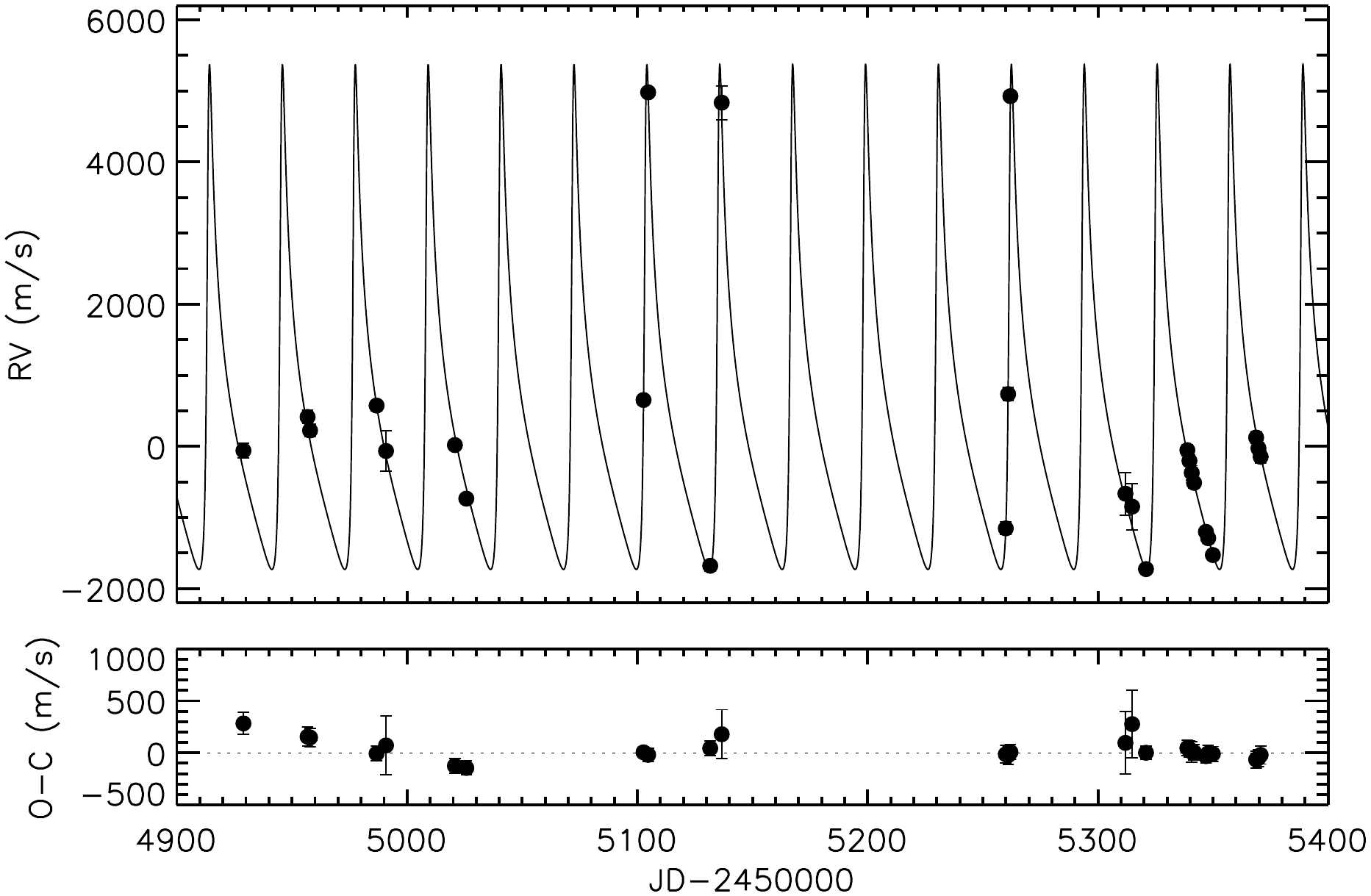}
	\hfill
	\centering
	\includegraphics[width=0.47\linewidth]{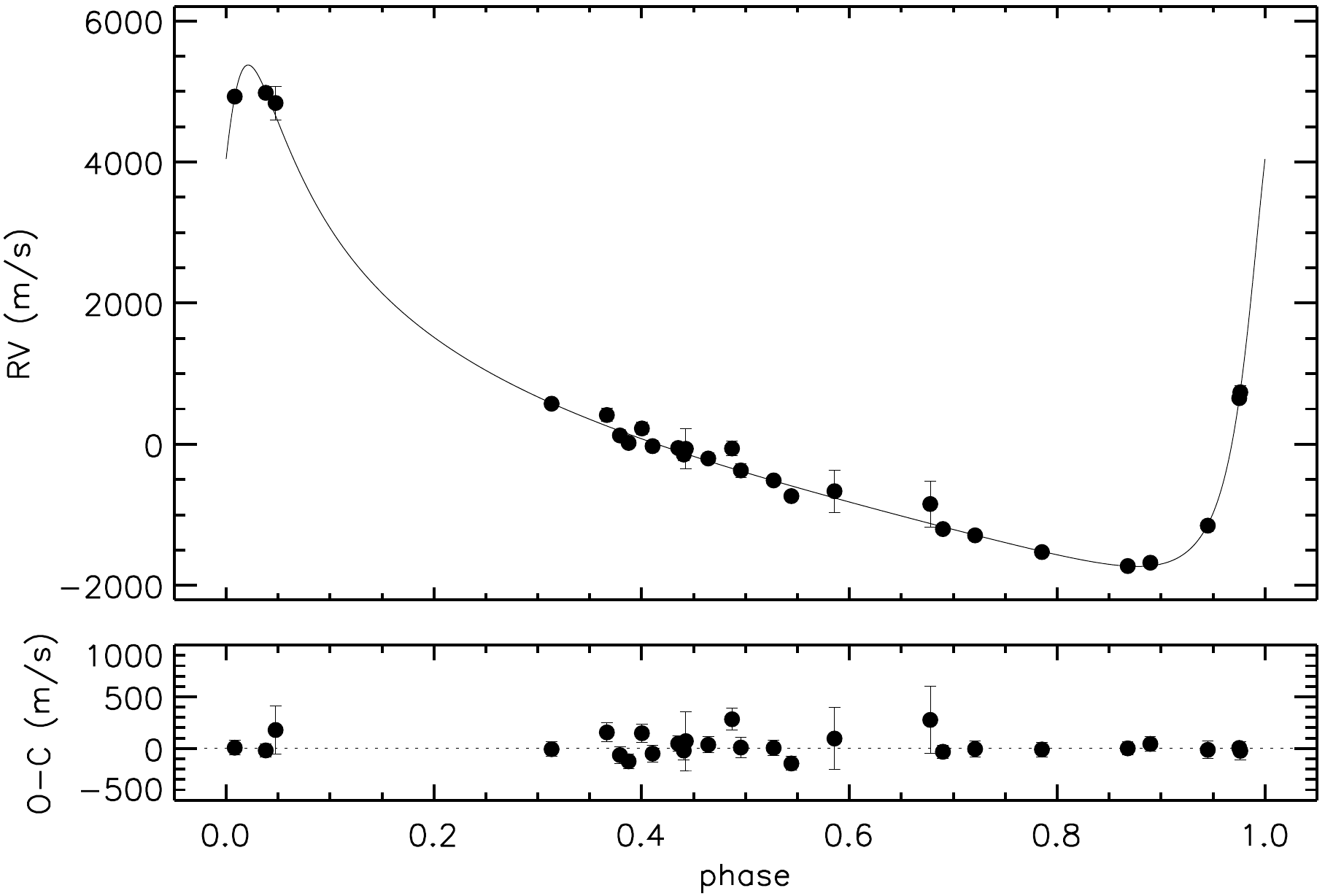}
	\hfill
	\centering
	\includegraphics[width=0.32\linewidth]{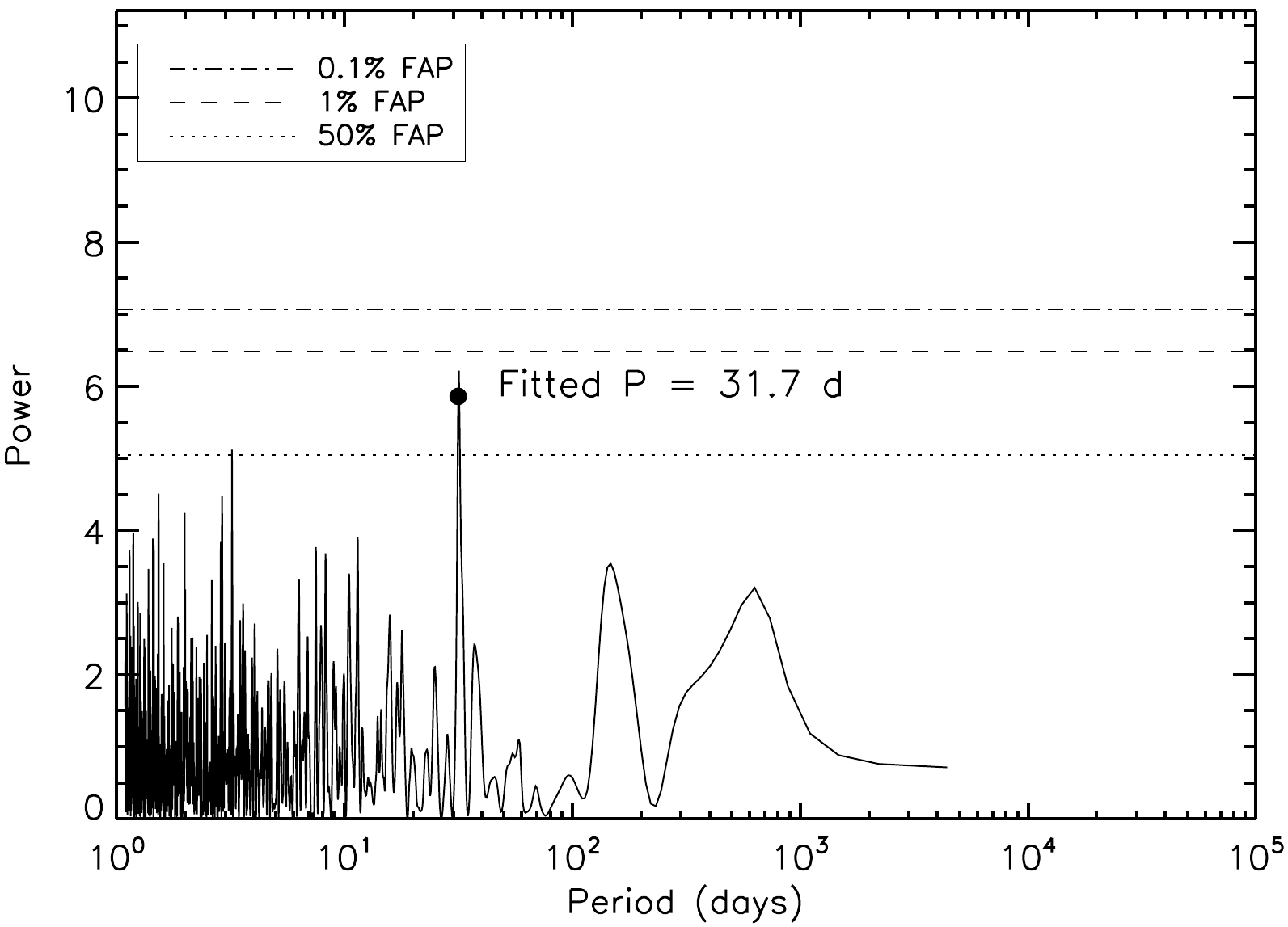}
	\hfill
	\centering
	\includegraphics[width=0.6\linewidth]{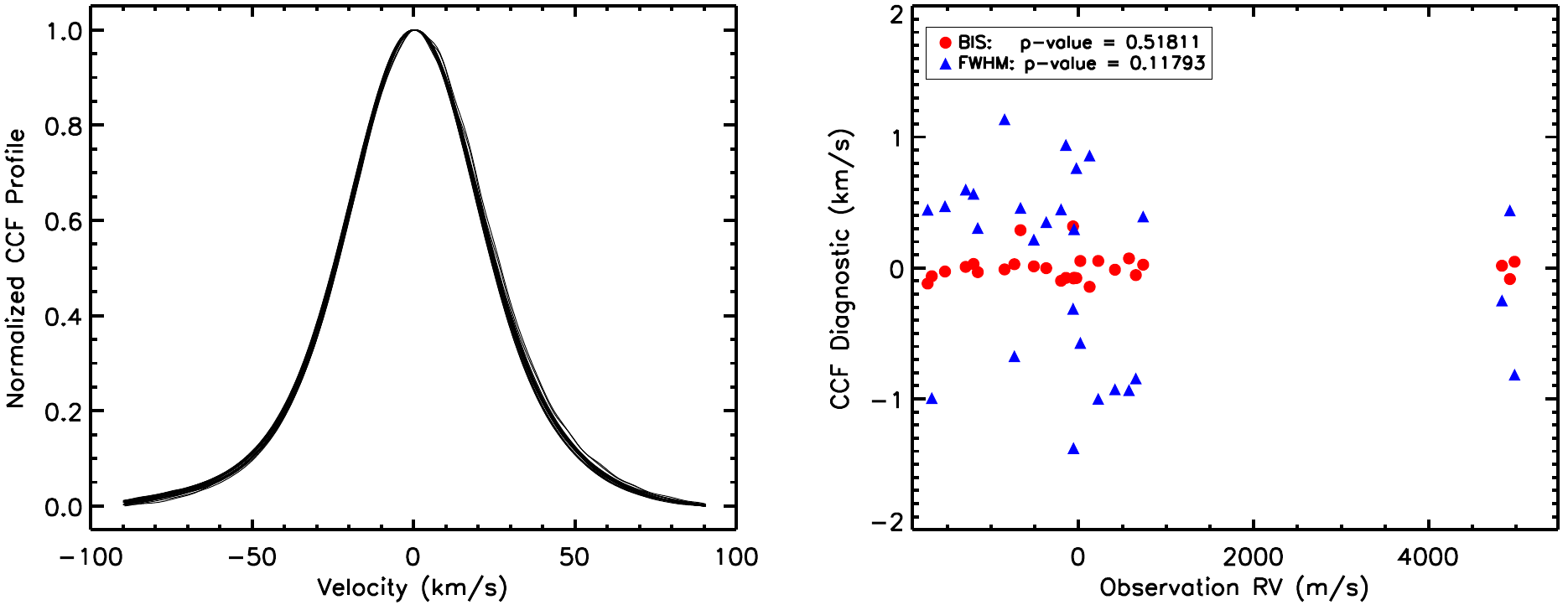}
	\hfill
\caption{\textbf{TYC 3130-00160-1 (MCUF1-11), MARVELS-16b}: M sin$\textit{i}$ $\sim$42.7 M$_{\text{Jup}}$, Period $\sim$31.656 days, ecc $\sim$0.690.}
\label{fig:marv-18}
\end{figure*}
\begin{figure*}
      	\centering
	\includegraphics[width=0.49\linewidth]{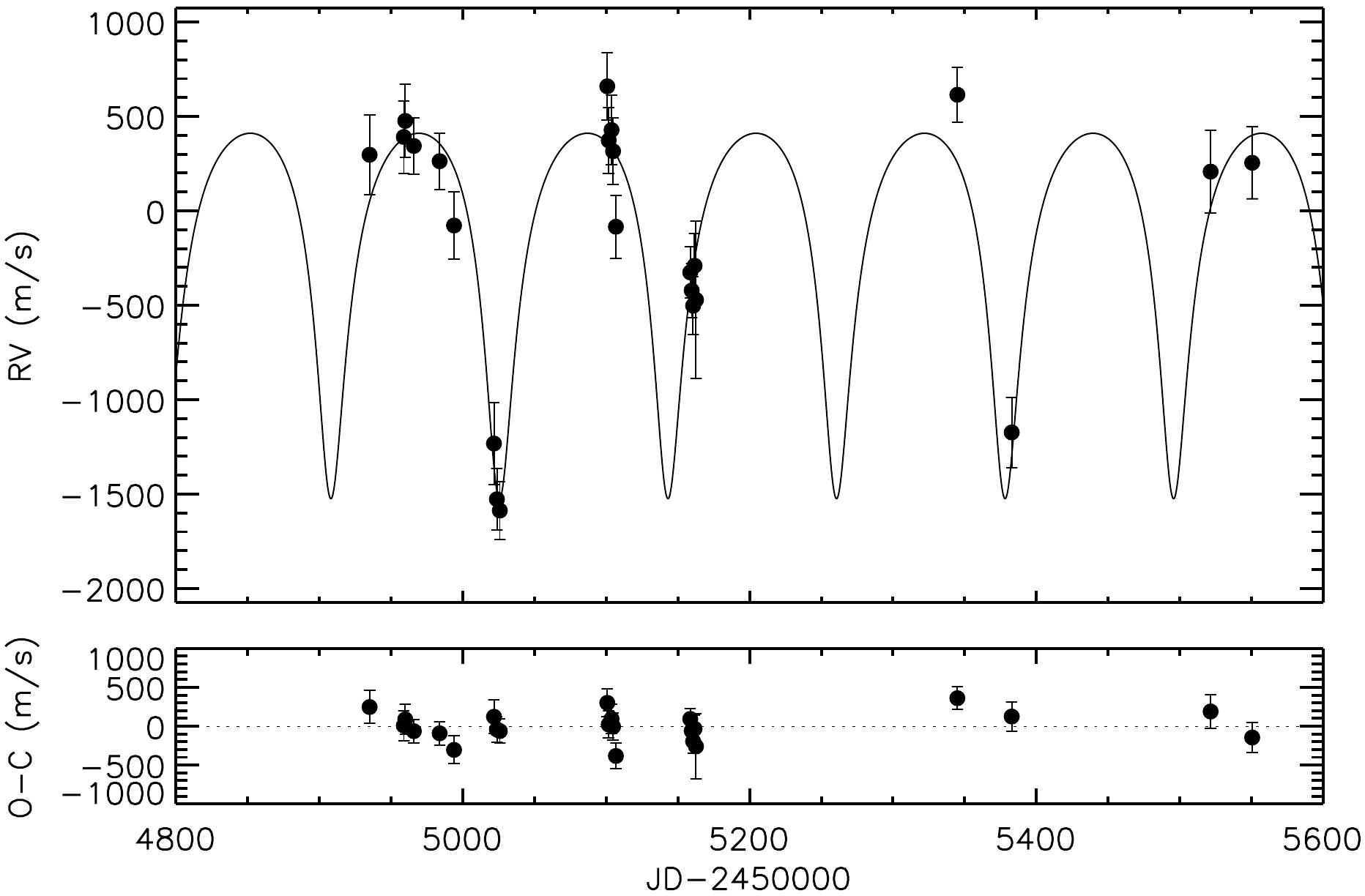}
	\hfill
	\centering
	\includegraphics[width=0.47\linewidth]{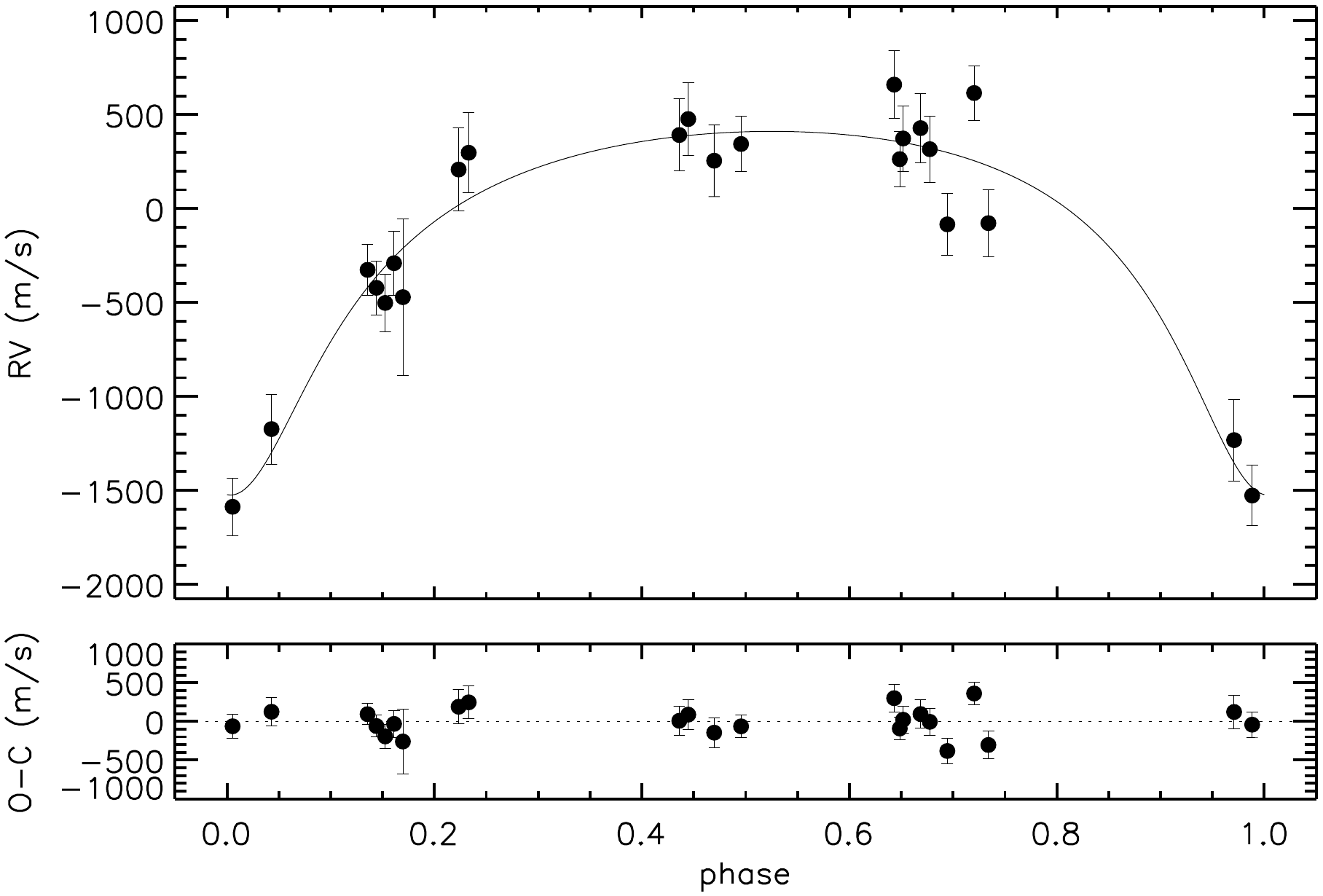}
	\hfill
	\centering
	\includegraphics[width=0.32\linewidth]{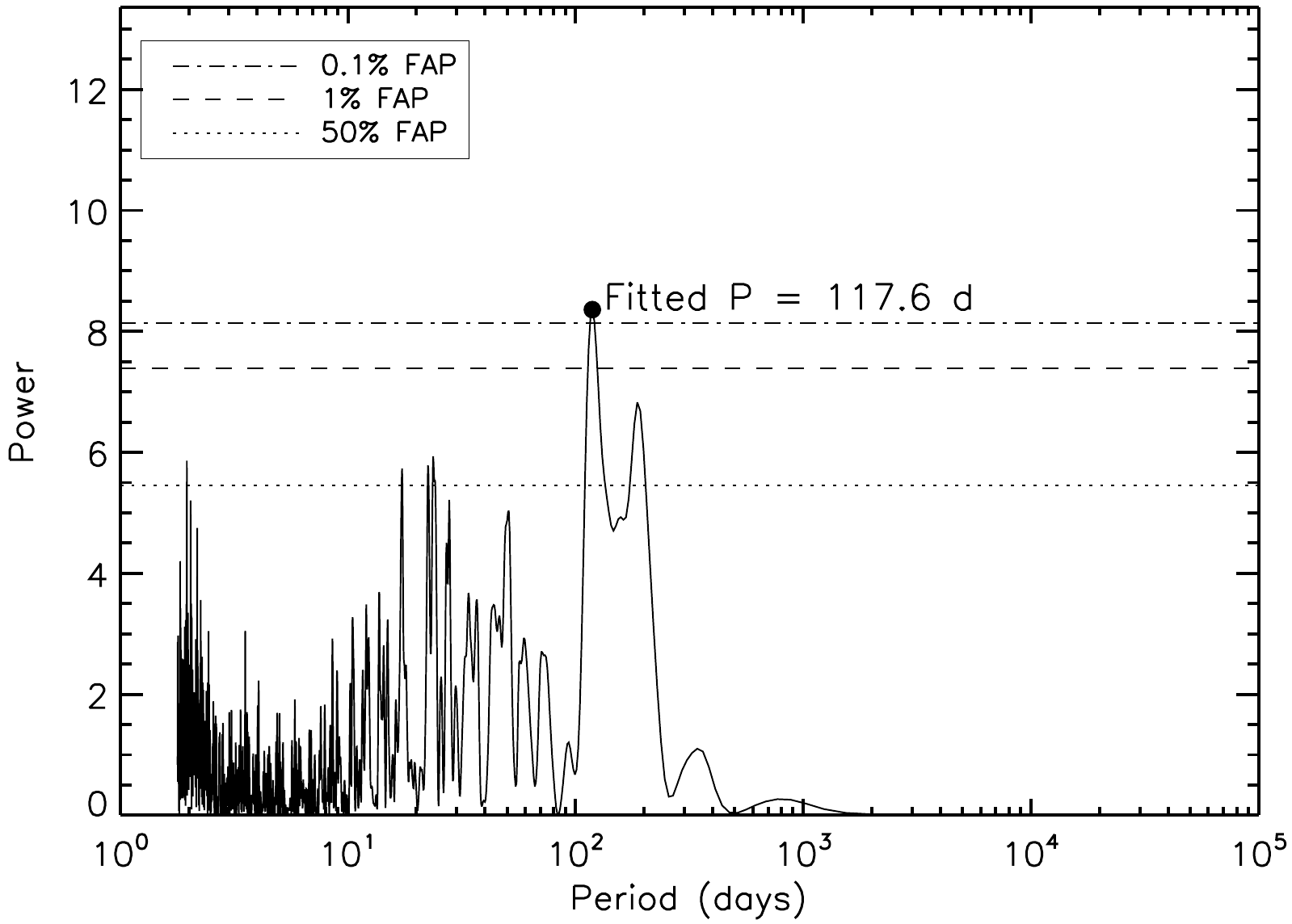}
	\hfill
	\centering
	\includegraphics[width=0.6\linewidth]{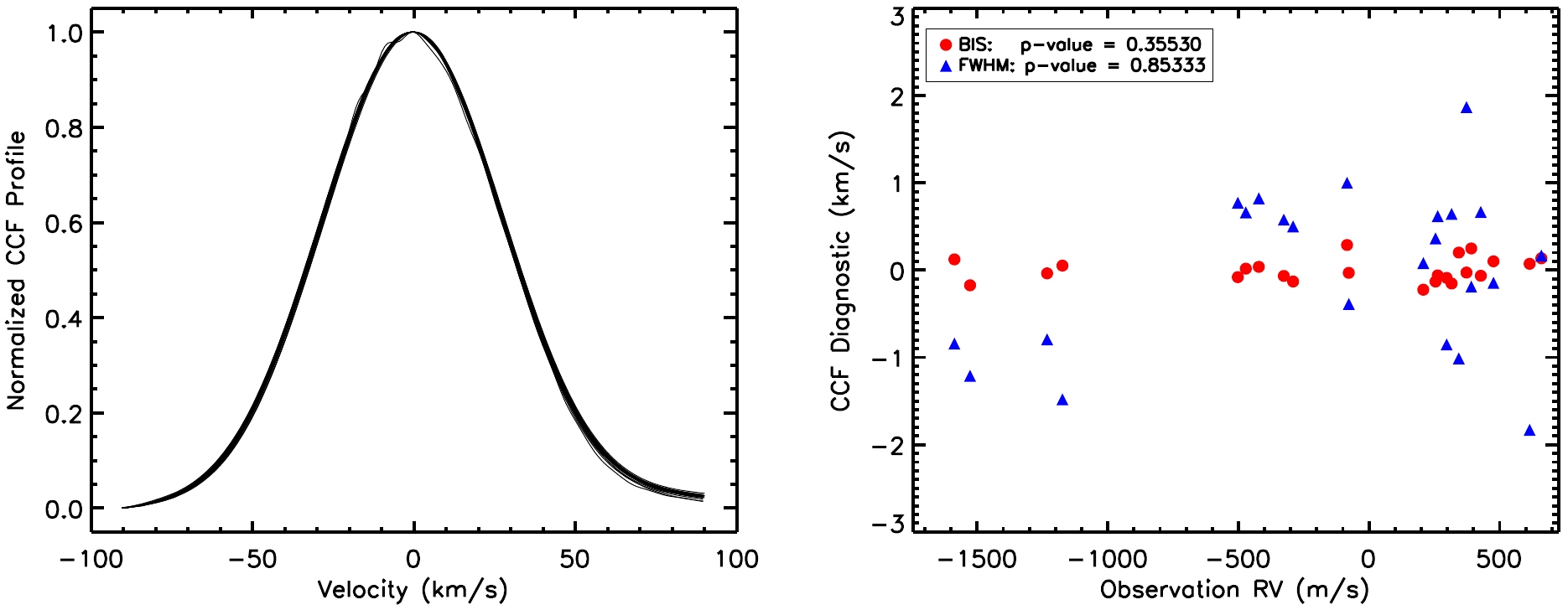}
	\hfill
\caption{\textbf{TYC 3556-03602-1, MARVELS-18b}: M sin$\textit{i}$ $\sim$24.4 M$_{\text{Jup}}$, Period $\sim$117.6 days, ecc $\sim$0.46.}
\label{fig:marv-20}
\end{figure*}
\begin{figure*}
      	\centering
	\includegraphics[width=0.49\linewidth]{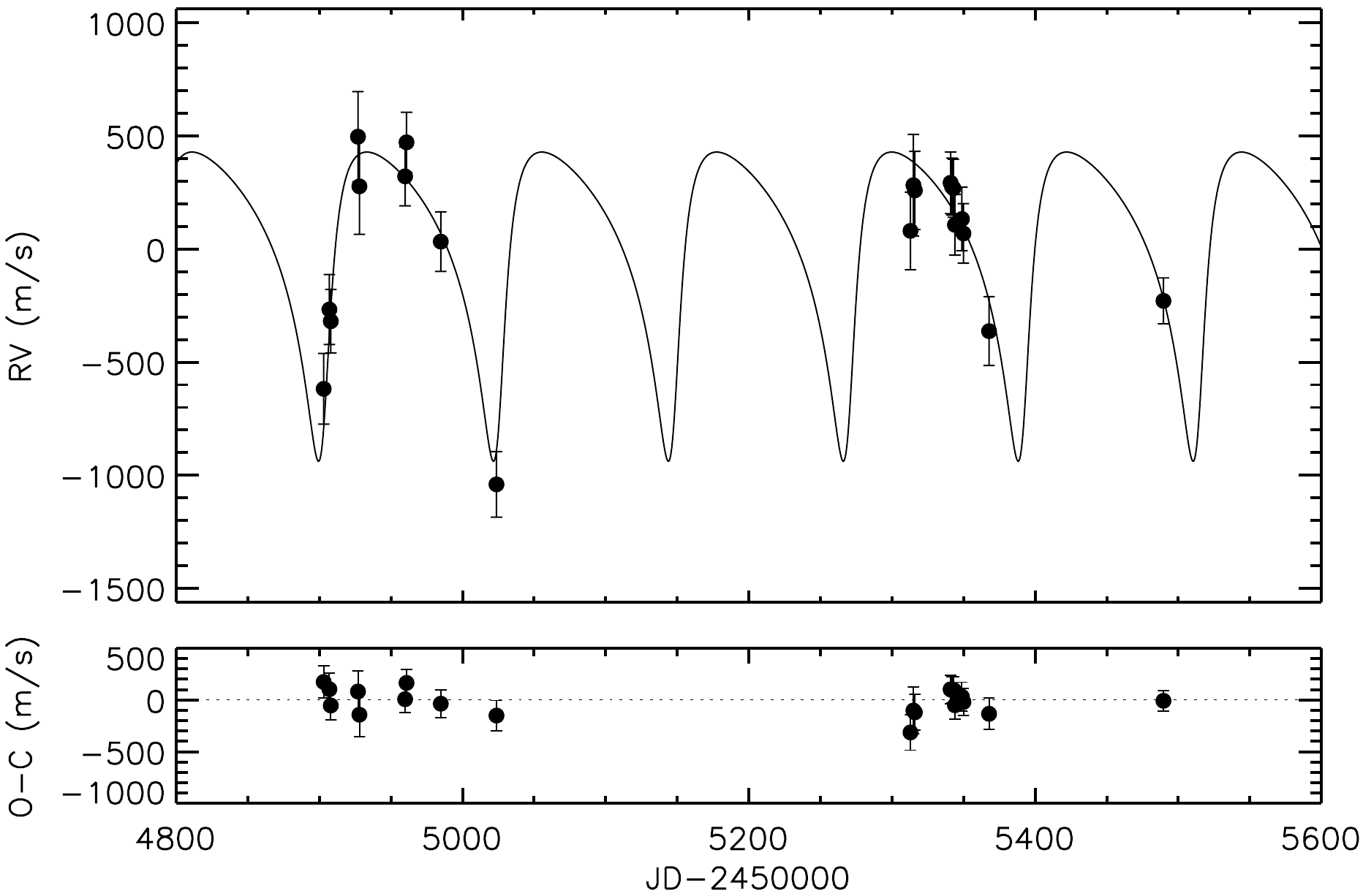}
	\hfill
	\centering
	\includegraphics[width=0.47\linewidth]{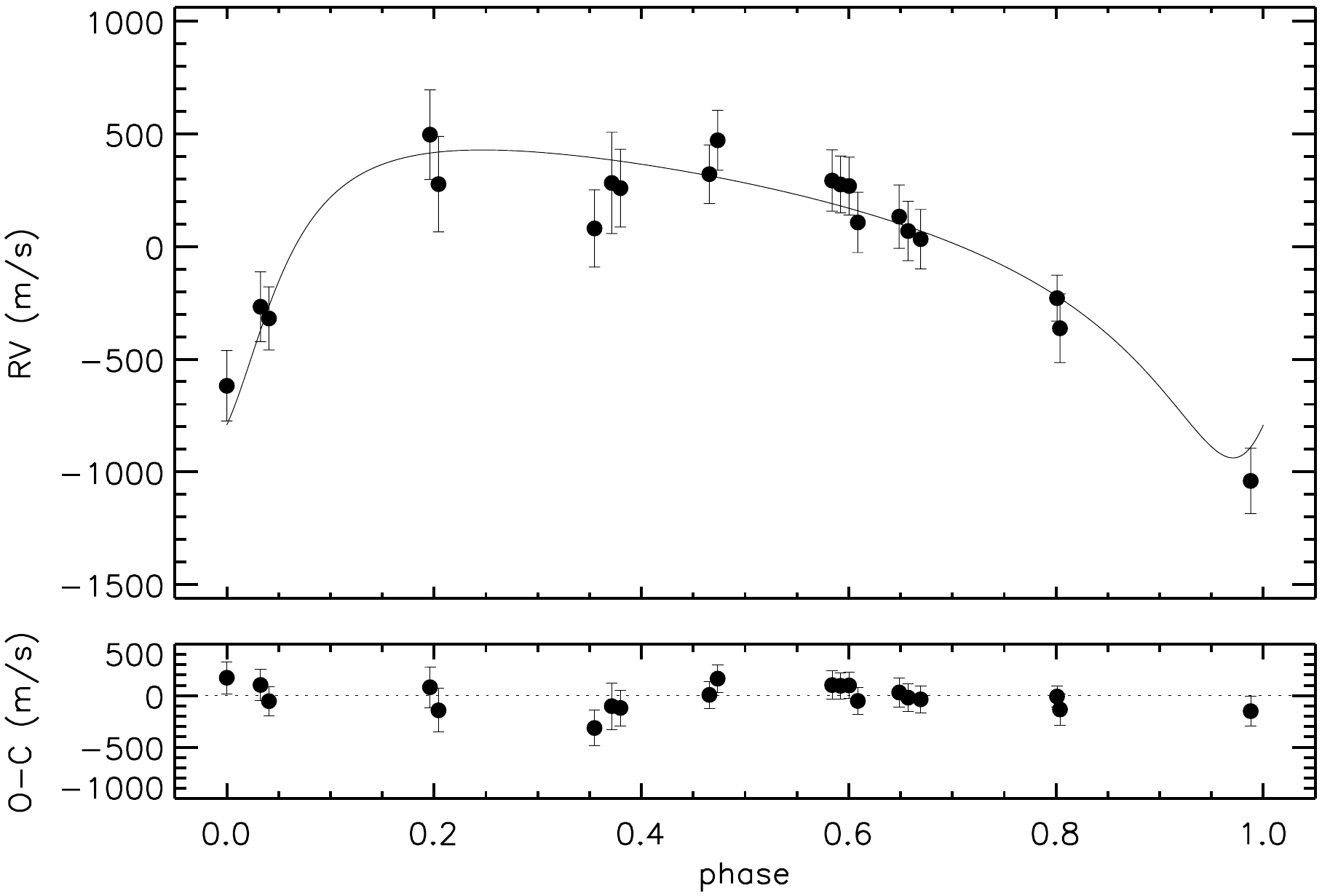}
	\hfill
	\centering
	\includegraphics[width=0.32\linewidth]{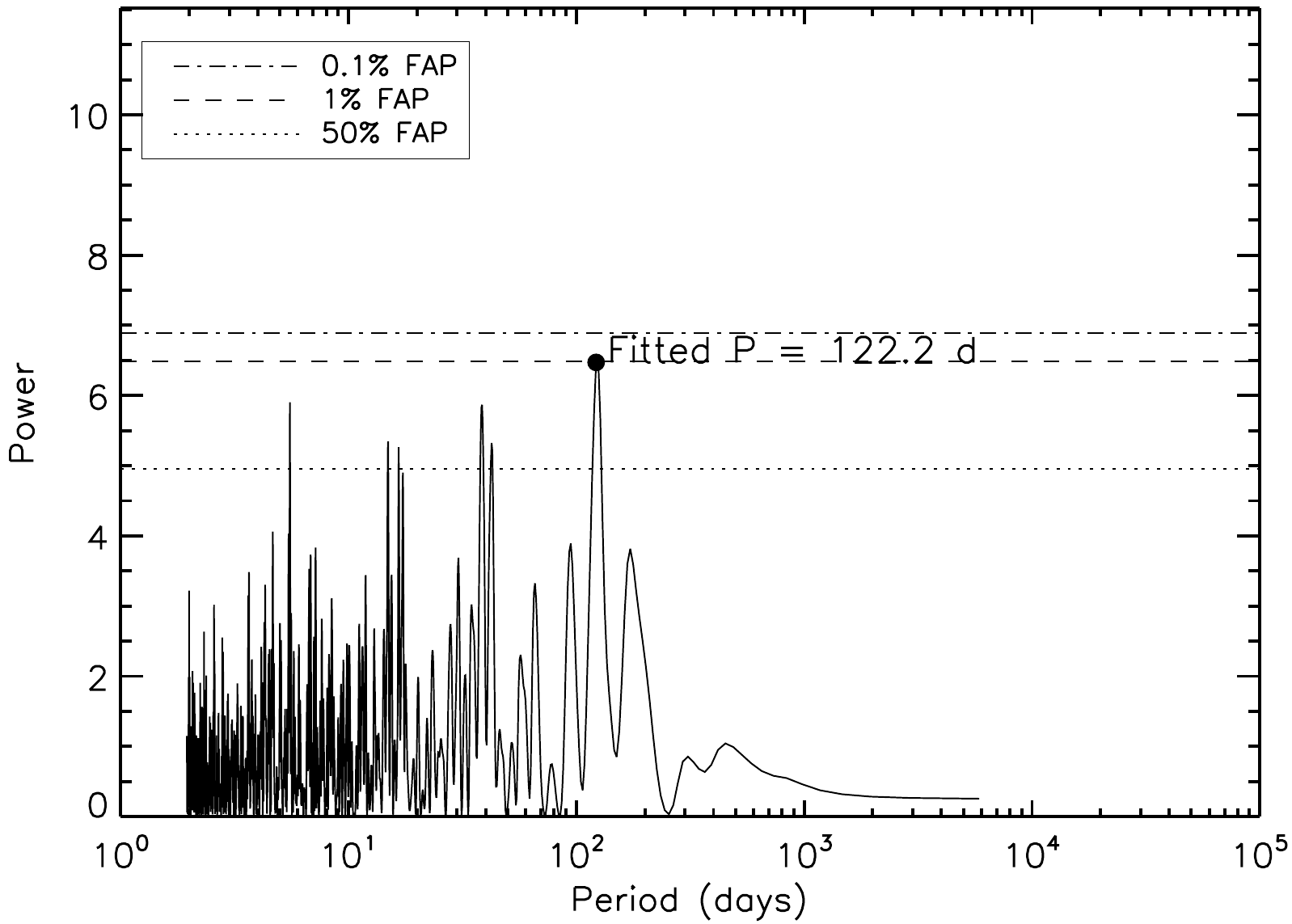}
	\hfill
	\centering
	\includegraphics[width=0.6\linewidth]{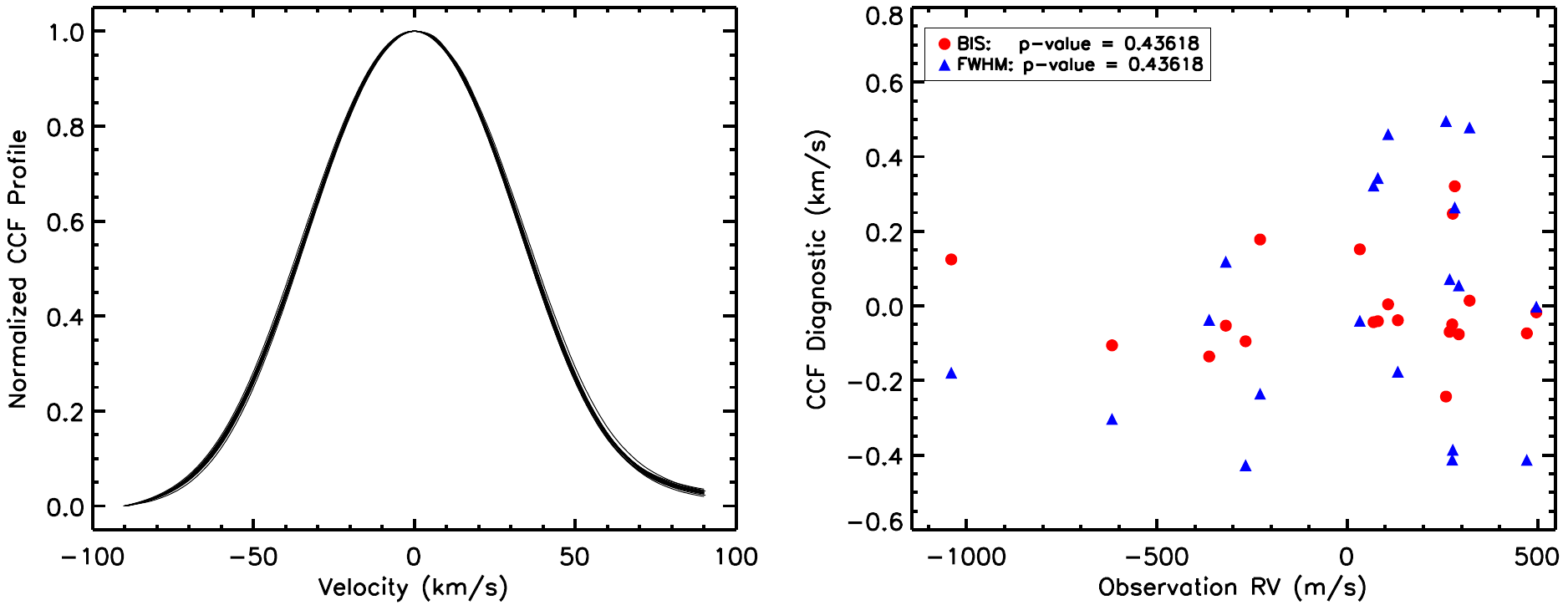}
	\hfill
\caption{\textbf{TYC 3148-02071-1, MARVELS-19b}: M sin$\textit{i}$ $\sim$17.3 M$_{\text{Jup}}$, Period $\sim$122.2 days, ecc $\sim$0.53.}
\label{fig:marv-21}
\end{figure*}


\bsp	
\label{lastpage}

\begin{thebibliography}{99}
\bibitem[Alam et al.(2015)]{Alam2015} Alam, S., Albareti, F.~D., Allende Prieto, C., et al.\ 2015, \apjs, 219, 12 
\bibitem[Boss(2001)]{Boss2001} Boss, A.~P.\ 2001, \apjl, 551, L167 
\bibitem[Boss et al.(2003)]{Boss2003} Boss, A.~P., Basri, G., Kumar, S.~S., et al.\ 2003, Brown Dwarfs, 211, 529
\bibitem[Carney et al.(2003)]{Carney2003} Carney, B.~W., Latham, D.~W., Stefanik, R.~P., Laird, J.~B., \& Morse, J.~A.\ 2003, \aj, 125, 293 
\bibitem[Chatterjee et al.(2008)]{Chatterjee2008} Chatterjee, S., Ford, E.~B., Matsumura, S., \& Rasio, F.~A.\ 2008, \apj, 686, 580-602
\bibitem[Collier Cameron et al.(2007)]{Collier2007} Collier Cameron, A., Wilson, D.~M., West, R.~G., et al.\ 2007, \mnras, 380, 1230 
\bibitem[Damiani \& D{\'{\i}}az(2016)]{DamianiDiaz2016} Damiani, C., \& D{\'{\i}}az, R.~F.\ 2016, \aap, 589, A55 
\bibitem[De Lee et al.(2013)]{DeLee2013} De Lee, N., Ge, J., Crepp, J.~R., et al.\ 2013, \aj, 145, 155
\bibitem[Duch{\^e}ne \& Kraus(2013)]{DucheneKraus2013} Duch{\^e}ne, G., \& Kraus, A.\ 2013, \araa, 51, 269  
\bibitem[Eisenstein et al.(2011)]{Eisenstein2011} Eisenstein, D.~J., Weinberg, D.~H., Agol, E., et al.\ 2011, \aj, 142, 72 
\bibitem[Fleming et al.(2010)]{Fleming2010} Fleming, S.~W., Ge, J., Mahadevan, S., et al.\ 2010, \apj, 718, 1186 
\bibitem[Fleming et al.(2011)]{Fleming2011} Fleming, S.~W., Maxted, P.~F.~L., Hebb, L., et al.\ 2011, \aj, 142, 50 
\bibitem[Fleming et al.(2012)]{Fleming2012} Fleming, S.~W., Ge, J., Barnes, R., et al.\ 2012, \aj, 144, 72 
\bibitem[Ford \& Rasio(2008)]{FordRasio2008} Ford, E.~B., \& Rasio, F.~A.\ 2008, \apj, 686, 621-636 
\bibitem[Ge(2002)]{Ge2002} Ge, J.\ 2002, \apjl, 571, L165 
\bibitem[Ge et al.(2006)]{Ge2006} Ge, J., van Eyken, J., Mahadevan, S., et al.\ 2006, \apj, 648, 683
\bibitem[Ge et al.(2008)]{Ge2008} Ge, J., Mahadevan, S., Lee, B., et al.\ 2008, Extreme Solar Systems, 398, 449 
\bibitem[Ge et al.(2009)]{Ge2009} Ge, J., Lee, B., de Lee, N., et al.\ 2009, \procspie, 7440, 74400L 
\bibitem[Ge et al.(2010)]{Ge2010} Ge, J., Zhao, B., Groot, J., et al.\ 2010, \procspie, 7735, 77350H 
\bibitem[Ghezzi et al.(2014)]{Ghezzi2014} Ghezzi, L., Dutra-Ferreira, L., Lorenzo-Oliveira, D., et al.\ 2014, \aj, 148, 105 
\bibitem[Grether \& Lineweaver(2006)]{GretherLineweaver2006} Grether, D., \& Lineweaver, C.~H.\ 2006, \apj, 640, 1051 
\bibitem[Guillot et al.(2014)]{Guillot2014} Guillot, T., Lin, D.~N.~C., Morel, P., Havel, M., \& Parmentier, V.\ 2014, EAS Publications Series, 65, 327 
\bibitem[Gunn et al.(2006)]{Gunn2006} Gunn, J.~E., Siegmund, W.~A., Mannery, E.~J., et al.\ 2006, \aj, 131, 2332 
\bibitem[Hennebelle \& Chabrier(2013)]{HennebelleChabrier2013} Hennebelle, P., \& Chabrier, G.\ 2013, \apj, 770, 150 
\bibitem[Hopkins(2012)]{Hopkins2012} Hopkins, P.~F.\ 2012, \mnras, 423, 2037
\bibitem[Ida \& Lin(2004)]{IdaLin2004} Ida, S., \& Lin, D.~N.~C.\ 2004, \apj, 616, 567 
\bibitem[Jiang et al.(2013)]{Jiang2013} Jiang, P., Ge, J., Cargile, P., et al.\ 2013, \aj, 146, 65  
\bibitem[Kumar(1962)]{Kumar1962} Kumar, S.~S.\ 1962, \aj, 67, 579 
\bibitem[Lasker et al.(2008)]{Lasker2008} Lasker, B.~M., Lattanzi, M.~G., McLean, B.~J., et al.\ 2008, \aj, 136, 735 
\bibitem[Lee et al.(2011)]{Lee2011} Lee, B.~L., Ge, J., Fleming, S.~W., et al.\ 2011, \apj, 728, 32 
\bibitem[Lomb(1976)]{Lomb1976} Lomb, N.~R.\ 1976, \apss, 39, 447 
\bibitem[Ma et al.(2013)]{Ma2013} Ma, B., Ge, J., Barnes, R., et al.\ 2013, \aj, 145, 20
\bibitem[Ma \& Ge(2014)]{MaGe2014} Ma, B., \& Ge, J.\ 2014, \mnras, 439, 2781 
\bibitem[Ma et al.(2016)]{Ma2016} Ma, B., Ge, J., Wolszczan, A., et al.\ 2016, \aj, 152, 112 
\bibitem[Mack et al.(2013)]{Mack2013} Mack, C.~E., III, Ge, J., Deshpande, R., et al.\ 2013, \aj, 145, 139 
\bibitem[Madore \& Freedman(2005)]{MadoreFreedman2005} Madore, B.~F., \& Freedman, W.~L.\ 2005, \apj, 630, 1054 
\bibitem[Majewski et al.(2015)]{Majewski2015} Majewski, S.~R., Schiavon, R.~P., Frinchaboy, P.~M., et al.\ 2015, arXiv:1509.05420 
\bibitem[Marcy \& Butler(2000)]{MarcyButler2000} Marcy, G.~W., \& Butler, R.~P.\ 2000, \pasp, 112, 137 
\bibitem[Markwardt(2009)]{Markwardt2009} Markwardt, C.~B.\ 2009, Astronomical Data Analysis Software and Systems XVIII, 411, 251 
\bibitem[Mata S{\'a}nchez et al.(2014)]{MataSanchez2014} Mata S{\'a}nchez, D., Gonz{\'a}lez Hern{\'a}ndez, J.~I., Israelian, G., et al.\ 2014, \aap, 566, A83 
\bibitem[Nakajima et al.(1995)]{Nakajima1995} Nakajima, T., Oppenheimer, B.~R., Kulkarni, S.~R., et al.\ 1995, \nat, 378, 463 
\bibitem[Paegert et al.(2015)]{Paegert2015} Paegert, M., Stassun, K.~G., De Lee, N., et al.\ 2015, \aj, 149, 186 
\bibitem[Palmer(2009)]{Palmer2009} Palmer, D.~M.\ 2009, \apj, 695, 496 
\bibitem[Patel et al.(2007)]{Patel2007} Patel, S.~G., Vogt, S.~S., Marcy, G.~W., et al.\ 2007, \apj, 665, 744 
\bibitem[Queloz et al.(2001)]{Queloz2001} Queloz, D., Henry, G.~W., Sivan, J.~P., et al.\ 2001, \aap, 379, 279 
\bibitem[Rebolo et al.(1995)]{Rebolo1995} Rebolo, R., Zapatero Osorio, M.~R., \& Mart{\'{\i}}n, E.~L.\ 1995, \nat, 377, 129 
\bibitem[Reipurth \& Clarke(2001)]{Reipurth2001} Reipurth, B., \& Clarke, C.\ 2001, \aj, 122, 432
\bibitem[Sahlmann et al.(2011)]{Sahlmann2011} Sahlmann, J., S{\'e}gransan, D., Queloz, D., et al.\ 2011, \aap, 525, A95 
\bibitem[Santerne et al.(2015)]{Santerne2015} Santerne, A., D{\'{\i}}az, R.~F., Almenara, J.-M., et al.\ 2015, \mnras, 451, 2337 
\bibitem[Santerne et al.(2016)]{Santerne2016} Santerne, A., Moutou, C., Tsantaki, M., et al.\ 2016, \aap, 587, A64 
\bibitem[Scargle(1982)]{Scargle1982} Scargle, J.~D.\ 1982, \apj, 263, 835 
\bibitem[Skrutskie et al.(2006)]{Skrutskie2006} Skrutskie, M.~F., Cutri, R.~M., Stiening, R., et al.\ 2006, \aj, 131, 1163 
\bibitem[Stamatellos et al.(2011)]{Stamatellos2011} Stamatellos, D., Maury, A., Whitworth, A., \& Andr{\'e}, P.\ 2011, \mnras, 413, 1787 
\bibitem[Stamatellos \& Whitworth(2009)]{StamatellosWhitworth2009} Stamatellos, D., \& Whitworth, A.~P.\ 2009, \mnras, 392, 413
\bibitem[Suveges(2012)]{Suveges2012} Suveges, M.\ 2012, Seventh Conference on Astronomical Data Analysis, 16 
\bibitem[Thomas (2015)]{Thomas2015UFphd} Thomas, N. \textit{Effect of metallicity on giant planet and brown dwarf formation}. PhD dissertation, University of Florida. 2015
\bibitem[Thomas et al.(2016)]{Thomas2016} Thomas, N., Ge, J., Grieves, N., Li, R., \& Sithajan, S.\ 2016, \pasp, 128, 045003 
\bibitem[Torres et al.(2010)]{Torres2010} Torres, G., Andersen, J., \& Gim{\'e}nez, A.\ 2010, \aapr, 18, 67 
\bibitem[Troup et al.(2016)]{Troup2016} Troup, N.~W., Nidever, D.~L., De Lee, N., et al.\ 2016, \aj, 151, 85
\bibitem[van Eyken et al.(2010)]{vanEyken2010} van Eyken, J.~C., Ge, J., \& Mahadevan, S.\ 2010, \apjs, 189, 156 
\bibitem[Vogt et al.(2002)]{Vogt2002} Vogt, S.~S., Butler, R.~P., Marcy, G.~W., et al.\ 2002, \apj, 568, 352  
\bibitem[Wang et al.(2011)]{Wang2011} Wang, J., Ge, J., Jiang, P., \& Zhao, B.\ 2011, \apj, 738, 132 
\bibitem[Wang, J. et al.(2012)]{Wang2012j} Wang, J., Ge, J., Wan, X., Lee, B., \& De Lee, N.\ 2012, \pasp, 124, 598 
\bibitem[Wang, X. et al.(2012)]{Wang2012x} Wang, X., Sharon, Wright, J.~T., Cochran, W., et al.\ 2012, \apj, 761, 46 
\bibitem[Wilson et al.(2016)]{Wilson2016} Wilson, P.~A., H{\'e}brard, G., Santos, N.~C., et al.\ 2016, \aap, 588, A144 
\bibitem[Wisniewski et al.(2012)]{Wisniewski2012} Wisniewski, J.~P., Ge, J., Crepp, J.~R., et al.\ 2012, \aj, 143, 107
\bibitem[Wittenmyer et al.(2009)]{Wittenmyer2009} Wittenmyer, R.~A., Endl, M., Cochran, W.~D., et al.\ 2009, \aj, 137, 3529 
\bibitem[Wright \& Howard(2012)]{WrightHoward2012} Wright, J., \& Howard, A.\ 2012, Astrophysics Source Code Library, ascl:1210.031 
\bibitem[Wright et al.(2013)]{Wright2013} Wright, J.~T., Roy, A., Mahadevan, S., et al.\ 2013, \apj, 770, 119 
\bibitem[York et al.(2000)]{York2000} York, D.~G., Adelman, J., Anderson, J.~E., Jr., et al.\ 2000, \aj, 120, 1579

\end{thebibliography}
\end{document}